\RequirePackage{ifpdf}
\ifpdf 
\documentclass[pdftex]{sigma}
\else
\documentclass{sigma}
\fi

\numberwithin{equation}{section}
\numberwithin{theorem}{section}
\numberwithin{proposition}{section}
\numberwithin{lemma}{section}
\numberwithin{corollary}{section}
\numberwithin{definition}{section}
\numberwithin{example}{section}
\numberwithin{remark}{section}
\numberwithin{note}{section}

\usepackage{xypic, xspace, enumerate}
\usepackage[mathscr]{eucal}

\newcommand{\Aut}{{\rm Aut}}

\newcommand{\Hom}{{\rm Hom}}

\newcommand{\IM}{{\rm Im}}

\newcommand{\ID}{{\rm id}}

\newcommand{\rO}{{\rm O}}

\newcommand{\SL}{{\rm SL}}

\newcommand{\Symb}{{\rm Symb}}
\newcommand{\SU}{{\rm SU}}

\newcommand{\U}{{\rm U}}

\newcommand{\A}{\mathcal A}

\newcommand{\D}{\mathcal D}
\newcommand{\E}{\mathcal E}
\newcommand{\F}{\mathcal F}
\renewcommand{\H}{\mathcal H}
\renewcommand{\cL}{\mathcal L}

\newcommand{\R}{\mathcal R}

\newcommand{\bC}{\mathbb{C}}

\newcommand{\bZ}{\mathbb{Z}}

\newcommand{\vp}{\varphi}
\newcommand{\vep}{\varepsilon}

\newcommand{\grp}{\mathcal G}

\newcommand{\ob}{{\rm Ob}}
\newcommand{\obg}{{\rm Ob(\mathsf{G)}}}

\newcommand{\wti}{\widetilde}
\newcommand{\what}{\widehat}

\renewcommand{\a}{\alpha}
\newcommand{\be}{\beta}

\newcommand{\del}{\partial}

\newcommand{\lra}{{\longrightarrow}}

\newcommand{\ovsetl}[1]{\overset {#1}{\lra}}

\newcommand{\midsqn}[1]{\ar@{}[dr]|{#1}}
\newcommand{\quadr}[4]
{\begin{pmatrix} & #1& \\[-1.1ex] #2 & & #3\\[-1.1ex]& #4&
 \end{pmatrix}}
\def\D{\mathsf{D}}
\def\GG{\Gamma}

\begin{document}

\allowdisplaybreaks

\renewcommand{\PaperNumber}{051}

\FirstPageHeading

\ShortArticleName{Algebraic Topology Foundations of Supersymmetry and Symmetry Breaking}

\ArticleName{Algebraic Topology Foundations of Supersymmetry\\
and Symmetry Breaking in
Quantum Field Theory \\ and Quantum Gravity: A Review}

\Author{Ion C. BAIANU~$^{\dag}$, James F. GLAZEBROOK~$^{\ddag}$ and Ronald BROWN~$^{\S}$}

\AuthorNameForHeading{I.C. Baianu, J.F. Glazebrook and R. Brown}

\Address{$^{\dag}$~FSHN and NPRE Departments, University of Illinois at Urbana--Champaign,\\
\hphantom{$^{\dag}$}~AFC-NMR {\rm \&} FT-NIR Microspectroscopy Facility,
  Urbana IL 61801 USA}
\EmailD{\href{mailto:ibaianu@illinois.edu}{ibaianu@illinois.edu}, \href{mailto:ibaianu@uiuc.edu}{ibaianu@uiuc.edu}}

\Address{$^{\dag}$~Department of Mathematics and Computer Science,
Eastern Illinois University, \\
\hphantom{$^{\dag}$}~600 Lincoln Avenue, Charleston IL 61920 USA\\[2mm]
\hphantom{$^{\dag}$}~Department of Mathematics, University
of Illinois at Urbana--Champaign,\\
\hphantom{$^{\dag}$}~1409 West Green Street, Urbana IL 61801 USA}
\EmailD{\href{mailto:jfglazebrook@eiu.edu}{jfglazebrook@eiu.edu}}

\Address{$^{\S}$~School of Computer Science,
University of Bangor,\\
\hphantom{$^{\S}$}~Dean Street, Bangor Gwynedd LL57 1UT UK}
\EmailD{\href{mailto:r.brown@bangor.ac.uk}{r.brown@bangor.ac.uk}}

\ArticleDates{Received November 25, 2008, in f\/inal form April 09,
2009; Published online April 23, 2009}

\Abstract{A novel algebraic topology approach to supersymmetry (SUSY) and
sym\-met\-ry breaking in quantum f\/ield and quantum gravity theories is
presented with a view to deve\-loping a wide range of physical
applications. These include: controlled nuclear fusion and other nuclear
reaction studies in quantum chromodynamics, nonlinear physics at
high energy densities, dynamic Jahn--Teller ef\/fects, superf\/luidity,
high temperature superconductors, multiple scattering by molecular
systems, molecular or atomic paracrystal structures, nanomaterials,
ferromagnetism in glassy materials, spin glasses, quantum phase
transitions and supergra\-vi\-ty. This approach requires a
unif\/ied conceptual framework that utilizes extended symmetries and
quantum groupoid, algebroid and functorial representations of
non-Abelian higher dimensional structures pertinent to quantized
spacetime topology and state space geometry of quantum operator
algebras. Fourier transforms, generalized Fourier--Stieltjes transforms,
and duality relations link, respectively, the quantum groups
 and quantum groupoids with their dual algebraic structures; quantum double constructions are also discussed
in this context in relation to quasi-triangular, quasi-Hopf algebras,
bialgebroids, Grassmann--Hopf algebras and higher dimensional algebra. On the one hand, this
quantum algebraic approach is known to provide solutions to the
quantum  Yang--Baxter equation. On the other hand, our novel approach to extended quantum symmetries
and their associated representations is shown to be relevant to locally covariant general relativity
theories that are consistent with either nonlocal quantum f\/ield theories or local bosonic (spin) models
with the extended quantum symmetry of entangled, `string-net condensed' (ground) states.}

\newpage


\Classification{81R40; 16W30; 22A22;  81R50; 43A05;  46L10}

{ \small
\tableofcontents

}

\renewcommand{\baselinestretch}{1.0}

\section{Introduction}\label{section1}

The theory of scattering by partially ordered, atomic or molecular,
structures in terms of \emph{paracrystals} and \emph{lattice convolutions}
was formulated by Hosemann and Bagchi in  \cite{Hosemann-Bagchi62}
using basic techniques of Fourier analysis and convolution products.
A natural generalization of such molecular, partial symmetries and
their corresponding analytical versions involves convolution
algebras -- a functional/distribution \cite{Schwartz45,Schwartz52} based theory that
 we will discuss in the context of a more general and original concept of
a \emph{convolution-algebroid of an extended symmetry
groupoid of a paracrystal}, of any molecular or nuclear system, or indeed, any
quantum system in general, including quantum f\/ields and local quantum net conf\/igurations that are endowed
with either partially disordered or `completely' ordered structures. Further specif\/ic applications
of the paracrystal theory to $X$-ray scattering, based on computer algorithms, programs and explicit numerical computations, were subsequently developed by the f\/irst author \cite{Baianu74} for one-dimensional paracrystals, partially ordered membrane lattices \cite{Baianu78} and other biological
structures with partial structural disorder \cite{Baianu80}. Such biological structures, `quasi-crystals', and the paracrystals, in general, provide rather interesting physical examples of such
extended symmetries (cf.~\cite{Hindeleh-Hosemann88}).

Further statistical analysis linked to structural symmetry and
scattering theory considerations shows that a real paracrystal
can be def\/ined by a three dimensional convolution polynomial
with a semi-empirically derived composition law, $*$,
\cite{Hosemann-etal81}. As was shown in \cite{Baianu74,Baianu78}~-- supported
with computed specif\/ic examples -- several systems of convolution
can be expressed analytically, thus allowing the numerical computation
of $X$-ray, or neutron, scattering by partially disordered layer lattices
via complex Fourier transforms of one-dimensional structural models
using fast digital computers.  The range of paracrystal theory applications is however much wider than the one-dimensional lattices with disorder, thus spanning very diverse non-crystalline systems, from metallic glasses and spin glasses to superf\/luids, high-temperature superconductors, and extremely hot anisotropic plasmas such as those encountered in controlled nuclear fusion (for example, JET) experiments. Other applications~-- as previously suggested in \cite{Baianu71}~-- may also include novel designs of `fuzzy' quantum machines and quantum computers with extended symmetries of quantum state spaces.

\subsection[Convolution product of groupoids and the convolution algebra of functions]{Convolution product of groupoids and the convolution algebra\\ of functions}\label{section1.1}

 A salient, and well-fathomed concept from the mathematical
perspective concerns that of a~$C^{\ast}$-algebra
of a (discrete) group (see, e.g.,~\cite{Connes94}). The
underlying vector space is that of complex valued functions
with f\/inite support, and the multiplication of the algebra is the fundamental
\emph{convolution product} which it is convenient for our
purposes to write slightly dif\/ferently from the common formula as
\begin{gather*}
(f \ast g )(z) = \sum_{xy=z} f(x)g(y),
\end{gather*}
and $\ast$-operation
\begin{gather*}
f^{\ast}(x)= \overline{f(x^{-1})} .
\end{gather*}

The more usual expression of these formulas has a sum over the
elements of the group. For topological groups, where
the underlying vector space consists of continuous complex valued
functions, this product requires the availability of some structure
of measure and of measurable functions, with the sum replaced by an
integral. Notice also that this algebra has an identity,
the distribution function~$\delta_1$, which has value~1
on the identity~1 of the group, and has zero value elsewhere.

Given this convolution/distribution representation that combines
crystalline (`perfect' or global-group, and/or group-like symmetries)
with partial symmetries of paracrystals and glassy
solids on the one hand, and also with non-commutative harmonic
analysis \cite{Mackey92} on the other hand, we propose that several
extended quantum symmetries can be represented algebraically in
terms of certain structured \emph{groupoids}, their
\emph{$C^*$-convolution quantum algebroids}, paragroup/\emph{quan\-tized groups} and/or other more general
mathematical structures that will be introduced in this
report. It is already known that such extensions to groupoid
and algebroid/coalgebroid symmetries require also a generalization of non-commutative
harmonic analysis which involves certain Haar measures, generalized Fourier--Stieltjes transforms and certain categorical duality relationships representing very general mathematical symmetries as well. Proceeding from the abstract structures endowed with extended symmetries to numerical applications in
quantum physics always involves representations through specif\/ication
 of concrete elements, objects and transformations.
Thus, groupoid and functorial representations that generalize
group representations in several, meaningful ways are key to
linking abstract, quantum operator algebras and symmetry
properties with actual numerical computations of quantum
eigenvalues and their eigenstates, as well as a wide variety of
numerical factors involved in computing quantum dynamics. The
well-known connection between groupoid convolution representations
and matrices~\cite{Weinstein96} is only one of the several
numerical computations made possible via groupoid
representations. A very promising approach to nonlinear (anharmonic) analysis of aperiodic quantum systems represented by rigged Hilbert space bundles may involve the computation of
representation coef\/f\/icients of Fourier--Stieltjes groupoid transforms that we will also discuss brief\/ly in Section~\ref{section7}.

Currently, however, there are important aspects of quantum
dynamics left out of the invariant, simplif\/ied picture provided by
group symmetries and their corresponding representations of
quantum operator algebras~\cite{Gilmore2k5}. An alternative approach proposed in
\cite{Harrison2k5} employs
dif\/ferential forms to f\/ind symmetries.

\looseness=-1
Often physicists deal with such problems in terms of either spontaneous symmetry
breaking or approximate symmetries that require underlying
explanations or ad-hoc dynamic restrictions that are
semi-empirical. A well-studied example of this kind is that of
the dynamic Jahn--Teller ef\/fect and the corresponding `theorem'
(Chapter~21 on pp.~807--831, as well as p.~735 of
\cite{Abragam-Bleaney70}) which in its
simplest form stipulates that a quantum state
with electronic non-Kramers degeneracy may be unstable against
small distortions of the surroundings, that would lower the
symmetry of the crystal f\/ield and thus lift the degeneracy (i.e.,
cause observable splitting of the corresponding energy levels);
this ef\/fect occurs in certain paramagnetic ion systems \textit{via} dynamic distortions of the crystal f\/ield symmetries around
paramagnetic or high-spin centers by moving ligands that are
diamagnetic. The established physical explanation is that the
Jahn--Teller coupling replaces a purely electronic degeneracy by a
vibronic degeneracy (of \textit{exactly the same} symmetry!).  The
dynamic, or spontaneous breaking of crystal f\/ield symmetry (for
example, distortions of the octahedral or cubic symmetry) results
in certain systems in the appearance of doublets of symmetry
$\gamma_3$ or singlets of symmetry $\gamma_1$ or $\gamma_2$. Such
dynamic systems could be locally expressed in terms of symmetry
representations of a Lie algebroid, or globally in terms of a
special Lie (or Lie--Weinstein) symmetry groupoid representations
that can also take into account the spin exchange interactions
between the Jahn--Teller centers exhibiting such quantum dynamic
ef\/fects. Unlike the simple symmetries expressed by group
representations, the latter can accommodate a much wider range of
possible or approximate symmetries that are indeed characteristic
of real, molecular systems with varying crystal f\/ield symmetry, as
for example around certain transition ions dynamically bound to
ligands in liquids where motional narrowing becomes very
important.  This well known example illustrates the importance of
the interplay between symmetry and dynamics in quantum processes
which is undoubtedly involved in many other instances including:
\emph{quantum chromodynamics, superfluidity, spontaneous symmetry
breaking, quantum gravity and Universe dynamics} (i.e., the inf\/lationary Universe).

Therefore, the various interactions and interplay between the
symmetries of quantum operator state space geometry and quantum
dynamics at various levels leads to both algebraic and topological
structures that are variable and complex, well beyond  symmetry
groups and well-studied group algebras (such as Lie algebras, see
for example~\cite{Gilmore2k5}). A unif\/ied treatment of quantum
phenomena/dynamics and structures may thus become possible with the
help of algebraic topology, non-Abelian treatments; such powerful
mathematical tools are capable of revealing novel, fundamental
aspects related to extended symmetries and quantum dynamics through
a detailed analysis of the variable geometry of (quantum) operator
algebra state spaces. At the center stage of non-Abelian algebraic
topology are groupoid and algebroid structures with their internal
and external symmetries \cite{Weinstein96} that allow one to treat
physical spacetime structures and dynamics within an unif\/ied
categorical, higher dimensional algebra frame\-work~\cite{Brown-etal2k7}.
As already suggested in our previous report, the interplay
between extended symmetries and dynamics generates higher dimensional structures of
quantized spacetimes that exhibit novel properties not found in lower dimensional representations
of groups, group algebras or Abelian groupoids.

It is also our intention here to explore, uncover, and then develop, new links
between several important but seemingly distinct mathematical approaches to extended quantum symmetries
that were not considered in previous reports.

\section[Quantum groups, quantum operator algebras, Ocneanu paragroups,\\ quantum groupoids
and related symmetries]{Quantum groups, quantum operator algebras, Ocneanu\\ paragroups, quantum groupoids
and related symmetries}\label{section2}

Quantum theories adopted a new lease of life post 1955 when von
Neumann beautifully re-formulated quantum mechanics (QM) in the
mathematically rigorous context of Hilbert spaces and operator
algebras. From a current physics perspective, von Neumann's
approach to quantum mechanics has done however much more: it has
not only paved the way to expanding the role of symmetry in
physics, as for example with the Wigner--Eckhart theorem and its
applications, but also revealed the fundamental importance in
quantum physics of the state space geometry of (quantum) operator
algebras.

The basic def\/inition of von Neumann and Hopf algebras (see for
example \cite{Majid95}), as well as the topological groupoid
def\/inition, are recalled in the Appendix to maintain a
self-contained presentation. Subsequent developments of the quantum
operator algebra were aimed at identifying more general quantum
symmetries than those def\/ined for example by symmetry groups, groups
of unitary operators and Lie groups, thus leading to the development
of theories based on various quantum groups \cite{Doebner-Hennig89}.
Several, related quantum algebraic concepts were also fruitfully developed, such as:
the Ocneanu \textit{paragroups}-later found to be represented by Kac--Moody
algebras, quantum groups represented either as Hopf algebras or
locally compact groups with Haar measure, `quantum' groupoids
represented as weak Hopf algebras, and so on. The Ocneanu paragroups
case is particularly interesting as it can be considered as an
extension through quantization of certain f\/inite group symmetries to
inf\/initely-dimensional von Neumann type $II_1$ algebras, and are, in
ef\/fect, \textit{quantized groups} that can be nicely constructed
as Kac algebras; in fact, it was recently shown that a paragroup can
be constructed from a crossed product by an outer action of a Kac(--Moody)
algebra. This suggests a relation to categorical aspects of
paragroups (rigid monoidal tensor categories \cite{Turaev-Viro92,Yetter93}). The
strict symmetry of the group of (quantum) unitary operators is thus
naturally extended through paragroups to the symmetry of the latter
structure's unitary representations; furthermore, if a subfactor of
the von Neumann algebra arises as a crossed product by a f\/inite
group action, the paragroup for this subfactor contains a very
similar group structure to that of the original f\/inite group, and
also has a unitary representation theory similar to that of the
original f\/inite group. Last-but-not least, a paragroup yields a
\emph{complete invariant} for irreducible inclusions of AFD von
Neumannn $II_1$ factors with f\/inite index and f\/inite depth (Theorem~2.6 of \cite{Sato-Wakui2k1}). This can be considered as a kind of internal,
`hidden' quantum symmetry of von Neumann algebras.

On the other hand, unlike paragroups, quantum locally compact
groups are not readily constructed as either Kac or Hopf
$C^*$-algebras. In recent years the techniques of Hopf symmetry and
those of weak Hopf $C^*$-algebras, sometimes called \emph{quantum
groupoids} (cf.\ B\"ohm et al.~\cite{Bohm-etal99}),
 provide important tools~-- in addition to the paragroups~--
 for studying the broader relationships of the
 Wigner fusion rules algebra, $6j$-symmetry~\cite{Rehren97},
 as well as the study of the noncommutative
 symmetries of subfactors within the Jones tower
 constructed from f\/inite index depth~2 inclusion of factors,
 also recently considered from the viewpoint of related
 Galois correspondences~\cite{Nikshych-Vainerman2k}.

We shall proceed at f\/irst by pursuing the relationships
between these mainly algebraic concepts and their extended
quantum symmetries, also including relevant computation examples;
then we shall consider several further extensions of symmetry
and algebraic topology in the context of local
 quantum physics/algebraic quantum f\/ield theory,
 symmetry breaking, quantum chromodynamics and the
 development of novel supersymmetry theories of quantum gravity.
 In this respect one can also take spacetime `inhomogeneity' as a
 criterion for the comparisons between physical, partial or local,
 symmetries: on the one hand, the example of paracrystals
reveals thermodynamic disorder (entropy) within its own spacetime
framework, whereas in spacetime itself, whatever the selected
model, the inhomogeneity arises through (super) gravitational
ef\/fects. More specif\/ically, in the former case one has the
technique of the generalized Fourier--Stieltjes transform (along
with convolution and Haar measure), and in view of the latter, we
may compare the resulting `broken'/paracrystal-type symmetry with
that of the supersymmetry predictions for weak gravitational
f\/ields (e.g., `ghost' particles) along with the broken
supersymmetry in the presence of intense gravitational f\/ields.
Another signif\/icant extension of quantum symmetries may result
from the superoperator algebra/algebroids of Prigogine's quantum
\textit{superoperators} which are def\/ined only for irreversible,
inf\/inite-dimensional systems~\cite{Prigogine80}.

\subsection[Solving quantum problems by algebraic methods: applications to
 molecular structure,\\ quantum chemistry and quantum theories]{Solving quantum problems by algebraic methods: applications\\ to
 molecular structure, quantum chemistry and quantum theories}\label{section2.1}

As already discussed in the Introduction, one often deals with continuity and continuous
transformations in natural systems, be they physical, chemical or self-organizing. Such
continuous `symmetries' often have a special type of underlying continuous group, called a
\emph{Lie group}. Brief\/ly, a \emph{Lie group} $G$ is generally
considered having a (smooth) $C^\infty$ manifold structure, and
acts upon itself smoothly. Such a \emph{globally smooth }structure
is surprisingly simple in two ways: it always admits an
\emph{Abelian fundamental group}, and seemingly also related to
this global property, it admits an associated, \emph{unique}~-- as
well as \emph{finite}~-- \emph{Lie algebra} that completely
specif\/ies \emph{locally} the properties of the Lie group
everywhere.

\subsubsection[The f\/inite Lie algebra of quantum commutators
and their unique (continuous)\\ Lie groups]{The f\/inite Lie algebra of quantum commutators\\
and their unique (continuous) Lie groups}\label{section2.1.1}

Lie algebras can greatly simplify quantum computations and the
initial problem of def\/ining the form and symmetry of the quantum
Hamiltonian subject to boundary and initial conditions in the
quantum system under consideration. However, unlike most regular
abstract algebras, a Lie algebra is \emph{not associative}, and it
is in fact a \emph{vector space} \cite{Heynman-Lifschitz58}. It
is also perhaps this feature that makes the Lie algebras somewhat
compatible, or `consistent', with quantum logics that are also
thought to have \emph{non-associative, non-distributive and
non-commutative lattice} structures. Nevertheless, the need for
`quantizing' Lie algebras in the sense of a certain \emph{non-commutative} `deformation' apparently remains for a quantum
system, especially if one starts with a `classical' \emph{Poisson
}algebra \cite{Landsman-Ramazan2k1}. This requirement remains apparently
even for the generalized version of a Lie algebra, called a
\emph{Lie algebroid} (see its def\/inition and related remarks in
Sections~\ref{section4} and~\ref{section5}).

The presence of Lie groups in many classical physics problems, in
view of its essential \emph{continuity }property and its
\emph{Abelian} fundamental group, is not surprising. However, what is
surprising in the beginning, is the appearance of
\emph{Lie groups and Lie algebras} in the context of commutators
of observable operators even in quantum systems \emph{with no
classical analogue observables such as the spin}, as~-- for
example~-- the $\SU(2)$ and its corresponding, unique
$\mathfrak{su}(2)$  algebra.

As a result of quantization, one would expect to deal with an
algebra such as the Hopf (quantum group) which is
\emph{associative}. On the other hand, the application of the
correspondence principle to the simple, classical harmonic
oscillator system leads to a quantized harmonic oscillator and
remarkably simple \emph{commutator} algebraic expressions, which
correspond precisely to the def\/inition of a Lie algebra.
Furthermore, this (\emph{Lie}) algebraic procedure of assembling
the quantum Hamiltonian from simple observable operator
commutators is readily extended to \emph{coupled, quantum harmonic
oscillators}, as shown in great detail by Fernandez and Castro in~\cite{Fernandez-Castro96}.

\subsection{Some basic examples}\label{section2.2}

\begin{example}[The Lie algebra of a quantum harmonic oscillator] \label{harmonic}
Here one aims to solve the time-independent Schr\"odinger
equations of motion in order to
determine the stationary states of the quantum harmonic
oscillator which has a quantum
Hamiltonian of the form:
\begin{gather*}
\mathbf{H} = \left(\frac {1}{2m}\right)\cdot P^2 + \frac{k}{2}\cdot X^2,
\end{gather*}
where $X$ and $P$ denote, respectively, the coordinate and
conjugate momentum operators. The terms $X$ and $P$
satisfy the Heisenberg commutation/uncertainty relations $[X,P]
= \iota \hbar I$, where the identity operator $I$
is employed to simplify notation.
 A simpler, equivalent form of the above Hamiltonian
is obtained by def\/ining physically dimensionless coordinate and momentum:
\begin{gather*}
\mathbf{x} = \left(\frac{X}{\alpha}\right), \qquad \mathbf{p}= \left(\frac{\alpha
P}{\hbar}\right) \qquad \text{and}\qquad \alpha = \sqrt {\frac{\hbar}{mk}}.
\end{gather*}
With these new dimensionless operators, $\mathbf{x}$ and
$\mathbf{p}$, the quantum Hamiltonian takes the form:
\begin{gather*}
\mathbf{H}= \left(\frac{\hbar \omega}{2}\right)\cdot \big(\mathbf{p}^2 +
\mathbf{x}^2\big),
\end{gather*}
which in units of $\hbar \cdot \omega$ is simply:
\begin{gather*}
\mathbf{H}' = \frac {1}{2} \big(\mathbf{p}^2 +
\mathbf{x}^2\big).
\end{gather*}
The commutator of $\mathbf{x}$ with its conjugate operator
$\mathbf{p}$ is simply $[\mathbf{x}, \mathbf{p}] = \iota$.

Next one def\/ines the superoperators $S_{Hx} = [H, x] = -\iota \cdot
p$, and $S_{Hp} = [H, p] = \iota \cdot \mathbf{x}$ that will lead to
new operators that act as generators of a Lie algebra for this
quantum harmonic oscillator. The eigenvectors $Z$ of these
superoperators are obtained by solving the equation $S_H \cdot Z =
\zeta Z$, where $\zeta$ are the eigenvalues, and $Z$ can be
written as $(c_1 \cdot x + c_2 \cdot p)$. The solutions are:
\begin{gather*}
\zeta = \pm 1 , \qquad \text{and} \qquad c_2 = \mp \iota  \cdot c_1.
\end{gather*}
Therefore, the two eigenvectors of $S_H$ can be written as:
\begin{gather*}
a^\dagger = c_1* (x- \iota p)\qquad  \text{and} \qquad a = c_1 (x+ \iota p),
\end{gather*}
respectively for $\zeta = \pm 1$. For $c_1 =\sqrt{2}$ one
obtains normalized operators $H$, $a$ and $a^\dagger$ that generate
a $4$-dimensional Lie algebra with commutators:
\begin{gather*}
[H,a] = -a, \qquad [H, a^\dagger ]= a^\dagger \qquad \text{and} \qquad [a,
a^\dagger]= I .
\end{gather*}
The term ${a}$ is called the \emph{annihilation} operator
and the term $a^\dagger$ is called the \emph{creation} operator.
This Lie algebra is solvable and generates after repeated
application of $a^\dagger$ all of the eigenvectors of the quantum
harmonic oscillator:
\begin{gather*}
\Phi_n = \frac{(a^\dagger)^n}{\sqrt{(n!)}} \Phi_0 .
\end{gather*}
The corresponding, possible eigenvalues for the energy, derived
then as solutions of the Schr\"o\-din\-ger equations for the quantum
harmonic oscillator are:
\begin{gather*}
E_n = \hbar \omega \left(n+ \tfrac{1}{2}\right) , \qquad \text{where} \qquad n =
0,1, \ldots, N.
\end{gather*}
The position and momentum eigenvector coordinates can
 be then also computed by iteration from
(\emph{finite}) matrix representations of the (\emph{finite}) Lie
algebra, using perhaps a simple computer programme to
calculate linear expressions of the annihilation and creation
operators. For example, one can show analytically that:
\begin{gather*}
[a, x^k] =  \frac{k}{\sqrt{2}} \, x_{k-1}.
\end{gather*}
One can also show by introducing a \emph{coordinate}
representation that the eigenvectors of
the harmonic oscillator can be expressed
 as \emph{Hermite polynomials} in terms of the
coordinates. In the coordinate representation the quantum
\emph{Hamiltonian} and \emph{bosonic} operators have,
respectively, the simple expressions:
\begin{gather*}
H = \frac{1}{2}\left[-\frac{d^2}{dx^2} + x^2\right], \qquad
a =\frac{1}{\sqrt{2}} \left(x + \frac{d}{dx}\right),\qquad  
a^\dagger = \frac{1}{\sqrt{2}} \left(x - \frac{d}{dx}\right).
\end{gather*}
The ground state eigenfunction normalized to unity is
 obtained from solving the simple
f\/irst-order dif\/ferential equation $a\Phi_0 (x) = 0$,
thus leading to the expression:
\begin{gather*}
\Phi_0 (x)= \pi^{-\frac{1}{4}} \exp\left(-\frac{x^2}{2}\right).
\end{gather*}
By repeated application of the creation operator written as
\begin{gather*}
a\dagger = -\frac{1}{\sqrt{2}}
\exp\left(\frac{x^2}{2}\right) \frac{d}{dx^2}
\exp\left(-\frac{x^2}{2}\right),
\end{gather*}
one obtains the $n$-th level eigenfunction:
\begin{gather*}
\Phi_n(x) = \frac{1}{\sqrt{\pi} 2^n n!}
\mathbf{He}_n (x),
\end{gather*}
where $\mathbf{He}_n(x)$ is \emph{the Hermite polynomial} of order
$n$. With the special generating function of the Hermite
polynomials
\begin{gather*}
F(t,x) = \pi^{-\frac{1}{4}} \left(\exp\left(-\frac{x^2}{2}\right) + tx
-\frac{t^2}{4}\right),
\end{gather*}
one obtains explicit analytical relations between the
 eigenfunctions of the quantum harmonic
oscillator and the above special generating
function:
\begin{gather*}
F(t,x) = \sum_{n=0} \frac{t^n}{\sqrt{2^n   n!}}
\Phi_n(x).
\end{gather*}
Such applications of the Lie algebra, and the related algebra of the
\emph{bosonic }operators as def\/ined above are quite numerous in
theoretical physics, and especially for various quantum f\/ield
carriers in QFT that are all \emph{bosons} (note also additional
examples of special Lie superalgebras for gravitational and other
f\/ields in Section~\ref{section6}, such as gravitons and Goldstone quanta that are
all \emph{bosons} of dif\/ferent spin values and `Penrose
homogeneity').

In the interesting case of a \emph{two-mode} bosonic quantum
system formed by the tensor (direct) product of \emph{one-mode}
bosonic states:
\begin{gather*}
\mid m,n\rangle  := \mid m \rangle \, \otimes \mid n\rangle ,
\end{gather*}
one can generate a $3$-dimensional Lie algebra in terms of
\emph{Casimir} operators. \emph{Finite}-dimensional Lie algebras
are far more tractable and easier to compute than those with an
inf\/inite basis set. For example, such a Lie algebra as the
$3$-dimensional one considered above for the two-mode, bosonic
states is quite useful for numerical computations of vibrational
(IR, Raman, etc.) spectra of two-mode, \emph{diatomic} molecules,
as well as the computation of scattering states. Other perturbative
calculations for more complex quantum systems, as well as
calculations of exact solutions by means of Lie algebras have also
been developed (see e.g.~\cite{Fernandez-Castro96}).
\end{example}

\begin{example}[The $\boldsymbol{\SU(2)}$ quantum group]\label{quantumgroup}
Let us consider the structure of the ubiquitous quantum group $\SU(2)$
\cite{Woronowicz1, Chaician-Demichev96}. Here $A$ is
taken to be a $C^*$-algebra generated by elements $\a$ and $\beta$
subject to the relations:
\begin{gather*}
\a \a^* + \mu^2 \beta \beta^*  = 1 ,\qquad  \a^* \a + \beta^* \beta =
1 , \nonumber\\
\beta \beta^* = \beta^* \beta ,\qquad \a \beta  = \mu \beta
\a , \qquad \a \beta^* = \mu \beta^* \a ,     \\ 
\a^* \beta = \mu^{-1} \beta \a^* ,\qquad \a^* \beta^*  = \mu^{-1} \beta^* \a^*,\nonumber
\end{gather*}
where $\mu \in [-1, 1]\backslash \{0\}$. In terms of the matrix
\begin{gather*}
u = \bmatrix \a & - \mu \beta^*  \\ \beta & \a^*
\endbmatrix
\end{gather*}
the coproduct $\Delta$ is then given via
\begin{gather*}
\Delta (u_{ij}) = \sum_k u_{ik} \otimes u_{kj}.
\end{gather*}
\end{example}

As will be shown in our later sections, such quantum groups and their associated algebras need be extended
to more general structures that involve \textit{supersymmetry}, as for example in the case of quantum gravity or \textit{supergravity} and superf\/ield theories. Another important example of such \textit{quantum supergroups} involves Drinfel'd 's quantum double construction and the $R$-matrix (e.g., as developed in~\cite{Khoroshkin-Tolstoy91} and subsequent reports related to quantum quasi-algebras \cite{Altintas-Arika2k8,Zhang91,Zhang-Gould99}).

Numerous quantum supergroup examples also emerge in the cases presented in the next subsection of either molecular groups of spins or nuclear quasi-particles coupled, respectively, by either dipolar (magnetic) or colour-charge and dipolar interactions. In such cases, the simple Lie algebras considered above in the f\/irst example need to be extended to Lie \textit{superalgebras} exhibiting supersymmetry that includes both fermionic and bosonic symmetries (as explained in Section~\ref{section6.2}). Furthermore~-- as discussed next~-- local bosonic models (or \textit{spin models}) were reported to lead to quantum gravity as well as the emergence of certain near massless fermions such as the electron.

\begin{example}[Quantum supergroups of dipolar-coupled spins]\label{groupsofdipolar-coupledspins:}
An important example for either nuclear magnetic or electron spin resonances in solids is that
of (magnetic) \textit{dipolar-coupled (molecular) groups of spins}. Among such systems in which an understanding of the dipole-dipole interactions is essential are molecular groups of dipolar-coupled spin-1/2 particles (fermions) with local symmetry, or symmetries, such as groups of dipolar-coupled protons or magnetically-coupled microdomains of unpaired electrons in solids~\cite{Baianu-etal78a,Baianu-etal78b, Baianu-etal78c,Baianu-etal79a,Baianu-etal79b,Baianu-etal81}. Although one might expect
such systems of fermions to follow the Fermi statistics, in fact, the dipolar-coupled groups of spin-1/2 particles behave much more like quasi-quadrupolar (quasi) particles of spin-1 for proton pairs (as for example in ice or dihydrate gypsum crystals), or as spin-3/2 quasi-particles in the case of hydrogen nuclei in methyl groups and hydronium ions in solids, or coupled $^{19}{\rm F}$ (spin-1/2) nuclei in ${-}{\rm CF}_3$ molecular groups in polycrystalline solids~\cite{Baianu-etal81}.
(Interestingly, quantum theories of fermions were also recently proposed that do not require the presence of fermion fields \cite{Ball2k5}.)
A partially symmetric local structure was reported from such $^1H$ NMR studies which involved the determination of both inter- and intra-molecular van Vleck second moments of proton dipolar interactions of water in strongly ionic ${\rm LiCl} \times n \,{\rm H}_2{\rm O}$ and $\times n\, {\rm D}_2{\rm O}$
electrolyte glasses (with $  2.6 < n \leq 12$) at low temperature. Thus, the local symmetry of the hydration sphere for ${\rm Li}^+$ cations in such glasses at low temperatures was reported to be \textit{quasi-tetrahedral}~\cite{Baianu-etal78c}, and this was subsequently conf\/irmed by independent neutron scattering and electron tunneling/spectroscopic studies of the same systems at low temperatures, down to 4K;
on the other hand, for $n \geq 4$ water molecules bridged between hydrated ${\rm Li}^+$ clusters and a ${\rm Cl}^-$ anion,
the local symmetry approached quasi-octahedral around the anion. Similar studies were carried out for ${\rm Ca}({\rm NO}_3)_2 \times n\, {\rm H}_2{\rm O}$, ${\rm Zn}({\rm NO}_3)_2 \times n \, {\rm H}_2{\rm O}$, ${\rm Cd}({\rm NO}_3)_2 \times m\, {\rm H}_2{\rm O}$ and ${\rm La}({\rm NO}_3)_3 \times k \, {\rm H}_2{\rm O}$ electrolyte glasses (with  $ 3 < m \leq 20$ and, respectively, $ 3 < k \leq 30$), and the hydration local symmetries were found, respectively, to be: quasi-octahedral for both ${\rm Ca}^{+2}$ and ${\rm Zn}^{+2}$ (divalent) hydrated cations, \textit{quasi-icosahedral} for the (divalent) ${\rm Cd}^{+2}$ hydrated cation, and quasi-dodecahedral for the (trivalent) ${\rm La}^{+3}$  hydrated cation. Interstitial water molecules between hydrated cation clusters exhibited however much lower local symmetry, and it was reported to be very close to that of water monomers and dimers in the vapor phase~\cite{Baianu-etal78c}.

 The NMR behaviour of such proton and $^{19}{\rm F}$ \textit{quasi-particles} in solids~\cite{Baianu-etal81} suggests therefore the use of an unif\/ied supersymmetry approach using Lie superalgebras~\cite{Weinberg95} and \textit{quantum supergroups}.

\looseness=-1
 \textit{Quasi-particles} were also recently reported for Anderson-localized electrons in solids with partial disorder, and in the case of suf\/f\/iciently strong disorder, ``the Mott--Anderson transition was characterized by a precisely def\/ined \textit{two-fluid} behaviour, in which only a fraction of the electrons undergo a `site selective' Mott localization'' \cite{Aguiar2k9} (see also related previous articles in~\cite{Mott-etal75,Mott77,Mott78,Mott-Davis78,Stradling78,Tsui-Allen2k7,Anderson58,Byczuk-Hofstetter-Vollhardt2k8}). Thus, in any non-crystalline system~-- such as a glass~-- the lowest states in the conduction band are ``localized'', or they act as traps, and ``on the energy scale there is a continuous range of such localized states leading from the bottom of the band up to a critical energy $E_c$, called the mobility edge, where states become non-localized or extended''~\cite{Mott77}. Recently, a concept of ``quantum glassiness'' was also
introduced~\cite{Chamon2k5}.

 Similarly to spin-1/2 dipolar-coupled pairs, dipolar-coupled linear chains of either spin-1 or spin-0 bosons exhibit most remarkable properties that also depend on the strength of \textit{dipole-dipole} interactions among the neighbour bosons in the chain, as well as the overall, \textit{extended quantum symmetry} (EQS) of the chain.  On the other hand, \textit{local bosonic models, or spin models,} may also provide a unif\/ied origin for identical particles, gauge interactions, Fermi statistics and near masslessness of certain fermions \cite{Levin-Wen2k6}. Gauge interactions and Fermi statistics were also suggested to be unif\/ied under the point of view of emergence of identical particles; furthermore, a local bosonic model was constructed from which gravitons also emerge \cite{Levin-Wen2k6}, thus leading to \textit{quantum gravity}. Spin-2 boson models on a lattice are therefore being studied in such theories of quantum gravity \cite{Gu-Wen2k6}.

 Examples of dipolar-coupled and colour charge-coupled spin-0 bosons may be very abundant in nuclear physics where quark pairs provide a better model than the often used, `quark bag' model. Such spin-0 boson models of coupled quark pairs may also provide new insights into how to achieve \textit{controlled thermonuclear fusion}~\cite{Kapitsa78}. An example of a system of dipolar-coupled spin-1 bosons is that of an array of deuterons ($^2{\rm H}$) in \textit{deuteriated} long chain molecules such as phospholipids or f\/luorinated aliphatic chains in liquid crystals (e.g., perf\/luorooctanoate). For such systems it is possible to set up an explicit form of the Hamiltonian and to digitally compute all the spin energy levels and the nuclear magnetic resonance properties (including the phase coherence and spin correlations) for the entire chain of \textit{dipolar-coupled spin-$1$ bosons} (see for example the simple ``bosonization'' computations in~\cite{Miranda2k3}).

  The case of dipolar-coupled and also ion-coupled (or phonon-coupled), spin-0 bosons is also most remarkable in its \textit{long-range} correlations/coherence properties, as well as the temperature dependent symmetry breaking behaviour which is well-established; consider, for example, the Cooper (electron) pairs in superconductors \cite{Lee-etal2k4,Ribeiro-Wen2k5} that share this behaviour with other superf\/luids (e.g., liquid $^3{\rm He}$). Somewhat surprisingly, \textit{long-range} magnetic correlations involving nonlinear magnon dispersion also occur in ferromagnetic metallic glasses that have only short-range (or local) atomic structures of very low, or \textit{broken} symmetry, but exhibit microwave resonance absorption spectra caused by (long-range, coupled electron) \textit{spin wave excitations} as reported in~\cite{Baianu-etal79b}. The corresponding, explicit form of the Hamiltonian for the latter systems~-- including magnetic dipolar coupling, exchange and magnon interactions~-- has also been specif\/ied in~\cite{Baianu-etal79b}, and the short-range local structure present in such metallic glasses~-- noncrystalline systems with broken, local symmetry~-- was reported from $X$-ray scattering and ferromagnetic resonance studies~\cite{Baianu-etal79a}. Such noncrystalline systems with long-range coupling may be therefore more amenable to descriptions in terms of topological order theories as pointed out in~\cite{Anderson77,Mott77,Mott78} rather than Landau symmetry-breaking models. Topological order theories and topological quantum computation were also recently reported to be of interest for the design of quantum computers~\cite{Nayak-etal2k7,Stern-Halperin2k6,Freedman-etal2k3,Kitaev2k3,Dennis-etal2k2}, and thus such fundamental topological order theories might, conceivably, also lead to practical applications in developing ultra-fast quantum supercomputers.
\end{example}

\subsection{Hopf algebras}\label{section2.3}

 Firstly, a unital associative algebra consists of a linear space $A$ together with two linear maps
\begin{gather*}
m  : \ A \otimes A \lra A \quad \text{(multiplication)},\qquad
\eta : \ \bC \lra A \quad  \text{(unity)} 
\end{gather*}
satisfying the conditions
\begin{gather*}
m(m \otimes {\mathbf 1})  = m ({\mathbf 1} \otimes m),  \qquad  m({\mathbf 1}
\otimes \eta)  = m (\eta \otimes {\mathbf 1}) = \ID.
\end{gather*}
This f\/irst condition can be seen in terms of a commuting diagram:
\begin{gather*} 
\begin{CD}
A \otimes A \otimes A @> m \otimes \ID>> A \otimes A
\\ @V \ID \otimes mVV   @VV m V
 \\ A \otimes A  @ > m >> A
\end{CD}
\end{gather*}
Next let us consider `reversing the arrows', and take an
algebra $A$ equipped with a linear homorphisms $\Delta : A \lra A
\otimes A$, satisfying, for $a,b \in A$:
\begin{gather} 
\Delta(ab) = \Delta(a) \Delta(b),
\qquad (\Delta \otimes \ID) \Delta  = (\ID \otimes \Delta) \Delta.
\end{gather}

We call $\Delta$ a \emph{comultiplication}, which is said to be
\emph{coassociative} in so far that the following diagram commutes
\begin{gather*}
\begin{CD}
A \otimes A \otimes A  @<\Delta \otimes \ID << A \otimes A \\
@A \ID \otimes \Delta AA  @AA \Delta A
 \\ A \otimes A  @ < \Delta << A
\end{CD}
\end{gather*}

There is also a counterpart to $\eta$, the \emph{counity} map
$\vep : A \lra \bC$ satisfying
\begin{gather*}
(\ID \otimes \vep) \circ \Delta = (\vep \otimes \ID) \circ \Delta
= \ID.
\end{gather*}
A \emph{bialgebra} $(A, m, \Delta, \eta,
\vep)$ is a linear space $A$ with maps $m$, $\Delta$, $\eta$, $\vep$
satisfying the above pro\-per\-ties.

Now to recover anything resembling a group structure, we must
append such a bialgebra with an antihomomorphism $S : A \lra A$,
satisfying $S(ab) = S(b) S(a)$, for $a,b \in A$. This map is
def\/ined implicitly via the property:
\begin{gather*}
m(S \otimes
\ID) \circ \Delta = m(\ID \otimes S) \circ \Delta = \eta \circ
\vep.
\end{gather*}
We call $S$ the \emph{antipode map}. A \emph{Hopf algebra} is then
a bialgebra $(A,m, \eta, \Delta, \vep)$ equipped with an antipode
map $S$.

Commutative and non-commutative Hopf algebras form the backbone of
quantum groups \cite{VC-AP94} and are essential to the generalizations of
symmetry. Indeed, in most respects a quantum group is identif\/iable
with a Hopf algebra. When such algebras are actually
associated with proper groups of matrices there is
considerable scope for their representations on both f\/inite
and inf\/inite dimensional Hilbert spaces.

\begin{example}[The $\boldsymbol{\SL_q (2)}$  Hopf algebra]\label{hopfalgebra}
 This algebra is def\/ined by the generators $a$, $b$, $c$, $d$ and the following
relations:
\begin{gather*}
ba = qab, \qquad db=qbd, \qquad ca = qac, \qquad dc = qcd
, \qquad bc = cb,
\end{gather*}
together with
\begin{gather*}
ad da = \left(q^{-1}- q\right)bc,  \qquad ad q^{-1}bc = 1,
\end{gather*}
and
\begin{gather*}
\Delta  \bmatrix a & b  \\ c & d
\endbmatrix = \bmatrix a & b  \\ c & d
\endbmatrix  \otimes  \bmatrix a & b  \\ c & d
\endbmatrix,\qquad
\epsilon  \bmatrix a & b  \\ c & d
\endbmatrix = \bmatrix 1 & 0  \\0 & 1
\endbmatrix ,\qquad
S \bmatrix a & b  \\ c & d
\endbmatrix = \bmatrix d & -qb  \\-q^{-1}c & a
\endbmatrix.
\end{gather*}
\end{example}

\subsection{Quasi-Hopf algebra}\label{section2.4}

A quasi-Hopf algebra is an extension of a Hopf algebra. Thus, a
quasi-Hopf algebra is a \emph{quasi-bialgebra} $\mathcal{B_H} =
(\mathcal{H}, \Delta, \varepsilon, \Phi) $ for which there exist
$\alpha, \beta \in \mathcal{H}$ and a bijective antihomomorphism~$S$ (the `antipode') of $\mathcal{H}$ such that
$\sum_i S(b_i) \alpha c_i = \varepsilon(a) \alpha$, $
\sum_i b_i \beta S(c_i) = \varepsilon(a) \beta$
for all $ a \in \mathcal{H}$, with
$\Delta(a) = \sum_i b_i \otimes c_i$, and the relationships
\begin{gather*}
\sum_i X_i \beta S(Y_i) \alpha Z_i = \mathbf{I}, \qquad
\sum_j S(P_j) \alpha Q_j \beta S(R_j) = \mathbf{I},
\end{gather*}
where the expansions for the quantities $\Phi$ and $\Phi^{-1}$ are
given by
\begin{gather*}
\Phi = \sum_i X_i \otimes Y_i \otimes Z_i, \qquad
\Phi^{-1}= \sum_j P_j \otimes Q_j \otimes R_j.
\end{gather*}
As in the general case of a quasi-bialgebra, the property of being
quasi-Hopf is unchanged by ``twisting''. Thus, twisting the comultiplication of a coalgebra
\begin{gather*}
\mathcal{C} = (C, \Delta , \epsilon)
\end{gather*}
over a f\/ield $k$ produces another coalgebra $\mathcal{C}^{\rm cop}$;
because the latter is considered as a vector space over the f\/ield $k$, the new comultiplication of
$\mathcal{C}^{\rm cop}$ (obtained by ``twisting'') is def\/ined by
\begin{gather*}
\Delta ^{\rm cop} (c) = \sum{c_{(2)} \otimes c_{(1)}},
\end{gather*}
with $c \in \mathcal{C}$ and
\begin{gather*}
\Delta (c) = \sum{c_{(1)} \otimes c_{(2)}}.
\end{gather*}
Note also that the linear dual $\mathcal{C}^*$ of $\mathcal{C}$ is an algebra with unit $\epsilon$
and the multiplication being def\/ined by
\begin{gather*}
\langle c^* {*} d^*, c \rangle = \sum \langle c^*, c_{(1)} \rangle
\langle d^*, c_{(2)} \rangle ,
\end{gather*}
 for $c^*, d^* \in \mathcal{C}^*$ and $c \in \mathcal{C}$ (see \cite{Lambe-Redford97}).

Quasi-Hopf algebras emerged from studies of Drinfel'd twists and
also from $F$-matrices associated with f\/inite-dimensional
irreducible representations of a quantum af\/f\/ine algebra. Thus,
$F$-matrices were employed to factorize the corresponding $R$-matrix.
In turn, this leads to several important applications in statistical quantum
mechanics, in the form of quantum \textit{affine} al\-geb\-ras; their
representations give rise to solutions of the quantum Yang--Baxter
equation. This provides solvability conditions for various quantum
statistics models, allowing characteristics of such models to be
derived from their corresponding quantum af\/f\/ine algebras. The
study of $F$-matrices has been applied to models such as the
so-called Heisenberg `$XXZ$ model', in the framework of the
algebraic \textit{Bethe ansatz}. Thus $F$-matrices and quantum groups
together with quantum af\/f\/ine algebras provide an ef\/fective
framework for solving two-dimensional integrable models by using
the quantum inverse scattering method as suggested by Drinfel'd
and other authors.

\subsection{Quasi-triangular Hopf algebra}\label{section2.5}

We begin by def\/ining the quasi-triangular Hopf algebra, and then discuss its usefulness for computing
the $R$-matrix of a quantum system.
\begin{definition}
A Hopf algebra, $H$, is called \emph{quasi-triangular} if there is
an invertible element~$R$, of $ H \otimes H $ such that:
\begin{itemize}\itemsep=0pt
\item[(1)]
$R \ \Delta(x) = (T \circ \Delta)(x) \ R $ for all $ x \in H $,
where $\Delta $ is the coproduct on $H$, and the linear map $ T : H
\otimes H \to H \otimes H $ is given by
\begin{gather*}
T(x \otimes y) = y \otimes x,
\end{gather*}

\item[(2)] $(\Delta \otimes 1)(R) = R_{13}   R_{23} $,

\item[(3)]
$ (\mathbf{1} \otimes
\Delta)(R) = R_{13}  R_{12} $, where $R_{12} = \phi_{12}(R)$,

\item[(4)]
$R_{13} = \phi_{13}(R) $, and $ R_{23} = \phi_{23}(R) $, where $
\phi_{12} : H \otimes H \to H \otimes H \otimes H $,

\item[(5)] $\phi_{13} : H \otimes H \to H \otimes H \otimes H $, and $
\phi_{23} : H \otimes H \to H \otimes H \otimes H $, are algebra
morphisms determined by
\begin{gather*}
\phi_{12}(a \otimes b)  = a \otimes b \otimes 1,\qquad \phi_{13}(a
\otimes b)  = a \otimes 1 \otimes b,\qquad \phi_{23}(a \otimes b)  = 1
\otimes a \otimes b.
\end{gather*}
$R$ is called the \emph{$R$-matrix}.
\end{itemize}
\end{definition}

An important part of the above algebra can be summarized in the
following commutative diagrams involving the algebra morphisms, the
coproduct on $H$ and the identity map $\ID$:
\begin{gather*}
\begin{CD}
H \otimes H \otimes H @< \phi_{12}, ~\phi_{13} << H \otimes H
\\ @A \ID \otimes \ID \otimes \Delta AA  @AA \Delta A
 \\ H \otimes H \otimes H @ < \phi_{23}, ~\ID \otimes \Delta << H \otimes H
\end{CD}
\end{gather*}
and
\begin{gather*}
\begin{CD}
H \otimes H \otimes H @< \Delta\otimes \ID<< H \otimes H
\\ @A \ID \otimes \Delta AA  @AA \Delta A
 \\ H \otimes H  @ < \Delta << H
\end{CD}
\end{gather*}

Because of this property of quasi-triangularity, the $R$-matrix,
$R$, becomes a solution of \emph{the Yang--Baxter equation}. Thus, a
module $M$ of $H$ can be used to determine quasi-invariants of
links, braids, knots and higher dimensional structures with
similar quantum symmetries. Furthermore, as a consequence of the
property of quasi-triangularity, one obtains:
\begin{gather*}
(\epsilon \otimes 1) R = (1 \otimes \epsilon) R = 1 \in H .
\end{gather*}
Finally, one also has:
\begin{gather*}
R^{-1} = (S \otimes 1)(R), \qquad
R = (1 \otimes S)\big(R^{-1}\big)  \qquad \text{and} \qquad
(S \otimes S)(R) = R.
\end{gather*}
One can also prove that the antipode $S$ is a linear
isomorphism, and therefore $S^2$ is an automorphism: $S^2$ is
obtained by conjugating by an invertible element, $S(x) = u x
u^{-1}$, with
\begin{gather*}
u = m (S \otimes 1)R^{21}.
 \end{gather*}
By employing Drinfel'd's quantum double construction
one can assemble a quasi-triangular
Hopf algebra from a Hopf algebra and its dual.

\subsubsection{Twisting a quasi-triangular Hopf algebra}\label{section2.5.1}

The property of being a quasi-triangular Hopf algebra is
invariant under twisting \emph{via} an invertible element $ F =
\sum_i f^i \otimes f_i \in \mathcal{A \otimes A} $ such that $
(\varepsilon \otimes {\rm id})F = ({\rm id} \otimes \varepsilon)F = 1 $, and also such
that the following cocycle condition is satisf\/ied:
\begin{gather*}
(F \otimes 1) \circ (\Delta \otimes {\rm id}) F = (1 \otimes F) \circ
({\rm id} \otimes \Delta) F .
\end{gather*}
Moreover, $ u = \sum_i f^i S(f_i) $ is
invertible and the twisted antipode is given by $ S'(a) = u
S(a)u^{-1} $, with the twisted comultiplication, $R$-matrix and
co-unit change according to those def\/ined for the quasi-triangular
quasi-Hopf algebra. Such a twist is known as \emph{an admissible, or
Drinfel'd, twist}.

\subsection{Quasi-triangular quasi-Hopf algebra (QTQH)}\label{section2.6}

\emph{A quasi-triangular quasi-Hopf algebra} as def\/ined by Drinfel'd in~\cite{Drinfeld92} is an extended form of a~quasi-Hopf algebra, and also of a quasi-triangular Hopf algebra. Thus, a quasi-triangular quasi-Hopf algebra is def\/ined as a \emph{quintuple} $\mathcal{B_H} =
(\mathcal{H}, R, \Delta, \varepsilon, \Phi)$ where the latter is
 a quasi-Hopf algebra, and $R \in \mathcal{H} \otimes \mathcal{H}$
referred to as the $R$-matrix  (as def\/ined above), which is an invertible
element such that:
\begin{gather*}
R \Delta(a)  = \sigma \circ \Delta(a) R, \qquad a \in \mathcal{H},\qquad
\sigma : \ \mathcal{H} \otimes \mathcal{H} \rightarrow \mathcal{H} \otimes
\mathcal{H},  \qquad x \otimes y  \rightarrow y \otimes x ,
\end{gather*}
so that $ \sigma $ is the switch map and
\begin{gather*}
(\Delta \otimes {\rm id})R  =
\Phi_{321}R_{13}\Phi_{132}^{-1}R_{23}\Phi_{123},\qquad ({\rm id} \otimes
\Delta)R  = \Phi_{231}^{-1}R_{13}\Phi_{213}R_{12}\Phi_{123}^{-1},
\end{gather*}
 where $\Phi_{abc}  = x_a \otimes x_b \otimes x_c$,  and
$\Phi_{123} = \Phi = x_1 \otimes x_2 \otimes x_3 \in
\mathcal{H} \otimes \mathcal{H} \otimes \mathcal{H}$.
The quasi-Hopf algebra becomes triangular if in addition one has
$R_{21}R_{12}=1$.

The twisting of $ \mathcal{B_H} $ by $F \in \mathcal{H \otimes H}$
is the same as for a quasi-Hopf algebra, with the additional
def\/inition of the twisted $R$-matrix. A quasi-triangular, quasi-Hopf
algebra with $ \Phi=1 $ is a quasi-triangular Hopf algebra because
the last two conditions in the def\/inition above reduce to the
quasi-triangularity condition for a Hopf algebra. Therefore, just
as in the case of the twisting of a quasi-Hopf algebra, the
property of being quasi-triangular of a quasi-Hopf algebra is
preserved by twisting.

\subsection[Yang--Baxter equations]{Yang--Baxter equations}\label{section2.7}

\subsubsection[Parameter-dependent Yang--Baxter equation]{Parameter-dependent Yang--Baxter equation}
\label{section2.7.1}

Consider $A$ to be an unital associative algebra. Then, \emph{the
parameter-dependent Yang--Baxter equation} is an equation for $R(u)
$, the parameter-dependent invertible element of the tensor
product $A \otimes A$ (here, $u$ is the parameter, which usually
ranges over all real numbers in the case of an additive parameter,
or over all positive real numbers in the case of a multiplicative
parameter; for the dynamic Yang--Baxter equation see also~\cite{Etingof-Schiffmann2k1}). The Yang--Baxter equation is usually stated as:
\begin{gather*}
R_{12}(u)   R_{13}(u+v)   R_{23}(v) = R_{23}(v)   R_{13}(u+v)
R_{12}(u),
\end{gather*}
for all values of $u$ and $v$, in the case of an additive
parameter, and
\begin{gather*}
R_{12}(u)   R_{13}(uv)   R_{23}(v) = R_{23}(v)   R_{13}(uv)
R_{12}(u),
\end{gather*}
for all values of $u$ and $v$, in the case of a multiplicative
parameter, where
\begin{gather*}
R_{12}(w)  = \phi_{12}(R(w)),\qquad  R_{13}(w)  = \phi_{13}(R(w)),\qquad
R_{23}(w)  = \phi_{23}(R(w))
\end{gather*}
for all values of the
parameter $ w $, and
\begin{gather*}
\phi_{12}  : \ H \otimes H \to H \otimes H \otimes H, \qquad  \phi_{13}  : \
H \otimes H \to H \otimes H \otimes H,  \\ \phi_{23}  : \ H \otimes H
\to H \otimes H \otimes H
\end{gather*}
are algebra
morphisms determined by the following (strict) conditions:
\begin{gather*}
\phi_{12}(a \otimes b)  = a \otimes b \otimes 1, \qquad  \phi_{13}(a
\otimes b)  = a \otimes 1 \otimes b,\qquad \phi_{23}(a \otimes b)  = 1
\otimes a \otimes b.
\end{gather*}

\subsubsection[The parameter-independent Yang--Baxter equation]{The parameter-independent Yang--Baxter equation}\label{section2.7.2}

Let $A$ be a unital associative algebra. The
parameter-independent Yang--Baxter equation is an equation for
$R$, an invertible element of the tensor product $A \otimes A$.
The Yang--Baxter equation is:
\begin{gather*}
R_{12}   R_{13}   R_{23}  = R_{23}   R_{13}   R_{12},
\end{gather*}
where $R_{12} = \phi_{12}(R)$, $R_{13} = \phi_{13}(R)$,
 and $R_{23} = \phi_{23}(R)$.

Let $V$ be a module over $A$. Let $ T : V \otimes V \to V
\otimes V $ be the linear map satisfying $ T(x \otimes y) = y
\otimes x $ for all $ x, y \in V$. Then a representation of the
braid group $B_n$, can be constructed on $V^{\otimes n}$ by $
\sigma_i = 1^{\otimes i-1} \otimes \check{R} \otimes 1^{\otimes
n-i-1} $ for $ i = 1,\dots,n-1 $, where $ \check{R} = T \circ R $
on $ V \otimes V $. This representation may thus be used to
determine quasi-invariants of braids, knots and links.

\subsubsection[Generalization of the quantum Yang--Baxter equation]{Generalization of the quantum Yang--Baxter equation}\label{section2.7.3}

The quantum Yang--Baxter equation was generalized in \cite{Lambe-Redford97}
to:
\begin{gather*}
R = qb  \sum_{i=1} ^n e_{ii} \otimes e_{ii}  + b \sum_{i > j} e_{ii}
\otimes e_{jj}  + c \sum_{i < j} e_{ii} \otimes e_{jj}  +
\left(qb-q^{-1}c\right) \sum_{i > j} e_{ij} \otimes e_{ji} ,
\end{gather*}
for $b,c \neq 0$. A solution of the quantum Yang--Baxter equation
has the form $ R: M \otimes M \to M \otimes M$, with $M$ being a
f\/inite dimensional vector space over a f\/ield $k$. Most of the
solutions are stated for a given ground f\/ield but in many cases a
commutative ring with unity  may instead be suf\/f\/icient. (See also the
classic paper by Yang and Mills~\cite{Yang-Mills54}.)

\subsection[$\SU(3)$, $\SU(5)$, $\SU(10)$ and $\rm{E}_6$ representations
in quantum chromodynamics\\ and unif\/ied theories involving
spontaneous symmetry breaking]{$\boldsymbol{\SU(3)}$, $\boldsymbol{\SU(5)}$, $\boldsymbol{\SU(10)}$ and $\boldsymbol{\rm{E}_6}$ representations\\
in quantum chromodynamics and unif\/ied theories\\ involving
spontaneous symmetry breaking}\label{section2.8}

There have been several attempts to take into consideration
extended quantum symmetries that would include, or embed, the
$\SU(2)$ and $\SU(3)$ symmetries in larger symmetry groups such as
$\SU(5)$, $\SU(10)$ and the exceptional Lie group $\rm{E}_6$, but so
far with only limited success as their representations make
several predictions that are so far unsupported by high energy
physics experiments \cite{Gilmore2k5}. To remove unobserved
particles from such predictions, one has invariably to resort to
ad-hoc spontaneous symmetry breaking assumptions that would
require still further explanations, and so on. So far the only
thing that is certain is the fact that the $\U(1) \times \SU(2)
\times \SU(3)$ symmetry is broken in nature, presumably in a
`spontaneous' manner. Due to the nonlocal character of quantum
theories combined with the restrictions imposed by relativity on the
`simultaneity' of events in dif\/ferent reference systems, a global
or universal, spontaneous symmetry breaking mechanism appears
contrived, with the remaining possibility that it does however
occur locally, thus resulting in quantum theories that use local approximations
for broken symmetries, and thus they are not
unif\/ied, as it was intended. Early approaches to space-time were
made in non-relativistic quantum mechanics~\cite{Feynman48}, and were
subsequently followed by relativistic and axiomatic approaches to
quantum f\/ield theory~\cite{Wightman56,Wightman-Garding64,Wightman76}.

 On the one hand, in GR all interactions are local, and therefore spontaneous, local sym\-met\-ry
breaking may appear not to be a problem for GR, except for the
major obstacle that it does severely limit the usefulness of the
Lorentz group of transformations which would have to be modif\/ied
accordingly to take into account the \textit{local} $\SU(2) \times
\SU(3)$ spontaneous symmetry brea\-king. This seems to cause
problems with the GR's equivalence principle for all reference
systems; the latter would give rise to an equivalence class, or
possibly a set, of reference systems. On the other hand, local,
spontaneous symmetry breaking generates a \emph{groupoid of
equivalence classes of reference systems}, and further, through
quantization, to a category of groupoids of such reference
systems, $\mathbf{Grpd}_{\Re}$, and their transformations def\/ined
as groupoid homomorphisms. Functor representations of
$\mathbf{Grpd}_{\Re}$ into the category $\mathbf{BHilb}$ of rigged
Hilbert spaces $\H_r$ would then allow the computation of
\textit{local} quantum operator eigenvalues and their eigenstates,
in a~manner invariant to the local, broken symmetry
transformations. One might call such a~theo\-ry, a locally
covariant~-- quantized GR (lcq-GR), as it would be locally, but
not necessarily, globally quantized. Obviously, such a locally
covariant GR theory is consistent with AQFT and its operator nets
of local quantum observables. Such an extension of the GR theory
to a locally covariant GR  in a~quantized form may not require the
`universal' or global existence of Higgs bosons as a~compelling
property of the expanding Universe; thus, any lcq-GR theory can
allow for the existence of inhomogeneities in spacetime caused by
distinct local symmetries in the presence of very intense
gravitational f\/ields, dark matter, or other condensed quantum
systems such as neutron stars and black holes (with or without
`hair'~-- cf.\ J.~Wheeler). The GR principle of equivalence is then
replaced in lcq-GR by the representations of the quantum
fundamental groupoid functor that will be introduced in Section~\ref{section9}.

In view of the existing problems and limitations encountered with
group quantum symmetries and their group (or group algebra)
representations, current research into the geometry of state spaces
of quantum operator algebras leads to extended symmetries
expressed as topological groupoid representations that were shown
to link back to certain $C^*$-algebra representations~\cite{Dixmier69} and the dual spaces of $C^*$-algebras~\cite{Fell60}. Such extended symmetries will be discussed in the next sections in terms of
quantum groupoid representations involving the notion of measure
Haar systems associated with locally compact quantum groupoids.

\section[Quantum groupoids and the groupoid $C^*$-algebra]{Quantum groupoids and the groupoid $\boldsymbol{C^*}$-algebra}\label{section3}

Quantum groupoid (e.g., weak Hopf algebras) and algebroid
symmetries f\/igure prominently both in the theory of dynamical
deformations of quantum groups \cite{Drinfeld87} (e.g., Hopf algebras) and the
quantum Yang--Baxter equations~\cite{Etingof-Varchenko98,Etingof-Varchenko99}. On the
other hand, one can also consider the natural extension of locally
compact (quantum) groups to locally compact (proper)
\emph{groupoids} equipped with a Haar measure and a corresponding
groupoid representation theory \cite{Buneci2k3} as a major,
potentially interesting source for locally compact (but generally
\emph{non-Abelian}) \emph{quantum groupoids.} The corresponding
quantum groupoid representations on bundles of Hilbert spaces extend
quantum symmetries well beyond those of quantum groups/Hopf
algebras and simpler operator algebra representations, and are also
consistent with the locally compact quantum group representations
that were recently studied in some detail by Kustermans and Vaes
(see~\cite{Kustermans-Vaes2k} and references cited therein). The latter quantum groups are
neither Hopf algebras, nor are they equivalent to Hopf algebras or
their dual coalgebras. As pointed out in the previous section,
quantum groupoid representations are, however, the next important
step towards unifying quantum f\/ield theories with general relativity
in a locally covariant and quantized form. Such representations need
not however be restricted to weak Hopf algebra representations, as
the latter have no known connection to any type of GR theory and
also appear to be inconsistent with GR.

We are also motivated here by the quantum physics examples
mentioned in the previous sections to introduce through several
steps of generality, a framework for quantum symmetry breaking in
terms of either locally compact quantum groupoid or related
algebroid representations, such as those of \emph{weak Hopf
$C^*$-algebroids with convolution} that are realized in the context
of \emph{rigged Hilbert spaces} \cite{Bohm-Gadella89}. A novel
extension of the latter approach is also now possible \textit{via}
generalizations of Grassman--Hopf algebras ($\mathcal{G}_H$), gebras \cite{Sweedler96,Fauser2k4} and co-algebra representations to those of graded Grassman--Hopf algebroids. Grassman--Hopf algebras and gebras not only are bi-connected in a manner somewhat similar to Feynman diagrams but also possess a unique left/right integral $\mu$  \cite[p.~288]{Fauser2k4}, whereas such integrals in general do not exist in
Clif\/ford--Hopf algebras \cite{Fauser2k2}. This unique integral
property of Grassman--Hopf algebras  makes them very interesting
candidates, for example, in physical applications that require
either \textit{generalized convolution} and measure concepts, or
generalizations of quantum groups/algebras to structures that are
more amenable than weak Hopf $C^*$-algebras. Another important point
made by Fauser \cite{Fauser2k4} is that~-- unlike Hopf and weak Hopf algebras
that have no direct physical visualization either in quantum
dynamics or in the Feynman interaction representation of Quantum
Electrodynamics~-- the duals, or tangles, of Grassman--Hopf
algebras, such as respectively G--H co-algebras and Grassman--Hopf
`algebras' \cite{Fauser2k4} provide direct visual representations of
physical interactions and quantum dynamics in Feynman-like
diagrams that utilize directly the dual/tangled, or `co-algebraic',
structure elements. Such visual representations can greatly
facilitate exact computations in quantum chromodynamics for the
dif\/f\/icult case of strong, nuclear interactions where approximate
perturbation methods usually fail. The mathematical def\/initions
and grading of Grassman--Hopf algebroids, (tangled/mirror) algebroids and co-algebroids then
follow naturally for supersymmetry, symmetry breaking, and other
physical theories. Furthermore, with regard to a unif\/ied and
global framework for symmetry breaking, as well as higher order
quantum symmetries, we look towards the \emph{double groupoid}
structures of Brown and Spencer~\cite{Brown-Spencer76a}, and introduce the concepts
of \emph{quantum and graded Lie bi-algebroids} which are expected
to carry a distinctive $C^*$-algebroid convolution structure. The
extension to \emph{supersymmetry} leads then naturally to
superalgebra, superf\/ield symmetries and their involvement in
supergravity or quantum gravity (QG) theories for intense
gravitational f\/ields in f\/luctuating, quantized spacetimes. Our
self-contained approach, leads to several novel concepts which
exemplify a certain \emph{non-reductionist} viewpoint and
theories of the nature of physical spacetime structure~\cite{Brown-etal2k7,Baianu-etal2k7}.

 A natural extension in higher dimensional algebra (HDA) of quantum symmetries may involve both \textit{quantum double groupoids} def\/ined as locally compact double groupoids equipped with Haar measures via convolution, and an extension to double algebroids, (that are naturally more general than the Lie double algebroids def\/ined in~\cite{Mackenzie2k5}).

 We shall now proceed to formally def\/ine several quantum algebraic
topology concepts that are needed to express the extended quantum
symmetries in terms of proper quantum groupoid and quantum algebroid
representations. Hidden, higher dimensional quantum symmetries
will then also emerge either \emph{via} generalized quantization
procedures from higher dimensional al\-geb\-ra representations or will be
determined as global or local invariants obtainable~-- at least in
principle~-- through non-Abelian algebraic topology (NAAT) methods \cite{Brown-etal2k9} (see also the
earlier classic paper by Fr{\"o}hlich~\cite{Frohlich61}).

\subsection{The weak Hopf algebra}\label{section3.1}

In this, and the following subsections, we proceed through several
stages of generality by relaxing the axioms for a Hopf algebra as def\/ined above.
The motivation begins with the more restrictive notion of a quantum group in relation
to a Hopf algebra where the former is often realized as an automorphism group for a quantum
space, that is, an object in a suitable category of generally
noncommutative algebras. One of the most common guises of a quantum
`group' is as the dual of a noncommutative, nonassociative Hopf
algebra. The Hopf algebras (cf.~\cite{Chaician-Demichev96,Majid95}),
and their generalizations~\cite{Karaali2k7}, are some of the fundamental building
blocks of quantum operator algebra, even though they cannot be generally `integrated' to groups
like the `integration' of Lie algebras to Lie groups, or the Fourier transformation of
certain commutative Hopf algebras to their dual, f\/inite commutative groups. However, Hopf algebras
are linked and limited only to certain quantum symmetries that
are represented by f\/inite compact quantum groups (CQGs).

In order to def\/ine a \emph{weak Hopf algebra}, one can relax
certain axioms of a Hopf algebra as follows:
\begin{itemize}\itemsep=0pt
\item[(1)] The comultiplication is not necessarily unit-preserving.
\item[(2)] The counit $\vep$ is not necessarily a homomorphism of algebras.
\item[(3)] The axioms for the antipode map $S : A \lra A$ with respect to the
counit are as follows. For all $h \in H$,
\begin{gather*} m(\ID \otimes S) \Delta (h)  = (\vep \otimes
\ID)(\Delta (1) (h \otimes 1)), \\ m(S \otimes \ID) \Delta (h)  =
(\ID \otimes \vep)((1 \otimes h) \Delta(1)), \qquad S(h)  = S(h_{(1)})
h_{(2)}  S(h_{(3)})  .
\end{gather*}
\end{itemize}

 These axioms may be appended by the following commutative diagrams
\begin{gather*}
{\begin{CD} A \otimes A @> S\otimes \ID >> A \otimes A
\\ @A \Delta AA   @VV m V
 \\ A @ > u \circ \vep >> A
\end{CD}} \qquad
{\begin{CD} A \otimes A @> \ID\otimes S >> A \otimes A
\\ @A \Delta AA   @VV m V
 \\ A @ > u \circ \vep >> A
\end{CD}}
\end{gather*}
along with the counit axiom:
\begin{gather*}
\xymatrix@C=3pc@R=3pc{ A \otimes A \ar[d]_{\vep \otimes 1} & A
\ar[l]_{\Delta} \ar[dl]_{\ID_A} \ar[d]^{\Delta}
\\ A  & A \otimes A \ar[l]^{1 \otimes \vep}}
\end{gather*}

Several mathematicians substitute the term \emph{quantum
groupoid} for a weak Hopf algebra, although this algebra in
itself is not a proper groupoid, but it may have a component
\emph{group} algebra as in the example of the quantum double
discussed next; nevertheless, weak Hopf algeb\-ras genera\-li\-ze Hopf
algebras~-- that with additional properties~-- were previously
introduced as` quantum group' by mathematical physicists. (The
latter are def\/ined in the Appendix~\ref{appendixA} and, as already discussed, are
not mathematical groups but algebras). As it will be shown in the
next subsection, quasi-triangular quasi-Hopf algebras are
directly related to quantum symmetries in conformal (quantum)
f\/ield theories. Furthermore, weak $C^*$-Hopf quantum algebras lead
to weak $C^*$-Hopf algebroids that are linked to quasi-group
quantum symmetries, and also to certain Lie algebroids (and their
associated Lie--Weinstein groupoids) used to def\/ine Hamiltonian
(quantum) algebroids over the phase space of (quantum)
$W_N$-gravity.

\subsubsection{Examples of weak Hopf algebras}\label{section3.1.1}

(1) We refer here to \cite{BSS2k2}. Let $G$ be a non-Abelian group
and $H \subset G$ a discrete subgroup. Let~$F(H)$ denote the space
of functions on $H$ and $\bC H$ the group algebra (which consists
of the linear span of group elements with the group structure).
\emph{The quantum double}~$D(H)$~\cite{Drinfeld87} is def\/ined by
\begin{gather*}
D(H) = F(H)  \, \wti{\otimes}  \, \bC H,
\end{gather*}
where, for $x \in H$, the `twisted tensor product' is specif\/ied by
\begin{gather*}
\wti{\otimes} \mapsto  (f_1 \otimes h_1) (f_2 \otimes h_2)(x) =
f_1(x) f_2\left(h_1 x h_1^{-1}\right) \otimes h_1 h_2 .
\end{gather*}
The physical interpretation is often to take $H$ as the `electric
gauge group' and $F(H)$ as the `magnetic symmetry' generated by $\{f
\otimes e\}$. In terms of the counit $\vep$, the double $D(H)$ has
a trivial representation given by $\vep(f \otimes h) = f(e)$. We
next look at certain features of this construction.

For the purpose of braiding relations there is an $R$ matrix, $R
\in D(H) \otimes D(H)$, leading to the operator
\begin{gather*}
\mathcal R \equiv \sigma \cdot \big(\Pi^A_{\a} \otimes \Pi^B_{\be}\big)
(R),
\end{gather*}
in terms of the Clebsch--Gordan series $\Pi^A_{\a} \otimes
\Pi^B_{\be} \cong N^{AB \gamma}_{\a \be C}  \Pi^C_{\gamma}$, and
where $\sigma$ denotes a f\/lip operator. The operator $\mathcal
R^2$ is sometimes called the \emph{monodromy} or
\emph{Aharanov--Bohm phase factor}. In the case of a condensate in
a state $\vert v \rangle$ in the carrier space of some
representation $\Pi^A_{\a}$ one considers the maximal Hopf
subalgebra $T$ of a Hopf algebra $A$ for which $\vert v \rangle$
is $T$-invariant; specif\/ically:
\begin{gather*}
\Pi^A_{\a} (P) \vert v \rangle = \vep(P) \vert v \rangle,\qquad
\forall \, P \in T.
\end{gather*}

(2)
For the second example, consider $A = F(H)$. The algebra of
functions on $H$ can be broken to the algebra of functions on $H/K$,
that is, to $F(H/K)$, where $K$ is normal in $H$, that is, $HKH^{-1}
=K$. Next, consider $A = D(H)$. On breaking a purely electric
condensate $\vert v \rangle$, the magnetic symmetry remains
unbroken, but the electric symmetry $\bC H$ is broken to $\bC N_v$,
with $N_v \subset H$, the stabilizer of $\vert v \rangle$. From
this we obtain $T = F(H) \wti{\otimes} \bC N_v$.

(3)
In \cite{Nikshych-Vainerman2k} quantum groupoids (considered as weak
$C^*$-Hopf algebras, see below) were studied in relationship to the
noncommutative symmetries of depth 2 von Neumann subfactors. If
\begin{gather*}
A \subset B \subset B_1 \subset B_2 \subset \cdots
\end{gather*}
is the Jones extension induced by a f\/inite index depth $2$
inclusion $A \subset B$ of $II_1$ factors, then $Q= A' \cap B_2$
admits a quantum groupoid structure and acts on $B_1$, so that $B
= B_1^{Q}$ and \mbox{$B_2 = B_1 \rtimes Q$}. Similarly, in~\cite{Rehren97}
`paragroups' (derived from weak $C^*$-Hopf algebras) comprise
(quantum) groupoids of equivalence classes such as those associated with
$6j$-symmetry groups (relative to a~fusion rules algebra). They
correspond to type~$II$ von Neumann algebras in quantum mechanics,
and arise as symmetries where the local subfactors (in the sense
of containment of observables within f\/ields) have depth~2 in the
Jones extension. A related question is how a~von Neumann algebra $N$, such as
of f\/inite index depth~2, sits inside a weak Hopf algebra formed as
the crossed product $N \rtimes A$~\cite{Bohm-etal99}.

(4)
Using a more general notion of the Drinfel'd construction, Mack and Schomerus developed in \cite{Mack-Schomerus92} the notion of a \emph{quasi-triangular quasi-Hopf algebra} (QTQHA) with the aim
of studying a~range of essential symmetries with special
properties, such as the quantum group algebra $\U_q (\rm{sl}_2)$ with
$\vert q \vert =1$. If $q^p=1$, then it is shown that a QTQHA is
canonically associated with $\U_q (\rm{sl}_2)$. Such QTQHAs are
claimed as the true symmetries of minimal conformal f\/ield
theories.

\subsubsection[The weak Hopf $C^*$-algebra in relation to quantum symmetry breaking]{The weak Hopf $\boldsymbol{C^*}$-algebra in relation to quantum symmetry breaking}\label{section3.1.2}

 In our setting, a \emph{weak $C^*$-Hopf algebra} is a weak $*$-Hopf
algebra which admits a faithful $*$-representation on a Hilbert
space. The weak $C^*$-Hopf algebra is therefore much more likely to
be closely related to a quantum groupoid representation than any
weak Hopf algebra. However, one can argue that locally compact
groupoids equipped with a Haar measure (after quantization) come
even closer to def\/ining quantum groupoids. There are already
several, signif\/icant examples that motivate the consideration of
weak $C^*$-Hopf algebras which also deserve mentioning in the
context of `standard' quantum theories. Furthermore, notions such
as (proper) \emph{weak $C^*$-algebroids} can provide the main
framework for symmetry breaking and quantum gravity that we are
considering here. Thus, one may consider the quasi-group
symmetries constructed by means of special transformations of the
coordinate space $M$. These transformations along with the
coordinate space $M$ def\/ine certain Lie groupoids, and also their
inf\/initesimal version~-- the Lie algebroids $\mathbf{A}$, when the
former are Weinstein groupoids. If one then lifts the algebroid
action from $M$ to the principal homogeneous space $\R$ over the
cotangent bundle~$T^*M \lra M$, one obtains a physically
signif\/icant algebroid structure. The latter was called the
Hamiltonian algebroid, ${\mathcal A}^H$, related to the Lie
algebroid,~$\mathbf{A}$. The Hamiltonian algebroid is an analog of
the Lie algebra of symplectic vector f\/ields with respect to the
canonical symplectic structure on~$\R$ or $T^*M$. In this recent
example, the Hamiltonian algebroid, $\mathcal{A}^H$ over $\R$, was
def\/ined over the phase space of $W_N$-gravity, with the anchor
map to Hamiltonians of canonical transformations~\cite{Levin-Olshanetsky08}. Hamiltonian algebroids thus generalize
Lie algebras of canonical transformations; canonical
transformations of the Poisson sigma model phase space def\/ine a
\emph{Hamiltonian algebroid} with the Lie brackets related to such
a Poisson structure on the target space. The Hamiltonian algebroid
approach was utilized to analyze the symmetries of generalized
deformations of complex structures on Riemann surfaces
$\sum_{g,n}$ of genus $g $ with $n$ marked points. However, its
implicit algebraic connections to von Neumann $*$-algebras and/or
\emph{weak $C^*$-algebroid representations} have not yet been
investigated. This example suggests that algebroid (quantum)
symmetries are implicated in the foundation of relativistic
quantum gravity theories and supergravity that we shall consider
in further detail in Sections~\ref{section6}--\ref{section9}.

\subsection{Compact quantum groupoids}\label{section3.2}

Compact quantum groupoids were introduced in \cite{Landsman98} as a
simultaneous generalization of a~compact groupoid and a quantum
group. Since this construction is relevant to the def\/inition of
locally compact quantum groupoids and their representations
investigated here, its  exposition is required before we can step up
to the next level of generality. Firstly, let $\mathfrak A$ and
$\mathfrak B$ denote $C^*$-algebras equipped with a $*$-homomorphism
$\eta_s : \mathfrak B \lra \mathfrak A$, and a $*$-antihomomorphism
$\eta_t : \mathfrak B \lra \mathfrak A$ whose images in $\mathfrak
A$ commute. A non-commutative Haar measure is def\/ined as a
completely positive map $P: \mathfrak A \lra \mathfrak B$ which
satisf\/ies $P(A \eta_s (B)) = P(A) B$. Alternatively, the
composition $\E = \eta_s \circ P : \mathfrak A \lra \eta_s (B)
\subset \mathfrak A$ is a faithful conditional expectation.

Next consider $\mathsf{G}$ to be a (topological) groupoid as def\/ined
in the Appendix~\ref{appendixA}. We denote by~$C_c(\mathsf{G})$ the space of smooth
complex-valued functions with compact support on $\mathsf{G}$. In
particular, for all $f,g \in C_c(\mathsf{G})$, the function def\/ined
via convolution
\begin{gather*}
(f  * g)(\gamma)
= \int_{\gamma_1 \circ \gamma_2 = \gamma} f(\gamma_1) g
(\gamma_2),
\end{gather*}
is again an element of $C_c(\mathsf{G})$, where the convolution product
def\/ines the composition law on~$C_c(\mathsf{G})$. We can turn
$C_c(\mathsf{G})$ into a $*$-algebra once we have def\/ined the involution
$*$, and this is done by specifying $f^*(\gamma) = \overline{f(\gamma^{-1})}$.
This $*$-algebra whose multiplication is the convolution becomes a~groupoid $C^*$-convolution algebra, or groupoid $C^*$-algebra, $G_{\rm CA}$, when $G$ is a~measured groupoid
and the $C^*$-algebra has a smallest $C^*$-norm which makes its representations continuous~\cite{Renault80, MRW87}.

We recall that following~\cite{Landsman2k} a \emph{representation}
 of a groupoid~$\grp$, consists of a
family (or f\/ield) of Hilbert spaces $\{\mathcal H_x \}_{x \in X}$
indexed by $X = \ob\, \grp$, along with a collection of maps $\{
U(\gamma)\}_{\gamma \in \grp}$, satisfying:
\begin{itemize}\itemsep=0pt
\item[1)]
$U(\gamma) : \mathcal H_{s(\gamma)} \lra  \mathcal H_{r(\gamma)}$,
is unitary;

\item[2)]
$U(\gamma_1 \gamma_2) = U(\gamma_1) U( \gamma_2)$, whenever
$(\gamma_1, \gamma_2) \in \grp^{(2)}$  (the set of arrows);

\item[3)]
$U(\gamma^{-1}) = U(\gamma)^*$, for all $\gamma \in \grp$.
\end{itemize}

Suppose now $\mathsf{G}_{lc}$ is a Lie groupoid. Then the isotropy group
$\mathsf{G}_x$ is a Lie group, and for a~(left or right) Haar
measure $\mu_x$ on $\mathsf{G}_x$, we can consider the Hilbert
spaces $\mathcal H_x = L^2(\mathsf{G}_x, \mu_x)$ as exemplifying the
above sense of a representation. Putting aside some technical
details which can be found in~\cite{Connes94,Landsman2k}, the
overall idea is to def\/ine an operator of Hilbert spaces
\begin{gather*}\pi_x(f) : \ L^2(\mathsf{G}_x,\mu_x)
 \lra L^2(\mathsf{G}_x, \mu_x),
\end{gather*}
given by
\begin{gather*}
(\pi_x(f) \xi)(\gamma) = \int f(\gamma_1) \xi (\gamma_1^{-1}
\gamma)  d\mu_x,
\end{gather*}
for all $\gamma \in \mathsf{G}_x$, and
$\xi \in \mathcal H_x$. For each $x \in X =\ob \,\mathsf{G}$, $\pi_x$
def\/ines an involutive representation $\pi_x : C_c(\mathsf{G}) \lra
\mathcal H_x$. We can def\/ine a norm on $C_c(\mathsf{G})$ given by
\begin{gather*}
\Vert f \Vert = \sup_{x \in X} \Vert \pi_x(f) \Vert,
\end{gather*}
whereby the completion of $C_c(\mathsf{G})$ in this norm, def\/ines
\emph{the reduced $C^*$-algebra $C^*_r(\mathsf{G})$ of $\mathsf{G}_{lc}$}.
It is
perhaps the most commonly used $C^*$-algebra for Lie groupoids
(groups) in noncommutative geometry~\cite{Varilly97, Connes79, Connes94}.

The next step requires a little familiarity with the theory of
Hilbert modules (see e.g.~\cite{Lance95}). We def\/ine a left
$\mathfrak B$-action $\lambda$ and a right $\mathfrak B$-action
$\rho$ on $\mathfrak A$ by $\lambda(B)A = A \eta_t (B)$ and~$\rho(B)A = A \eta_s(B)$. For the sake of localization of the
intended Hilbert module, we implant a $\mathfrak B$-valued inner
product on $\mathfrak A$ given by $\langle A, C \rangle_{\mathfrak
B} = P(A^* C)$. Let us recall that $P$ is def\/ined as a
\emph{completely positive map}.
Since $P$ is faithful, we f\/it a new norm on $\mathfrak A$ given
by $\Vert A \Vert^2 = \Vert P(A^* A)
\Vert_{\mathfrak B}$. The completion of $\mathfrak A$ in this new
norm is denoted by $\mathfrak A^{-}$ leading then to a Hilbert
module over $\mathfrak B$.

The tensor product $\mathfrak A^{-} \otimes_{\mathfrak B}\mathfrak
A^{-}$ can be shown to be a Hilbert bimodule over $\mathfrak B$,
which for $i=1,2$, leads to $*$-homorphisms $\vp^{i} : \mathfrak A
\lra \mathcal L_{\mathfrak B}(\mathfrak A^{-} \otimes \mathfrak
A^{-})$. Next is to def\/ine the (unital) $C^*$-algebra $\mathfrak A
\otimes_{\mathfrak B} \mathfrak A$ as the $C^*$-algebra contained in
$ \mathcal L_{\mathfrak B}(\mathfrak A^{-} \otimes \mathfrak
A^{-})$ that is generated by $\vp^1(\mathfrak A)$ and~$\vp^2(\mathfrak A)$. The last stage of the recipe for def\/ining a
compact quantum groupoid entails considering a certain coproduct
operation $\Delta : \mathfrak A \lra \mathfrak A
\otimes_{\mathfrak B} \mathfrak A$, together with a coinverse $Q :
\mathfrak A \lra \mathfrak A$ that it is both an algebra and
bimodule antihomomorphism. Finally, the following axiomatic
relationships are observed:
\begin{gather*}
(\ID \otimes_{\mathfrak B} \Delta) \circ \Delta  = (\Delta
\otimes_{\mathfrak B} \ID) \circ \Delta, \qquad (\ID \otimes_{\mathfrak
B} P) \circ \Delta = P , \qquad \tau \circ (\Delta \otimes_{\mathfrak
B} Q) \circ \Delta = \Delta \circ Q,
\end{gather*}
where $\tau$ is a f\/lip map: $\tau(a \otimes b) = (b \otimes a)$.

There is a natural extension of the above def\/inition of
a quantum compact groupoid
to a~\textit{locally compact} quantum groupoid by
taking $\mathsf{G}_{lc}$ to be a locally compact groupoid
(instead of a~compact groupoid), and then following the steps
in the above construction
with the topological groupoid $\mathsf{G}$ being replaced
by $\mathsf{G}_{lc}$.  Additional integrability and Haar
measure system conditions need however be also satisf\/ied
as in the general case of locally compact groupoid \textit{representations}
(for further details, see for example the monograph \cite{Buneci2k3},
the Appendix, and also our subsequent sections on groupoid and
categorical/functor representations).
In the last three sections we shall tacitly
consider quantum groupoids to be, in general,
\textit{locally compact} quantum groupoids that are endowed with a
Haar measure system (as described in  \cite{Buneci2k3} and
references cited therein), and also generated through the construction method recalled in this subsection
following Landsman~\cite{Landsman98}.

\section{Algebroids and their symmetries}\label{section4}

By an \emph{algebroid structure} $A$ we shall specif\/ically mean
also a ring, or more generally an algebra, but \emph{with several
objects} (instead of a single object), in the sense of Mitchell~\cite{Mitchell65}. Thus, an algebroid has been def\/ined by Mosa in~\cite{Mosa86}
and by Brown and Mosa \cite{Brown-Mosa86} as follows.

An \textit{$R$-algebroid } $A$ on a set of `objects' $A_0$ is a
directed graph over $A_0$ such that for each $x,y \in A_0$,
$A(x,y)$ has an $R$-module structure and there is an $R$-bilinear
function
\begin{gather*}
\circ : \ A(x,y) \times A(y,z) \lra A(x,z),
\end{gather*}
where $(a,b) \mapsto a\circ b$ is the composition, that satisf\/ies
the associativity condition, and the existence of identities. A
{\em pre-algebroid} has the same structure as an algebroid and the
same axioms except for the fact that the existence of identities
$1_x \in A(x,x)$ is not assumed. For example, if $A_0$ has exactly
one object, then an $R$-algebroid $A$ over $A_0$ is just an
$R$-algebra. An ideal in $A$ is then an example of a
pre-algebroid. Let now $R$ be a commutative ring.

An $R$-\textit{category } $\mathcal{A}$ is a category equipped with an
$R$-module structure on each $\Hom$ set such that the composition
is $R$-bilinear. More precisely, let us assume for instance that
we are given a commutative ring $R$ with identity. Then a small
$R$-category~-- or equivalently an \emph{$R$-algebroid}~-- will be
def\/ined as a category enriched in the monoidal category of
$R$-modules, with respect to the monoidal structure of tensor
product. This means simply that for all objects $b$, $c$ of $\mathcal{A}$, the
set $\mathcal{A}(b,c)$ is given the structure of an $R$-module, and
composition $\mathcal{A}(b,c) \times \mathcal{A}(c,d) \lra \mathcal{A}(b,d)$ is $R$-bilinear,
or is a morphism of $R$-modules $\mathcal{A}(b,c) \otimes_R \mathcal{A}(c,d) \lra
\mathcal{A}(b,d)$.

If $\grp$ is a groupoid  (or, more generally, a category) then we
can construct an \emph{$R$-algebroid} $R\grp$ as follows. The
object set of $R\grp$ is the same as that of $\grp$ and
$R \grp(b,c)$ is the free $R$-module on the set $\grp(b,c)$, with
composition given by the usual bilinear rule, extending the
composition of~$\grp$.

Alternatively, we can def\/ine $\bar{R}\grp(b,c)$ to be the set of
functions $\grp(b,c)\lra R$ with f\/inite support, and then we
def\/ine the \emph{convolution product} as follows:
\begin{gather}\label{(4.2)}
(f*g)(z)= \sum \{(fx)(gy)\mid z=x\circ y \}.
\end{gather}

As is well known, it  is the second construction which is natural
for the topological case, when we need to replace `function' by
`continuous function with compact support' (or \emph{locally compact
support} for the QFT extended symmetry sectors), and in this case $R
\cong \mathbb{C}$. The point we are making here is that to make the
usual construction and end up with an algebra rather than an
algebroid, is a procedure analogous to replacing a groupoid $\grp$
by a semigroup $G'= \grp \cup \{0\}$ in which the compositions not
def\/ined in $\grp$ are def\/ined to be $0$ in $G'$. We argue that this
construction removes the main advantage of groupoids, namely the
spatial component given by the set of objects.

At present, however, the question of how one can use
categorical duality in order to f\/ind the analogue of the diagonal of
a Hopf algebra remains open. Such questions require further work and
also future development of the theoretical framework proposed here
for extended symmetries and the related fundamental aspects of
quantum f\/ield theories. Nevertheless, for Fourier--Stieltjes
groupoid representations, there has already been substantial
progress made~\cite{Paterson2k3a} with the specif\/ication of their dual
Banach algebras (but not algebroids), in a manner similar to the
case of locally compact groups and their associated Fourier
algebras. Such progress will be further discussed in Section~\ref{section7}.

 A related problem that we are addressing next is how the much studied
theory of $C^*$-algebras and their representations would be naturally
extended to carefully selected $C^*$-algebroids so that novel applications in quantum physics become possible.
This is indeed a moot point because the classif\/ication problem for $C^*$-algebra representations is more complex and appears much more dif\/f\/icult to solve in the general case than it is in the case of von Neumann algebra representations. On the other hand, the extended symmetry links that we shall also discuss next, between locally compact groupoid \textit{unitary} representations and their induced $C^*$-algebra representations, also warrant further careful consideration.

\subsection[The weak $C^*$-Hopf algebroid and its symmetries]{The weak $\boldsymbol{C^*}$-Hopf algebroid and its symmetries}\label{section4.1}

Progressing to the next level of generality, let $A$ denote an
algebra with local identities in a~commutative subalgebra~$R
\subset A$. We adopt the def\/inition of a \emph{Hopf algebroid
structure on~$A$ over~$R$} following~\cite{Mrcun2k2}. Relative to
a ground f\/ield $\mathbb F$ (typically $\mathbb F = \mathbb C$ or
$\mathbb R$), the def\/inition commences by taking three $\mathbb
F$-linear maps, the \emph{comultiplication}
$\Delta : A \lra A \otimes_R A$, the \emph{counit} $\vep : A \lra R$,
and the \emph{antipode} $S : A \lra A$,
such that:
\begin{itemize}\itemsep=0pt
\item[$(i)$] $\Delta$ and $\vep$ are homomorphisms of left
$R$-modules satisfying $(\ID \otimes \vep) \circ \Delta = \ID$
and $(\vep \otimes \ID) \circ \Delta = \ID$.

\item[$(ii)$]
$\vep \vert_R = \ID,$ $\Delta \vert_R$ is the canonical embedding
$R \cong R \otimes_R R \subset A \otimes_R A$, and the two right
$R$-actions on $A \otimes_R A$ coincide on $\Delta A$.

\item[$(iii)$] $\Delta (ab) = \Delta (a) \Delta (b)$ for any $a,b
\in A$.

\item[$(iv)$] $S \vert_R = \ID$ and $S \circ S = \ID$.

\item[$(v)$]
$S(ab) = S(a) S(b)$ for any $a,b \in A$.

\item[$(vi)$] $\mu \circ (S \otimes \ID) \circ \Delta = \vep \circ
S$, where $\mu : A \otimes_R A \lra A$ denotes the multiplication.
\end{itemize}

If $R$ is a commutative subalgebra with local identities, then a
\emph{Hopf algebroid over $R$} is a~\textit{quadruple} $(A, \Delta, \vep,
S)$ where $A$ is an algebra which has $R$ for a subalgebra and has
local identities in $R$, and where $(\Delta, \vep, S)$ is a~Hopf algebroid structure on $A$ over $R$. Our interest lies in the fact that a Hopf-algebroid comprises a (universal) enveloping algebra for a quantum groupoid, thus hinting either at an adjointness situation or duality between the Hopf-algebroid and such a quantum groupoid.

\begin{definition}
Let $(A, \Delta, \vep, S)$ be a Hopf algebroid as above. We say
that $(A, \Delta, \vep, S)$ is a~\emph{weak $C^*$-Hopf algebroid}
when the following axioms are satisf\/ied:
\begin{itemize}\itemsep=0pt
\item[$(w1)$] $A$ is a unital $C^*$-algebra.
We set $\mathbb F = \mathbb C$.
\item[$(w2)$] The comultiplication $\Delta : A \lra A \otimes A$ is a
coassociative $*$-homomorphism. The counit is a~positive linear
map $\vep : A \lra R$ satisfying the above compatibility
condition. The antipode~$S$ is a complex-linear
anti-homomorphism and anti-cohomorphism $S : A \lra A$ (that is,
it reverses the order of the multiplication and comultiplication),
and is inverted under the $*$-structure: $S^{-1}(a) = S(a^*)^*$.

\item[$(w3)$]
\begin{gather*}
\Delta (\mathbf 1)  \equiv \mathbf 1_{(1)} \otimes \mathbf 1_{(2)}
= \text{projection}, \qquad \vep(ap)  = \vep(a \mathbf 1_{(1)}) \cdot
\vep(\mathbf 1_{(2)} p), \\
S(a_{(1)})a_{(2)} \otimes a_{(3)}  = (\mathbf 1 \otimes a) \cdot \Delta (\mathbf 1).
\end{gather*}
Here $a_{(1)} \otimes a_{(2)}$ is shorthand notation for the
expansion of $\Delta(a)$.

\item[$(w4)$]
The dual $\what{A}$ is def\/ined by the linear maps $\hat{a} : A
\lra \bC$. The structure of $\what{A}$ is canonically dualized
via the pairing and $\what{A}$ is endowed with a dual $*$-structure
via $\langle \hat{a}^*, a \rangle_A = \overline{\langle \hat{a},
S(a)^* \rangle_A}$. Further, $(\what{A}, \what{\Delta},
\hat{\vep}, \what{S})$ with $*$ and $\vep = \what{\mathbf 1}$, is
a weak $C^*$-Hopf algebroid.
\end{itemize}
\end{definition}

\section[Comparing groupoid and algebroid quantum symmetries:\\
weak Hopf $C^*$-algebroid vs. locally compact quantum groupoid symmetry]{Comparing groupoid and algebroid quantum symmetries:\\
weak Hopf $\boldsymbol{C^*}$-algebroid vs. locally compact quantum\\ groupoid symmetry}\label{section5}

At this stage, we make a comparison between the Lie group `classical'
symmetries discussed in Section~\ref{section2} and a schematic representation for
the extended groupoid and algebroid symmetries considered in
Sections~\ref{section3} and~\ref{section4}, as follows:
\begin{enumerate}\itemsep=0pt
\item[] \emph{Standard classical and quantum group/algebra symmetries}:

\item[] Lie groups ~
$\Longrightarrow$ ~ Lie algebras ~ $\Longrightarrow$ ~ universal
enveloping algebra ~ $\Longrightarrow$ ~ quantization $\rightarrow$
quantum group symmetry (or noncommutative (quantum) geometry).

\item[] \emph{Extended quantum, groupoid and algebroid, symmetries}:

\item[] Quantum groupoid/algebroid~ $\leftarrow$ ~ weak Hopf algebras  ~
$\longleftarrow$ ~ representations ~$\leftarrow$ ~ quantum groups.
\end{enumerate}

Our intention here is to view the latter scheme in terms of
\emph{weak Hopf $C^*$-algebroid}~-- and/or other~-- extended
symmetries, which we propose to do, for example, by incorporating
the concepts of \emph{rigged Hilbert spaces} and \emph{sectional
functions for a small category}. We note, however, that an
alternative approach to quantum groupoids has already been
reported in~\cite{Maltsiniotis92} (perhaps also related to
non-commutative geometry); this was later expressed in terms of
deformation-quantization:  the Hopf algebroid deformation of the
universal enveloping algebras of Lie algebroids \cite{Xu97} as the
classical limit of a quantum `groupoid'; this also parallels the
introduction of quantum `groups' as the deformation-quantization
of Lie bialgebras. Furthermore, such a Hopf algebroid approach~\cite{Lu96} leads to categories of Hopf algebroid modules~\cite{Xu97} which are monoidal, whereas the links between Hopf
algebroids and monoidal bicategories were investigated in~\cite{Day-Street97}.

As def\/ined in Section~\ref{section7} and the Appendix~\ref{appendixA}, let $(\mathsf{G}_{lc}, \tau)$
be a locally compact groupoid endowed with a (left) Haar system,
and let $A= C^*(\mathsf{G}_{lc}, \tau)$ be the convolution
$C^*$-algebra (we append $A$ with $\mathbf 1$ if necessary, so
that $A$ is unital). Then consider such a \textit{groupoid
representation}
\begin{gather}\label{(5.1)}
\Lambda : \ (\mathsf{G}_{lc}, \tau) \lra \{\mathcal
H_x, \sigma_x \}_{x \in X},
\end{gather}
that respects a compatible measure
$\sigma_x$ on $\mathcal H_x$ \cite{Buneci2k3}. On taking a state
$\rho$ on $A$, we assume a~parametrization
\begin{gather}\label{(5.2)} (\mathcal H_x, \sigma_x) := (\mathcal H_{\rho},
\sigma)_{x \in X}.
\end{gather}

Furthermore, each $\mathcal H_x$ is considered as a \emph{rigged Hilbert
space} \cite{Bohm-Gadella89}, that is, one also
 has the following nested inclusions:
\begin{gather*}
\Phi_x \subset (\mathcal H_x, \sigma_x) \subset
\Phi^{\times}_x,
\end{gather*}
in the usual manner, where $\Phi_x$ is a dense subspace of
$\mathcal H_x$ with the appropriate locally convex topology, and
$\Phi_x^{\times}$ is the space of continuous antilinear
functionals of $\Phi$. For each $x \in X$, we require $\Phi_x$ to
be invariant under $\Lambda$ and $\IM~ \Lambda \vert \Phi_x$ is a
continuous representation of $\mathsf{G}_{lc}$ on~$\Phi_x$. With these
conditions, representations of (proper) quantum groupoids that are
derived for weak $C^*$-Hopf algebras (or algebroids) modeled on
rigged Hilbert spaces could be suitable generalizations in the
framework of a Hamiltonian generated semigroup of time evolution
of a quantum system via integration of Schr\"odinger's equation
$\iota \hslash \frac{\del \psi}{\del t} = H \psi$ as studied in
the case of Lie groups \cite{Wickramasekara-Bohm2k2}. The
adoption of the rigged Hilbert spaces is also based on how the
latter are recognized as reconciling the Dirac and von Neumann
approaches to quantum theories~\cite{Bohm-Gadella89}.

Next let $\mathsf{G}_{lc}$ be a locally compact Hausdorf\/f
groupoid and $X$ a locally compact Hausdorf\/f space. In order to
achieve a small $C^*$-category we follow a suggestion of A.~Seda
(private communication) by using a general principle in the context
of Banach bundles \cite{Seda76,Seda82}. Let
\begin{gather*}
q= (q_1, q_2) : \
\mathsf{G}_{lc}\lra X \times X,
\end{gather*}
be a continuous, open and surjective
map. For each $z = (x,y) \in X \times X$, consider the f\/ibre
$\mathsf{G_z} = \mathsf{G}_{lc}(x,y) = q^{-1}(z)$, and set $\A_z =
C_0(\mathsf{G_z}) = C_0(\mathsf{G}_{lc})$ equipped with a uniform
norm $\Vert ~ \Vert_z$. Then we set $\A = \bigcup_z \A_z$. We form
a Banach bundle $p : \A \lra X \times X$ as follows. Firstly, the
projection is def\/ined via the typical f\/ibre $p^{-1}(z) = \A_z =
\A_{(x,y)}$. Let $C_c(\mathsf{G}_{lc})$ denote the continuous
complex valued functions on $\mathsf{G}_{lc}$ with compact support.
We obtain a \textit{sectional function} $\wti{\psi} : X \times X \lra \A$
def\/ined via restriction as $\wti{\psi}(z) = \psi \vert
\mathsf{G_z} = \psi \vert \mathsf{G}_{lc}$. Commencing from the
vector space $\gamma = \{\wti{\psi} : \psi \in C_c(\mathsf{G}_{lc})
\}$, the set $\{\wti{\psi}(z) : \wti{\psi} \in \gamma \}$ is dense
in $\A_z$. For each $\wti{\psi} \in \gamma$, the function $\Vert
\wti{\psi}(z) \Vert_z$ is continuous on $X$, and each $\wti{\psi}$
is a continuous section of $p : \A \lra X \times X$. These facts
follow from \cite[Theorem~1]{Seda82}. Furthermore, under the
convolution product $f* g$, \textit{the space $C_c(\mathsf{G}_{lc})$
forms an associative algebra over $\bC$}
(cf.\ \cite[Theorem~3]{Seda82}).

\begin{definition}\label{definition5.1}
The data proposed for a \emph{weak $C^*$-Hopf symmetry} consists of:
\begin{itemize}\itemsep=0pt
\item[(1)] A weak $C^*$-Hopf algebroid $(A, \Delta, \vep, S)$, where as above, $A = C^*(\grp, \tau)$ is constructed via sectional functions over a small category.
\item[(2)] A family of GNS representations
\begin{gather*}
(\pi_{\rho})_x : \  A \lra (\mathcal H_{\rho})_x := \mathcal H_x,
\end{gather*}
where for each, $x \in X$, $\mathcal H_x$ is a rigged Hilbert
space.
\end{itemize}
\end{definition}

\subsection[Grassmann--Hopf algebra and the Grassmann--Hopf algebroid]{Grassmann--Hopf algebra and the Grassmann--Hopf algebroid}\label{section5.1}

Let $V$ be a (complex) vector space ($\dim_{\bC} V = n$) and let
$\{e_0, e_1, \ldots, \}$ with identity $e_0 \equiv 1$, be the
generators of a Grassmann (exterior) algebra
\begin{gather*}
\Lambda^*V = \Lambda^0 V \oplus \Lambda^1 V \oplus \Lambda^2 V
\oplus  \cdots
\end{gather*}
subject to the relation $e_i e_j + e_j e_i = 0$. Following
\cite{Wickramasekara-Bohm2k2,Fauser2k4} we append this algebra with a Hopf structure to obtain a
`co-gebra' based on the interchange (or \emph{tangled
duality}):
\[
\text{(\textit{objects/points}, \textit{morphisms})}
\mapsto \text{(\textsl{morphisms}, \textsl{objects/points})}.
\] This
leads to a \emph{tangled duality} between
\begin{itemize}\itemsep=0pt
\item[(i)] the binary product $A \otimes A \ovsetl{m} A$, and

\item[(ii)] the coproduct $C \ovsetl{\Delta} C \otimes C$.
\end{itemize}
where the Sweedler notation \cite{Sweedler96}, with respect to an
arbitrary basis is adopted:
\begin{gather*}
\Delta (x) = \sum_r a_r \otimes b_r = \sum_{(x)} x_{(1)} \otimes
x_{(2)} = x _{(1)} \otimes x_{(2)},\\ \Delta (x^i) = \sum_i
\Delta^{jk}_i = \sum_{(r)} a^j_{(r)} \otimes b^k_{(r)} = x _{(1)}
\otimes x_{(2)}.
\end{gather*}
Here the $\Delta^{jk}_i$ are called `section coef\/f\/icients'. We
have then a generalization of associativity to coassociativity
\begin{gather*}
\begin{CD}
C  @> \Delta >> C \otimes C
\\ @VV \Delta V   @VV \ID \otimes \Delta V  \\ C \otimes C
 @> \Delta \otimes \ID >> C \otimes C \otimes C
\end{CD}
\end{gather*}
inducing a tangled duality between an associative (unital algebra
$\mathcal A = (A,m)$, and an associative (unital) `co-gebra'
$\mathcal C = (C, \Delta)$. The idea is to take this structure
and combine the Grassmann algebra $(\Lambda^*V, \wedge)$ with the
`co-gebra' $(\Lambda^*V, \Delta_{\wedge})$ (the `tangled dual')
along with the Hopf algebra compatibility rules: 1) the product
and the unit are `co-gebra' morphisms, and 2) the coproduct and
counit are algebra morphisms.

Next we consider the following ingredients:
\begin{itemize}\itemsep=0pt
\item[(1)]
the graded switch $\hat{\tau} (A \otimes B) = (-1)^{\del
A \del B} B \otimes A$;
\item[(2)]
the counit $\varepsilon$ (an algebra morphism) satisfying
$(\varepsilon \otimes \ID) \Delta = \ID = (\ID \otimes
\varepsilon) \Delta$;
\item[(3)] the antipode $S$.
\end{itemize}

The \textit{Grassmann--Hopf algebra} $\widehat{H}$ thus consists of the
\textit{septet}  $\widehat{H} = (\Lambda^*V, \wedge, \ID, \varepsilon,
\hat{\tau}, S)$.

 Its generalization to a \textit{Grassmann--Hopf algebroid} $H^{\wedge}$ is straightforward by def\/ining the $H^{\wedge}$-\textit{al\-gebroid} over $\widehat{H}$ as a~\textit{quadruple} $(\widehat{H}, \Delta, \vep, S)$, with $H^{\wedge}$ subject to the Hopf algebroid def\/ining axioms, and with $\widehat{H} = (\Lambda^*V, \wedge, \ID, \varepsilon, \hat{\tau}, S)$ subject to the standard Grassmann--Hopf algebra axioms stated above. We may also def\/ine $(\widehat{H}_w, \Delta, \vep, S)$ as a \emph{weak $C^*$-Grassmann--Hopf algebroid}, $H^{\wedge}_w$,  when $\widehat{H}_w$ is selected as a unital $C^*$-algebra, and axioms $(w2)$--$(w4)$ of the weak $C^*$-Hopf algebroid are also satisf\/ied by $H^{\wedge}_w$. We thus set $\mathbb F = \mathbb C$. Note however that the tangled-duals of Grassman--Hopf algebroids retain the intuitive interactions/dynamic diagram advantages of their physical, extended symmetry representations exhibited by the Grassman--Hopf algebras, gebras and co-gebras over those of either weak $C^*$-Hopf algebroids or weak Hopf $C^*$-algebras.

Alternatively, if $\grp$ is a groupoid (or, more generally, a category) then we
can construct a~\textit{Grassmann--Hopf algebroid} $H^{\wedge}$ as a special case of an \emph{$R$-algebroid} $H^{\wedge}\grp$. The object set of~$H^{\wedge}\grp$ is the same as that of $\grp$ and
$H^{\wedge}\grp(b,c)$ is the free $H^{\wedge}$-module on the set~$\grp(b,c)$, with
composition given by the usual bilinear rule, extending the composition of $\grp$. Furthermore, can def\/ine also def\/ine as above $\bar{H^{\wedge}}\grp(b,c)$ to be the set of functions $\grp(b,c)\lra H^{\wedge}$ with f\/inite support, and then we def\/ine the \emph{convolution product} as in equation~\eqref{(4.2)}
\begin{gather*}
(f*g)(z)= \sum \{(fx)(gy)\mid z=x\circ y \}.
\end{gather*}

As already pointed out, this second, convolution construction is natural for the topological~$\grp$ case, when we need to replace `function' by `continuous function with compact support'~-- or with \emph{locally compact
support} in the case of QFT extended symmetry sectors~-- and in this case one also has that
 $H^{\wedge}\cong \mathbb{C},$ or the def\/inition of a \textit{convolution Grassmann--Hopf algebroid} $H^{\wedge}_c$.  By also making $H^{\wedge}_c$ subject to axioms $(w1)$--$(w4)$ one obtains
a \textit{weak $C^*$-convolution Grassmann--Hopf algebroid}. Its duals are the corresponding co-algebroid,
$H^{\wedge *}_{c}$ and also the \textit{tangled weak $C^*$-convolution Grassmann--Hopf gebroid}, $\widetilde{H}^{\wedge}_c$ with distinct mathematical properties and physical signif\/icance.

\section[Non-Abelian algebroid representations of quantum state space geometry\\ in
quantum supergravity f\/ields]{Non-Abelian algebroid representations of quantum state\\ space geometry in
quantum supergravity f\/ields}\label{section6}

Supergravity, in essence, is an extended supersymmetric theory of
both matter and gravitation~\cite{Weinberg95}. A f\/irst approach to
supersymmetry relies on a curved `superspace' \cite{Wess-Bagger83},
and is analogous to supersymmetric gauge theories
(see, for example, Sections~27.1--27.3 of \cite{Weinberg95}).
Unfortunately, a complete non-linear supergravity theory might
be forbiddingly complicated and furthermore,
the constraints that need be made on
the graviton superf\/ield appear somewhat subjective,
according to \cite{Weinberg95}. On the other hand, the
second approach to supergravity is much more transparent
than the f\/irst, albeit theoretically less elegant.
The physical components of the gravitational superf\/ield
can be identif\/ied in this approach based on f\/lat-space
superf\/ield \mbox{methods} (Chapters~26 and~27 of~\cite{Weinberg95}).
By implementing the weak-f\/ield approximation one obtains several
of the most important consequences of supergravity theory,
including masses for the hypothetical gravitino and gaugino `particles'
whose existence may be expected from supergravity theories.
Furthermore, by adding on the higher order terms in the gravitational
constant to the supersymmetric transformation, the general coordinate
transformations form a closed algebra and the Lagrangian that
describes the interactions of the physical f\/ields is invariant
under such transformations. Quantization of such a f\/lat-space
superf\/ield would obviously involve its `deformation' as discussed
in Section~\ref{section2} above, and as a result its corresponding
\emph{supersymmetry algebra} would become \emph{non-commutative}.

\subsection[The metric superf\/ields]{The metric superf\/ields}\label{section6.1}

Because in supergravity both spinor and tensor f\/ields are being
considered, the gravitational f\/ields are represented in terms of
\emph{tetrads}, $e^a_\mu(x),$ rather than in terms of the general
relativistic metric $g_{\mu \nu}(x)$. The connections between
these two distinct representations are as follows:
\begin{gather*}
g_{\mu\nu}(x)=\eta_{ab}  e^a_\mu (x)e^b_\gamma(x),
\end{gather*}
with the general coordinates being indexed by $\mu$, $\nu,$ etc.,
whereas local coordinates that are being def\/ined in a locally
inertial coordinate system are labeled with superscripts $a$, $b$,
etc.;   $ \eta_{ab}$ is the diagonal matrix with elements $+1$, $+1$,
$+1$ and $-1$. The tetrads are invariant to two distinct types of
symmetry transformations~-- the local Lorentz transformations:
\begin{gather*}
e^a_\mu (x)\longmapsto \Lambda^a_b (x) e^b_\mu (x),
\end{gather*}
(where $\Lambda^a_b$ is an arbitrary real matrix), and the general
coordinate transformations:
\begin{gather*}
x^\mu \longmapsto (x')^\mu(x) .
\end{gather*}
In a weak gravitational f\/ield the tetrad may be represented as:
\begin{gather*}
e^a_\mu (x)=\delta^a_\mu(x)+ 2\kappa \Phi^a_\mu (x),
\end{gather*}
where $\Phi^a_\mu(x)$ is small compared with $\delta^a_\mu(x)$ for
all $x$ values, and $\kappa= \sqrt{8}\pi G$, where $G$ is Newton's
gravitational constant. As it will be discussed next, the
supersymmetry algebra (SA) implies that the graviton has a
fermionic superpartner, the hypothetical \emph{gravitino}, with
helicities $\pm 3/2$. Such a self-charge-conjugate massless
particle as the gravitiono with helicities $\pm 3/2$ can only have
\emph{low-energy} interactions if it is represented by a Majorana
f\/ield $\psi _\mu(x)$ which is invariant under the gauge
transformations:
\begin{gather*}
\psi _\mu(x)\longmapsto \psi _\mu(x)+\delta _\mu \psi(x) ,
\end{gather*}
with $\psi(x)$ being an arbitrary Majorana f\/ield as def\/ined in
\cite{Grisaru-Pendleton77}.
The tetrad f\/ield $\Phi _{\mu
\nu}(x)$ and the graviton f\/ield $\psi _\mu(x)$ are then
incorporated into a term $H_\mu (x,\theta)$ def\/ined as the
\emph{metric superfield}. The relationships between $\Phi _{\mu _
\nu}(x)$ and $\psi _\mu(x)$, on the one hand, and the components
of the metric superf\/ield $H_\mu (x,\theta)$, on the other hand,
can be derived from the transformations of the whole metric
superf\/ield:
\begin{gather*}
H_\mu (x,\theta)\longmapsto H_\mu (x,\theta)+ \Delta _\mu
(x,\theta),
\end{gather*}
by making the simplifying~-- and physically realistic~-- assumption
of a weak gravitational f\/ield (further details can be found, for
example, in Chapter~31 of Vol.~3 of~\cite{Weinberg95}). The interactions
of the entire superf\/ield $H_\mu (x)$ with matter would be then
described by considering how a weak gravitational f\/ield,
$h_{\mu \nu}$ interacts with an energy-momentum tensor $T^{\mu
\nu}$ represented as a linear combination of components of a real
vector superf\/ield $\Theta^\mu$. Such interaction terms would,
therefore, have the form:
\begin{gather*}
 I_{\mathcal M}= 2\kappa \int dx^4 [H_\mu \Theta^\mu]_D ,
\end{gather*}
($\mathcal M$ denotes `matter') integrated over a four-dimensional
(Minkowski) spacetime with the metric def\/ined by the superf\/ield
$H_\mu (x,\theta)$. The term $\Theta^\mu$, as def\/ined above, is
physically a \emph{supercurrent} and satisf\/ies the conservation
conditions:
\begin{gather*}
\gamma^\mu \mathbf{D} \Theta _\mu = \mathbf{D} ,
\end{gather*}
where $\mathbf{D}$ is the four-component super-derivative and $X$
denotes a real chiral scalar superf\/ield. This leads immediately to
the calculation of the interactions of matter with a weak
gravitational f\/ield as:
\begin{gather*}
I_{\mathcal M} = \kappa \int d^4 x T^{\mu \nu}(x)h_{\mu \nu}(x) ,
\end{gather*}
It is interesting to note that the gravitational actions for the
superf\/ield that are invariant under the generalized gauge
transformations $H_\mu \longmapsto H _\mu  + \Delta _\mu$ lead to
solutions of the Einstein f\/ield equations for a homogeneous,
non-zero vacuum energy density $\rho _V$ that correspond to either
a~de Sitter space for $\rho _V>0$, or an anti-de Sitter space \cite{Witten98} for
$\rho _V <0$. Such spaces can be represented in terms of the
hypersurface equation
\begin{gather*}
x^2_5 \pm \eta _{\mu,\nu} x^\mu x^\nu = R^2 ,
\end{gather*}
in a quasi-Euclidean f\/ive-dimensional space with the metric
specif\/ied as:
\begin{gather*}
ds^2 = \eta _{\mu,\nu} x^\mu x^\nu \pm dx^2_5 ,
\end{gather*}
with ``$+$' for de Sitter space and `$-$' for anti-de Sitter space,
respectively.

The spacetime symmetry groups, or groupoids~-- as the case may
be~-- are dif\/ferent from the `classical' Poincar\'e symmetry group
of translations and Lorentz transformations. Such spacetime
symmetry groups, in the simplest case, are therefore the
$\rO(4,1)$ group for the de Sitter space and the $\rO(3,2)$ group
for the anti-de Sitter space. A detailed calculation indicates
that the transition from ordinary f\/lat space to a bubble of
anti-de Sitter space is \emph{not} favored energetically and,
therefore, the ordinary (de Sitter) f\/lat space is stable (cf.\
\cite{Coleman-DeLuccia80}), even though quantum f\/luctuations
might occur to an anti-de Sitter bubble within the limits
permitted by the Heisenberg uncertainty principle.

\subsection[Supersymmetry algebras and Lie ($\mathbb{Z}_2$-graded)
superalgebras]{Supersymmetry algebras and Lie ($\boldsymbol{\mathbb{Z}_2}$-graded)
superalgebras}\label{section6.2}

It is well known that \emph{continuous symmetry transformations}
can be represented in terms of a \emph{Lie algebra} of linearly
independent \emph{symmetry generators} $t_j$ that satisfy the
commutation relations:
\begin{gather*}
[t_j,t_k] = \iota \Sigma_l C_{jk} t_l ,
\end{gather*}

Supersymmetry is similarly expressed in terms of the symmetry
generators $t_j$ of a \textit{graded (`Lie') algebra}~-- which is in
fact def\/ined as a \textit{superalgebra}~-- by satisfying relations of the
general form:
\begin{gather*}
t_j t_k - (-1)^{\eta _j \eta _k} t_k  t_j = \iota \Sigma_l C_{jk}
^l t_l .
\end{gather*}
The generators for which $\eta _j =1$ are fermionic whereas those
for which $\eta _j =0$ are bosonic. The coef\/f\/icients $C^l_{jk}$
are structure constants satisfying the following conditions:
\begin{gather*}
C _{jk} ^l = -(-1)^{\eta _j \eta _k} C _{jk} ^l .
\end{gather*}
If the generators $t_j$ are quantum Hermitian operators, then the
structure constants satisfy the reality conditions $C_{jk}^* = -
C_{jk}$.  Clearly, such a graded algebraic structure is a superalgebra
and not a proper Lie algebra; thus graded Lie algebras are often called
\textit{Lie superalgebras} \cite{Kac77}.

The standard computational approach in QM utilizes the $S$-matrix
approach, and therefore, one needs to consider the general,
\emph{graded} `Lie algebra' of \emph{supersymmetry generators} that
commute with the $S$-matrix. If one denotes the fermionic generators
by~$Q$, then $U^{-1}(\Lambda)Q U(\Lambda)$ will also be of the
same type when $U(\Lambda)$ is the quantum operator corresponding
to arbitrary, homogeneous Lorentz transformations
$\Lambda^{\mu \nu}$. Such a group of generators provide therefore a
representation of the homogeneous Lorentz group of transformations
$ \mathbb{L}$. The irreducible representation of the homogeneous
Lorentz group of transformations provides therefore a
classif\/ication of such individual generators.

\subsubsection{Graded `Lie' algebras and superalgebras}\label{section6.2.1}

 A set of quantum operators $Q^{AB}_{jk}$ form an $\mathbf A$,
$\mathbf B$ representation of the group $\mathbb{L}$ def\/ined above
which satisfy the commutation relations:
\begin{gather*}
[\mathbf{A},Q^{AB}_{jk}] = -[\Sigma _j' J^A _{j j'}, Q^{AB}_{j'k}],
\end{gather*}
and
\begin{gather*}
[\mathbf{B},Q^{AB}_{jk}] = -[\Sigma _{j'} J^A _{k k'},
Q^{AB}_{jk'}] ,
\end{gather*}
with the generators $\mathbf{A}$ and $\mathbf{B}$ def\/ined by
$\mathbf{A}\equiv (1/2)(\mathbf{J} \pm i\mathbf{K})$ and
$\mathbf{B} \equiv (1/2)(\mathbf{J }- i\mathbf{K})$, with
$\mathbf{J}$ and $\mathbf{K}$ being the Hermitian generators of
rotations and `boosts', respectively.

In the case of the two-component Weyl-spinors $Q _{jr}$ the
Haag--Lopuszanski--Sohnius (HLS) theorem applies, and thus the
fermions form a \emph{supersymmetry algebra} def\/ined by the
anti-commutation relations:
\begin{gather*}
[Q _{jr}, Q _{ks}^*] = 2\delta _{rs} \sigma^\mu _{jk} P _\mu ,
\qquad [Q _{jr}, Q _{ks}] = e _{jk} Z _{rs} ,
\end{gather*}
where $P _\mu$ is the 4-momentum operator, $Z_{rs} = -Z _{s r}$
are the bosonic symmetry generators, and~$\sigma _\mu$ and
$\mathbf{e}$ are the usual~$2 \times 2$ Pauli matrices.
Furthermore, the fermionic generators commute with both energy and
momentum operators:
\begin{gather*}
[P _\mu,Q _{jr}] = [P _\mu, Q^* _{jr}] = 0 .
\end{gather*}
The bosonic symmetry generators $Z _{ks}$ and $Z^* _{ks}$
represent the set of \emph{central charges} of the supersymmetric
algebra:
\begin{gather*}
~[Z _{rs}, Z^* _{tn}] = [Z^* _{rs}, Q _{jt}]=  [Z^* _{rs}, Q^*
_{jt}]= [Z^* _{rs}, Z^* _{tn}]=0 .
\end{gather*}
From another direction, the Poincar\'e symmetry mechanism of
special relativity can be extended to new algebraic systems
\cite{Tanasa2k6}.
In \cite{Moultaka-etal2k5} in view of such
extensions, are considered invariant-free Lagrangians and bosonic
multiplets  constituting a symmetry that interplays with (Abelian)
$\U(1)$-gauge symmetry that may possibly be described in
categorical terms, in particular, within the notion of a
\emph{cubical site} \cite{Grandis-Mauri2k3}.

We shall proceed to introduce in the next section generalizations
of the concepts of Lie algebras and graded Lie algebras to the
corresponding Lie \emph{algebroids} that may also be regarded as
$C^*$-convolution representations of \emph{quantum gravity
groupoids} and superf\/ield (or supergra\-vi\-ty) supersymmetries. This
is therefore a novel approach to the proper representation of the
\emph{non-commutative geometry of quantum spacetimes}~-- that are
\emph{curved} (or `deformed') by the presence of \emph{intense}
gravitational f\/ields~-- in the framework of \emph{non-Abelian,
graded Lie algebroids}. Their correspondingly \emph{deformed
quantum gravity groupoids} (QGG) should, therefore, adequately
represent supersymmetries modif\/ied by the presence of such intense
gravitational f\/ields on the Planck scale. Quantum f\/luctuations
that give rise to quantum `foams' at the Planck scale may be then
represented by \emph{quantum homomorphisms} of such QGGs. If the
corresponding graded Lie algebroids are also \emph{integrable},
then one can reasonably expect to recover in the limit of $\hbar
\rightarrow 0$ the Riemannian geometry of general relativity and
the \emph{globally hyperbolic spacetime} of Einstein's classical
gravitation theory (GR), as a result of such an integration to the
\emph{quantum gravity fundamental groupoid} (QGFG). The following
subsection will def\/ine the precise mathematical concepts
underlying our novel quantum supergravity and extended
supersymmetry notions.

\subsection[Extending supersymmetry in relativistic quantum supergravity:\\
Lie bialgebroids and a novel graded Lie algebroid concept]{Extending supersymmetry in relativistic quantum supergravity:\\
Lie bialgebroids and a novel graded Lie algebroid concept}\label{section6.3}

Whereas not all Lie algebroids are integrable to Lie groupoids, there is
a subclass of the latter called sometimes `Weinstein groupoids' that are
in a one-to-one correspondence with their Lie algebroids.

\subsubsection{Lie algebroids and Lie bialgebroids}\label{section6.3.1}

One can think of a Lie algebroid as generalizing the idea of a
tangent bundle where the tangent space at a point is ef\/fectively
the equivalence class of curves meeting at that point (thus
suggesting a groupoid approach), as well as serving as a site on
which to study inf\/initesimal geometry (see, e.g.,~\cite{Mackenzie2k5}).
Specif\/ically, let~$M$ be a manifold and let $\mathfrak X(M)$
denote the set of vector f\/ields on~$M$. Recall that a Lie
algebroid over $M$ consists of a vector bundle $E \lra M$,
equipped with a~Lie bracket $[~,~]$ on the space of sections
$\gamma(E)$, and a~bundle map $\Upsilon : E \lra TM$, usually
called the \emph{anchor}. Further, there is an induced map $
\Upsilon : \gamma (E) \lra \mathfrak X(M)$, which is required to
be a~map of Lie algebras, such that given sections $\a, \beta \in
\gamma(E)$ and a dif\/ferentiable function $f$, the following
Leibniz rule is satisf\/ied:
\begin{gather*}
[ \a, f \beta] = f [\a, \beta] + (\Upsilon (\a)) \beta.
\end{gather*}
A typical example of a Lie algebroid is when $M$ is a Poisson
manifold and $E=T^*M$ (the cotangent bundle of $M$).

Now suppose we have a Lie groupoid $\mathsf{G}$:
\begin{gather*}
r,s: \ \xymatrix{ \mathsf{G} \ar@<1ex>[r]^r \ar[r]_s & \mathsf{G}^{(0)}}=M.
\end{gather*}
There is an associated Lie algebroid $\A = \A( \mathsf{G})$, which in the
guise of a vector bundle, is in fact the restriction to $M$ of the
bundle of tangent vectors along the f\/ibers of $s$ (i.e.\ the
$s$-vertical vector f\/ields). Also, the space of sections $\gamma
(\A)$ can be identif\/ied with the space of $s$-vertical,
right-invariant vector f\/ields $\mathfrak X^s_{\rm inv} (\mathsf{G})$ which
can be seen to be closed under $[~,~]$, and the latter induces a
bracket operation on $\gamma (A)$ thus turning $\A$ into a Lie
algebroid. Subsequently, a Lie algebroid $\A$ is integrable if
there exists a Lie groupoid $\mathsf{G}$ inducing $\A$.

\subsubsection{Graded Lie bialgebroids and symmetry breaking}\label{section6.3.2}

A \emph{Lie bialgebroid} is a Lie algebroid $E$ such that $E^*
\lra M$ also has a Lie algebroid structure. Lie bialgebroids are
often thought of as the inf\/initesimal variations of Poisson
groupoids. Specif\/ically, with regards to a Poisson structure
$\Lambda$, if $\xymatrix{(G \ar@<0.5ex>[r] \ar@<-0.5ex>[r] & M, \Lambda
)}$ is a Poisson groupoid and if $E\mathsf{G}$ denotes the Lie algebroid
of $\mathsf{G}$, then $(E\mathsf{G}, E^*\mathsf{G})$ is a Lie bialgebroid.
Conversely, a~Lie bialgebroid structure on the Lie algebroid of a
Lie groupoid can be integrated to a Poisson groupoid structure.
Examples are Lie bialgebras which correspond bijectively with
simply connected Poisson Lie groups.

\subsection{Graded Lie algebroids and bialgebroids}\label{section6.4}

 A grading on a Lie algebroid follows by endowing a graded Jacobi
bracket on the smooth functions $C^{\infty}(M)$ (see
\cite{Grabowski-Marmo2k1}). A Graded Jacobi bracket of degree $k$ on a
$\bZ$-graded associative commutative algebra $\A = \bigoplus_{i
\in \bZ} \A$ consists of a graded bilinear map
\begin{gather*}
\{\cdot ,  \cdot \} : \ \A \times \A \lra \A,
\end{gather*}
of degree $k$ (that is, $\vert \{a,b\} \vert = \vert a \vert +
\vert b \vert + k$) satisfying:
\begin{itemize}\itemsep=0pt
\item[1)] $\{a,b\} = - (-1)^{\langle a + k,  b + k \rangle}
\{b, a\}$~(graded anticommutativity);

\item[2)] $\{a,bc \} = \{a, b\}c  + (-1)^{\langle a + k, b\rangle}~b \{a, c\} -
 \{a, \mathbf 1\}bc $ (graded generalized Leibniz rule);

\item[3)] $\{\{a,b\}, c \} =\{a, \{b, c\}\} -
(-1)^{\langle a + k, b + k \rangle} \{b, \{a, c \}\}$ (graded
Jacobi identity),
\end{itemize}
where $\langle\ \cdot, \cdot \rangle$ denotes the usual pairing in
$\bZ^n$. Item 2) says that $\{~,~ \}$ corresponds to a
f\/irst-order bidif\/ferential operator on $\A$, and an odd Jacobi
structure corresponds to a generalized \textit{graded Lie bialgebroid}.

Having considered and also introduced several extended quantum
symmetries, we are summarizing in the following diagram the key
links between such quantum symmetry related concepts; included
here also are the groupoid/algebroid representations of quantum
symmetry and QG supersymmetry breaking.  Such interconnections
between quantum symmetries and supersymmetry are depicted in the
following diagram in a manner suggestive of novel physical
applications that will be reported in further detail in a
subsequent paper \cite{Baianu-etal2k9b}
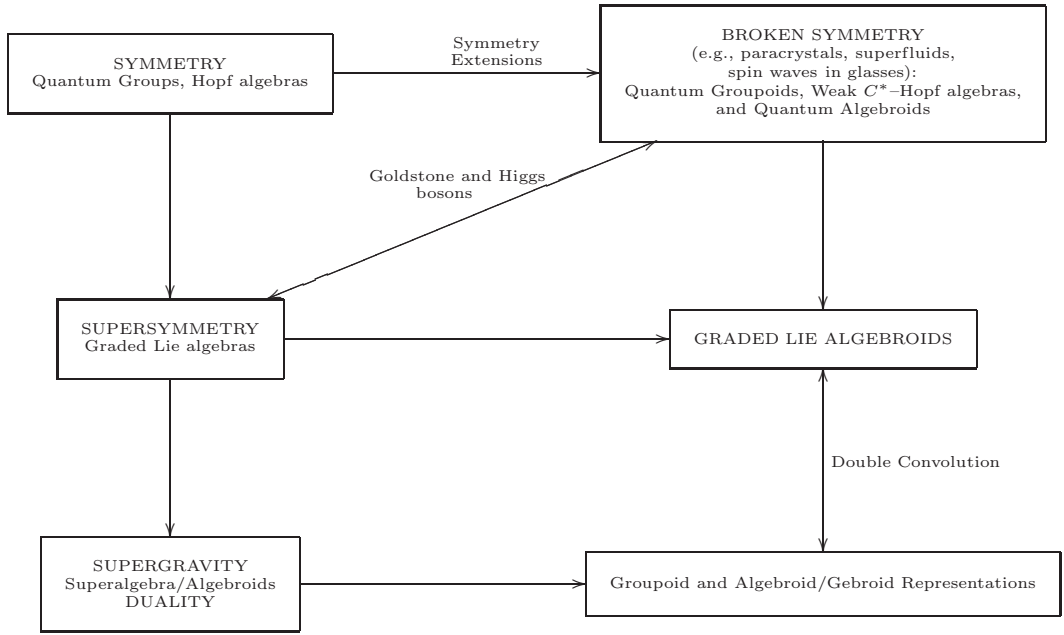
\begin{gather*}
\mbox{\tiny$
\xymatrix@C=8pc@R=5pc{ *+++[F]{\txt{SYMMETRY\\Quantum Groups, Hopf
algebras}} \ar[d] \ar[r]^{\txt{Symmetry \\ Extensions}} &
*+++[F]{\txt{BROKEN SYMMETRY \\ (e.g., paracrystals, superf\/luids, \\ spin waves in glasses):\\
Quantum Groupoids, Weak $C^*$--Hopf algebras,\\ and Quantum Algebroids}} \ar[dl] \ar[d]
\\ *+++[F]{\txt{SUPERSYMMETRY\\
Graded Lie algebras}} \ar[d] \ar[ur]^{\txt{Goldstone and Higgs \\ bosons~~~~~}}
 \ar[r]  &
*+++[F]{\txt{GRADED LIE ALGEBROIDS}} \ar[d]^{\txt{Double Convolution}}
\\ *+++[F]{\txt{SUPERGRAVITY \\ Superalgebra/Algebroids \\DUALITY}} \ar[r] &
*+++[F]{\txt{Groupoid and Algebroid/Gebroid Representations}}
\ar[u]}$}
\end{gather*}

The extended quantum symmetries formalized in the next
section are def\/ined as representations of the groupoid, algebroid
and categorical structures considered in the above sections.


\section[Extended quantum symmetries as algebroid and groupoid  representations]{Extended quantum symmetries as algebroid\\ and groupoid  representations}\label{section7}

\subsection{Algebroid representations}\label{section7.1}

A def\/inition of a \textit{vector bundle representation} $(VBR)$,
$(\rho,\emph{V})$, of a Lie algebroid $\Lambda$ over a manifold~$M$ was given in \cite{Levin-Olshanetsky08}
as a vector
bundle $V \lra M$ and a bundle map $\rho$ from $\Lambda$ to the
bundle of order $\leq 1$ dif\/ferential operators $\mathsf{D}:
\Gamma (V) \lra \Gamma (V)$ on sections of $V$ compatible with the
anchor map and commutator such that:
\begin{itemize}\itemsep=0pt
\item[$(i)$] for any $\epsilon_1,\epsilon_2 \in \gamma $ the symbol
$\Symb(\rho(\epsilon))$ is a scalar equal to the anchor of
$\epsilon$:
\begin{gather*}
\Symb(\rho(\epsilon)) = \delta_\epsilon {\rm Id}_V,
\end{gather*}
\item[$(ii)$] for any $\epsilon_1,\epsilon_2 \in \gamma (\Lambda)$ and
$f \in C^{\infty}(M)$ we have $[\rho(\epsilon_1),\rho(\epsilon_2)]
= \rho([\epsilon_1,\epsilon_2])$.
\end{itemize}

In $(ii)$ $C^{\infty}(M)$ is the algebra of $\Re$-valued functions on $M$.

\subsection[Hopf and weak Hopf $C^*$-algebroid representations]{Hopf and weak Hopf $\boldsymbol{C^*}$-algebroid representations}\label{section7.2}

We shall begin in this section with a consideration of the Hopf
algebra representations that are known to have additional
structure to that of a Hopf algebra. If $H$ is a Hopf algebra and~$A$ is an algebra with the product operation $\mu: A \otimes A
\longrightarrow A$, then a linear map $\rho: H \otimes A
\longrightarrow A$ is an \emph{algebra representation} of $H$ if
in addition to being a (vector space) representation of $H$, $\mu$
is also an $H$-intertwiner. If A happens to be unital, it will
also be  required that there is an $H$-intertwiner from
$\epsilon_H$ to $A$ such that the unity of $\epsilon_H$ maps to the
unit of~$A$.

On the other hand, the Hopf-algebroid $H_A$ over
$C_c^{\infty}(M)$, with $M$ a smooth manifold, is sometimes
considered as a quantum groupoid because one can construct its
spectral $\acute{e}tale$ Lie groupoid $\mathsf{G}_s \rightarrow
p(H_A)$ representation beginning with the groupoid algebra
$C_c(\mathsf{G})$ of smooth functions with compact support on
$\mathsf{G}_{lc}$; this is an $\acute{e}tale$ Lie groupoid for $M$'s
that are not necessarily Hausdorf\/f (cf.~\cite{Mrcun2k2,Mrcun07}).
Recently, Konno~\cite{Konno2k8} reported a systematic construction of
both f\/inite and
inf\/inite-dimensional \textit{dynamical representations of a
H-Hopf algebroid}(introduced in \cite{Etingof-Varchenko98}),
and their parallel structures to the \textit{quantum affine
algebra} $U_q(\hat{sl}_2)$. Such generally non-Abelian structures
are constructed in terms of the Drinfel'd generators of the
quantum af\/f\/ine algebra $U_q(\hat{sl}_2)$ and a Heisenberg algebra.
The structure of the tensor product of two evaluation representations
was also provided by Konno~\cite{Konno2k8}, and an elliptic analogue of
the Clebsch--Gordan coef\/f\/icients was expressed by using
certain balanced elliptic hypergeometric
series ${}_{12} V_{11}$.

\subsection{Groupoid representations}\label{section7.3}

Whereas group representations of quantum unitary operators are
extensively employed in standard quantum mechanics, the applications
of groupoid representations are still under development. For
example, a description of stochastic quantum mechanics in curved
spacetime \cite{Drechsler-Tuckey96} involving a Hilbert bundle is
possible in terms of \textit{groupoid representations} which can
indeed be def\/ined on such a Hilbert bundle $(X* \H,\pi)$, but cannot
be expressed as the simpler group representations on a Hilbert space
$\H$. On the other hand, as in the case of group representations,
unitary groupoid representations induce associated $C^*$-algebra
representations. In the next subsection we recall some of the basic
results concerning groupoid representations and their associated
groupoid $*$-algebra representations. For further details and recent
results in the mathematical theory of groupoid representations one
has also available the succint monograph \cite{Buneci2k3} and
references cited therein
(\url{www.utgjiu.ro/math/mbuneci/preprint.html},
\cite{Buneci2k2a,Buneci2k2b,Buneci2k3a,Buneci2k3b,Buneci2k3c,Buneci2k3d,Buneci2k4,Buneci2k5a,Buneci2k5b}).

\subsection[Equivalent groupoid and algebroid representations:
the correspondence between\\ groupoid unitary representations  and the
associated $C^*$-algebra representations]{Equivalent groupoid and algebroid representations:\\
the correspondence between groupoid unitary representations\\  and the
associated $\boldsymbol{C^*}$-algebra representations}\label{section7.4}

We shall brief\/ly consider here a main result due to Hahn \cite{Hahn78a}
that relates groupoid and associated groupoid algebra representations \cite{Hahn78b}:

\begin{theorem}[Theorem 3.4 on p.~50 in \cite{Hahn78a}]\label{theorem7.1}
Any representation of a groupoid $\mathsf{G}_{lc}$ with Haar measure $(\nu,
\mu)$ in a separable Hilbert space $\H$ induces a $*$-algebra
representation $f \mapsto X_f$ of the associated groupoid algebra
$ \Pi (\mathsf{G}_{lc}, \nu)$ in $L^2 (U_{\mathsf{G}_{lc}} , \mu, \H )$ with the
following properties:
\begin{itemize}\itemsep=0pt
\item[$(1)$]
For any $l,m \in \H $ ,
one has that
$| \langle X_f(u \mapsto l), (u \mapsto m)\rangle | \leq \left\|f_l\right\|
\left\|l \right\| \left\|m \right\|,$
and

\item[$(2)$]  $M_r (\alpha) X_f = X_{f \alpha \circ r}$,  where
$M_r: L^\infty (U_{\grp}, \mu, \H)   \longrightarrow \mathcal L (
L^2 (U_{\grp}, \mu, \H))$, with $M_r (\alpha)j = \alpha \cdot j$.
\end{itemize}
Conversely, any $*$-algebra representation with the above two
properties induces a groupoid representation, $X$, as follows:
\begin{gather*}
\left\langle X_f , j, k\right\rangle   =   \int
f(x)[X(x)j(d(x)),k(r(x))d \nu (x)].
\end{gather*}
\end{theorem}

Furthermore, according to Seda  \cite[p.~116]{Seda86}  the continuity of a
Haar system is equivalent to the continuity of the convolution
product $f*g$ for any pair $f$, $g$ of continuous functions with
compact support. One may thus conjecture that similar results
could be obtained for functions with \textit{locally compact}
support in dealing with convolution products of either locally
compact groupoids or quantum groupoids. Seda' s result also implies
that the convolution algebra $C_c (\mathsf{G})$ of a groupoid $\mathsf{G}$ is
closed with respect to convolution if and only if the f\/ixed Haar
system associated with the measured groupoid $\mathsf{G}$ is
\textit{continuous}~\cite{Buneci2k3}.

In the case of groupoid algebras of transitive groupoids,
\cite{Buneci2k3} and in related~\cite{Buneci2k3, Buneci2k2b, Buneci2k3a,
Buneci2k3b,Buneci2k3c,Buneci2k3d,Buneci2k4,Buneci2k5a,Buneci2k5b} showed that
representations of a measured groupoid $({\mathsf{G}, [\int \nu ^u d
\tilde{ \lambda} (u)] = [\lambda]})$ on a separable Hilbert space
$\H$ induce  \textit{non-degenerate}
$*$-representations  $f \mapsto  X_f$ of the associated groupoid
algebra $ \Pi (\mathsf{G}, \nu,\tilde{ \lambda})$ with properties
formally similar to (1) and (2) above~\cite{Buneci2k6}.
 Moreover, as in the case
of groups, \textit{there is a correspondence between the unitary
representations of a~groupoid and its associated $C^*$-convolution
algebra representations} of~\cite[p.~182]{Buneci2k3}, the latter
involving however f\/iber bundles of Hilbert spaces instead of
single Hilbert spaces. Therefore, groupoid representations appear
as the natural construct for algebraic quantum f\/ield theories in
which nets of local observable operators in Hilbert space f\/iber
bundles were introduced by Rovelli in~\cite{Rovelli98}.

\subsection[Generalized Fourier--Stieltjes transforms of groupoids: Fourier--Stieltjes algebras\\ of locally compact groupoids and quantum groupoids; left regular groupoid\\ representations and the Fourier algebra of a measured groupoid]{Generalized Fourier--Stieltjes transforms of groupoids:\\ Fourier--Stieltjes algebras of locally compact groupoids\\ and quantum groupoids; left regular groupoid representations\\ and the Fourier algebra of a measured groupoid}

We shall recall f\/irst that the \emph{Fourier--Stieltjes algebra}
$\mathbf{B}(G_{lc})$ of a locally compact group $G_{lc}$ is def\/ined by the
space of coef\/f\/icients $(\xi, \eta)$ of \textit{Hilbert space
representations} of $G_{lc}$. In the special case of left regular
representations and a measured groupoid, $\mathsf{G}$, the
Fourier--Stieltjes algebra $\mathbf{B}(\mathsf{G}, {\nu^u}, \mu)$~-- def\/ined as an involutive subalgebra of $L^\infty (\mathsf{G})$~-- becomes
the \textit{Fourier algebra} $\textbf{A}(\mathsf{G})$ def\/ined by Renault
\cite{Renault97}; such algebras are thus def\/ined as a set of
\textit{representation coefficients} $(\mu, U_{\mathsf{G}}*H, L)$, which
are ef\/fectively realized as a function $(\xi , \eta ): \mathsf{G}
\longrightarrow  \mathbb C$, def\/ined by
\begin{gather*}
(\xi , \eta )(x):= \left\langle{\xi(r(x)), \hat{L}(x) \eta(d(x))}\right\rangle,
\end{gather*}
(see~\cite[pp.~196--197]{Buneci2k3}).

The Fourier--Stieltjes (FS) and Fourier (FR) algebras, respectively, $\mathbf{B}(G_{lc})$, $\textbf{A}(G_{lc})$, were f\/irst studied by P.~Eymard for a general locally compact group $G_{lc}$ in~\cite{Eymard64}, and have since played ever increasing roles in harmonic analysis and in the study of the operator algebras generated by~$G_{lc}$.

 Recently, there is also a considerable interest in developing
extensions of these two types of algebras for \textit{locally
compact groupoids} because, as in the group case, such algebras
play a~useful role both in the study of the theory of quantum
operator algebras and that of groupoid operator algebras.
Furthermore, as discussed in the Introduction, there are new links
between (physical) scattering theories for paracrystals, or other
systems with local/partial ordering such as glasses/
`non-crystalline' solids, and the generalizations of Fourier
transforms that realize the well-established duality between the
physical space,~$S$, and the `dif\/fraction', or \textit{reciprocal},
space, $\R = \tilde{S}$. On the other hand, the duality between
the real time of quantum dynamics/resonant processes, $T$,  and
the `spectral space', $\F = \tilde{T}$, of resonance frequencies
(and the corresponding quanta of energies, $h\nu$) for electrons,
nucleons and other particles in bound conf\/igurations is just as
well-established by comparison with that occurring between the
`real' and reciprocal spaces in the case of electrons, neutron or
emf/$X$-ray dif\/fraction and scattering by periodic and aperiodic
solids. The deep quantum connection between these two fundamental
dualities, or symmetries, that seem to be ubiquitous in nature,
can possibly lead to an unif\/ied quantum theory of
\textit{dispersion} in solids, liquids, superf\/luids and plasmas.

Let $X$ be a locally compact Hausdorf\/f space and $C(X)$  the
algebra of bounded, continuous, complex-valued functions on $X$.
Then denote the space of continuous functions in $C(X)$ that vanish
at inf\/inity by~$C_0(X)$, while $C_c(X)$ is the space of functions in
$C(X)$ with compact support. The space of complex, bounded, regular
Borel measures on~$X$ is then denoted by~$M(X)$. The Banach spaces
$\mathbf{B}(\mathsf{G}_{lc})$, $\mathbf{A}(\mathsf{G}_{lc})$ (where
$\mathsf{G}_{lc}$ denotes a locally compact groupoid) as considered
here occur naturally in the group case in both non-commutative
harmonic analysis and duality theory.  Thus, in the case when $G$ is
a locally compact group, $\mathbf{B}(G_{lc})$ and
$\mathbf{A}(G_{lc})$ are just the well known Fourier--Stieltjes and
Fourier algebras discussed above. The need to have available
generalizations of these Banach algebras for the case of a
\emph{locally compact groupoid} stems from the fact that many of the
operator algebras of current interest~-- as for example in
non-commutative geometry and quantum operator algebras~-- originate
from \textit{groupoid}, rather than group, representations, so that
one needs to develop the notions of $\mathbf{B}(\mathsf{G}_{lc})$,
$\mathbf{A}(\mathsf{G}_{lc})$ in the groupoid case for groupoid
operator algebras (or indeed for \emph{algebroids}) that are much
more general than $\mathbf{B}(G_{lc})$, $\mathbf{A}(G_{lc})$. One
notes also that in the operator space context, $\mathbf{A}(G_{lc})$
is regarded as the \emph{convolution algebra of the dual quantum
group} \cite{Paterson2k3a}.

However, for groupoids and more general structures (e.g.,
categories and toposes of {\L}M--algebras), such an extension of
Banach space duality still needs further investigation. Thus, one
can also conceive the notion of a measure theory based on
\L ukasiewicz--Moisil ({\L}M)  $N$-valued logic algebras
(see \cite{Georgescu2k6} and references cited therein), and a
corresponding {\L}M-topos generalization of harmonic (or anharmonic) analysis by
def\/ining extended Haar--LM measures, {\L}M-topos
representations and $\F_{\text{S--L--M}}$ transforms. This raises the
natural question of duality for the catgeory of {\L}M-algebras
that was introduced by Georgescu and Vraciu \cite{Georgescu-Vraciu70}.
An appropriate framework for such logic {\L}M-algebras is provided by algebraic categories~\cite{Georgescu-Popescu68}.

Let us consider f\/irst the algebra involved in the simple example
of the Fourier transform and then note that its extension to the
Fourier--Stieltjes transform involves a convolution, just as it did
in the case of the paracrystal scattering theory.

Thus, consider as in \cite{Paterson2k3a} the Fourier algebra in the
locally compact group case and further assume that $G_{lc}$ is a
locally compact Abelian group with character space $\hat{G_{lc}}$;
then an element of~$\hat{G_{lc}}$ is a continuous homomorphism $t :
G_{lc} \rightarrow T$ , with $\hat{G_{lc}}$ being a locally compact
abelian group with pointwise product and the topology of uniform
convergence on compacta. Then, the Fourier transform $f \rightarrow
\hat{f}$ takes $f \in L^1(G_{lc})$ into $C_0( \hat{G_{lc}})$, with
 $\hat{f}(t) = \int f(x) \overline{t(x)} dx $,
where $dx$ is def\/ined as a left Haar measure on $G_{lc}$. On the other
hand, its inverse Fourier transform $\mu \lra \check{\mu}$
reverses the process by taking $M(\hat{G_{lc}})$ back into $C(G_{lc})$, with
$\check{\mu}$ being def\/ined by the (inverse Fourier transform)
integral: $\check{\mu}(x) = \int \check{x}(t) d \mu(t)$. For
example, when $G_{lc} = \Re$, one also has that $\hat{G_{lc}} = \Re$ so that
$t \in \hat{G_{lc}}$ is associated with the character $x \mapsto e^{ixt}$.
Therefore, one obtains in this case the usual Fourier transform
$\hat{f}(t) = \int f(x)e^{-ixt} dx $ and its inverse (or dual)
$\check{\mu}(x) = \int e^{itx} d \mu(t)$. By considering
$M(\hat{G_{lc}})$ as a \textit{convolution} Banach algebra (which
contains $L^1( \hat{G_{lc}})$ as a closed ideal) one can def\/ine the
Fourier--Stieltjes algebra $\mathbf{B}(G_{lc})$ by  $M(\hat{G_{lc}}){~
\check{}}$, whereas the simpler Fourier algebra, $\mathbf{A}(G_{lc})$,
is def\/ined as $L^1(\hat{G_{lc}})~ \check{}$.

\begin{remark}\label{remark7.1}
In the case of a \textit{discrete} Fourier transform, the integral
is replaced by summation of the terms of a Fourier series. The
discrete Fourier (transform) summation has by far the widest and
most numerous applications in digital computations in both science
and engineering. Thus, one represents a continuous function by an
inf\/inite Fourier series of `harmonic' components that can be either
real or complex, depending on the \textit{symmetry} properties of
the represented function; the latter is then approximated to any
level of desired precision by truncating the Fourier series to a
f\/inite number of terms and then neglecting the remainder. To avoid
spurious `truncation errors' one then applies a `smoothing'
function, such as a negative exponential, that is digitized at
closely spaced sample points so that the Nyquist's theorem criterion
is met in order to both obtain the highest possible resolution and
to drastically reduce the noise in the f\/inal, computed fast Fourier
transform (FFT). Thus, for example, in the simpler case of a
centrosymmetric electron density of a unit cell in a crystalline
lattice, the dif\/fracted $X$-ray, electron or neutron intensity can be
shown to be proportional to the modulus squared of the \textit{real}
Fourier transform of the (centrosymmetric) electron density of the
lattice. In a (digital) FFT computation, the approximate electron
density reconstruction of the lattice structure is obtained through
truncation to the highest order(s) of dif\/fraction observed, and thus
the spatial resolution obtained is limited to a corresponding value
in real 3-D space.
\end{remark}

\begin{remark}[Laplace vs 1-D and 2-D Fourier transforms]\label{remark7.2}
On the other hand, although Laplace transforms are being used in some
engineering applications to calculate transfer functions, they are much
less utilized in the experimental sciences than the Fourier
transforms even though the former may have advantages over FFT for
obtaining both improved resolution and increased signal-to-noise. It
seems that the major reason for this strong preference for FFT is
the much shorter computation time on digital computers, and perhaps
also FFT's relative simplicity when compared with Laplace
transforms; the latter may also be one of the main reasons for the
presence of very few digital applications in experimental science of
the Fourier--Stieltjes transforms which generalize Fourier
transforms. Somewhat surprising, however, is the use of FFT also in
\textit{algebraic quantum field computations on a lattice} where
both FS or Laplace transforms could provide superior results, albeit
at the expense of increased digital computation time and
substantially more complex programming. On the other hand, one also
notes the increasing use of `two-dimensional' FFT in comparison
with one-dimensional FFT in both experimental science and medicine
(for example, in 2D-NMR, 2D-chemical (IR/NIR) imaging and MRI
cross-section computations, respectively), even though the former
require both signif\/icantly longer computation times and more complex
programming.
\end{remark}

\subsubsection[Fourier--Stieltjes transforms as generalizations of the classical
Fourier transforms\\ in harmonic analysis to extended anharmonic analysis in quantum theories]{Fourier--Stieltjes transforms as generalizations\\ of the classical
Fourier transforms in harmonic analysis\\ to extended anharmonic analysis in quantum theories}\label{section7.5.1}

Not surprisingly, there are several versions of the near-`harmonic' F--S
algebras for the locally compact groupoid case that
appear at least in three related theories:
\begin{enumerate}\itemsep=0pt
\item[1)]  the \emph{measured groupoid} theory of J.~Renault \cite{Renault80,Renault87,Renault97},
\item[2)] a Borel theory of A.~Ramsay and M.~Walter \cite{Ramsay-Walter97}, and
\item[3)] a continuity-based theory of A.~Paterson \cite{Paterson2k3a,Paterson2k3b}.
\end{enumerate}

Ramsay and Walter \cite{Ramsay-Walter97} made a f\/irst step
towards extending the
theory of Fourier--Stieltjes algebras from groups to groupoids,
thus paving the way to the extension of F--S applications to
generalized anharmonic analysis in Quantum theories \emph{via}
quantum algebra and quantum groupoid representations. Thus, if
$\mathsf{G}_{lc}$ is a locally compact (second countable) groupoid,
Ramsay and Walter showed that $\mathbf{B}(\mathsf{G}_{lc})$, which
was def\/ined as the linear span of the Borel positive def\/inite
functions on $\mathsf{G}_{lc}$, is a \emph{Banach algebra} when
represented as an algebra of completely bounded maps on a
$C^*$-algebra associated with the $\mathsf{G}_{lc}$ that involves equivalent
elements of~$\mathbf{B}(\mathsf{G}_{lc})$; po\-si\-tive def\/inite
functions will be def\/ined in the next paragraph using the notation
of~\cite{Paterson2k3a}. Corresponding to the universal $C^*$-algebra,
$C^*(G)$, in the group case is the univer\-sal~$C^*_\mu (\mathsf{G})$ in
the measured groupoid $\mathsf{G}$ case. The latter is the
completion of $C_c(\mathsf{G}_{lc})$ under the largest $C^*$-norm
coming from some measurable $\mathsf{\mathsf{G}_{lc}}$-Hilbert
bundle $(\mu, \Re,L)$. In the group case, it is known that
$\mathbf{B}(G)$ is isometric to the Banach space \textit{dual} of
$C^*(G)$. On the other hand, for groupoids, one can consider a
representation of $\mathbf{B}(\mathsf{G}_{lc})$ as a Banach space of
completely bounded maps from a $C^*$-algebra associated with
$\mathsf{G}_{lc}$ to a $C^*$-algebra associated with the equivalence
relation induced by $\mathsf{G}_{lc}$. Obviously, any Hilbert space
$\H$ can also be regarded as an operator space by identifying it
with a subspace of $\mathbf{B}(\mathbb C,\H)$: each $\xi \in \H$ is
identif\/ied with the map ${a} \lra {a} \xi$ for $a \in \mathbb C $;
thus, $\H^*$ is an operator space as a subspace of $\mathbf{B}(\H,
\mathbb C )$. Renault showed for measured groupoids that the
operator space $C^*_\mu(\mathsf{G}_{lc})$ is a completely
contractive left $L^\infty (\mathsf{G}_{lc}^0)$ module. If~$E$ is a
right, and~$F$ is a left, $A$-operator module, with $A$ being a
$C^*$-algebra, then a Haagerup tensor norm is determined on the
algebraic tensor product $E \otimes_A F$ by setting  $\left\|
\textit{ u}  \right\| = \sum \limits_{i=1}^n \left\|  e_i\right\| \left\|
f_i\right\|$ over all representations $ \textit{u} = \sum \limits_{i=1}^n
e_i \otimes_A f_i $.

According to \cite{Paterson2k3a}, the completion  $E \otimes_A F$ of
$E $ is called the \emph{module Haagerup tensor product of $E$ and
$F$ over $A$}. With this def\/inition, the module Haagerup tensor
product is:
\begin{gather*}
X(\mathsf{G}_{lc}) = L^2(\mathsf{G}_{lc}^0)^* \otimes C^*_{\mu}
(\mathsf{G}_{lc}) \otimes
L^2(\mathsf{G}_{lc}^0),
 \end{gather*}
taken over $L^\infty (\mathsf{G}_{lc}^0)$. Then, with this tensor product
construction, Renault was able to prove that
\begin{gather*}
X(\mathsf{G}_{lc})^* = \mathbf{B}_\mu(\mathsf{G}_{lc}) .
\end{gather*}

Thus, each $\phi = (\xi, \eta)$ can be expressed by the
linear functional $a^* \otimes f \otimes b \lra \int \overline{a
\circ r}(\phi f)b \circ s \, d \nu$ with $f \in C_c(\mathsf{G}_{lc})$.

We shall also brief\/ly discuss here Paterson's generalization to
the groupoid case in the form of a \textit{Fourier--Stieltjes
algebra of a groupoid}, $\mathbf{B}_\mu(\mathsf{G}_{lc})$, which was def\/ined
(e.g., in \cite{Paterson2k3a}) as the space of coef\/f\/icients $\phi = (\xi,
\eta)$,  where $\xi, \eta$ are $L^{\infty}$-sections for some
measurable $\mathsf{G}$-Hilbert bundle $(\mu, \Re,L)$). Thus, for $x \in
\mathsf{G}_{lc}$,
\begin{gather*}
\phi(x) = (L(x) \xi (s(x)), \eta (r(x))).
\end{gather*}

Therefore, $\phi$ belongs to
$L^\infty(\mathsf{G}_{lc}) = L^\infty(\mathsf{G}_{lc},\nu)$.

 Both in the groupoid and group case, the set
$P_{\mu}(\mathsf{G}_{lc})$ of \textit{positive definite functions}
in $L^\infty(\mathsf{G}_{lc})$ plays the central role. Thus, a
function $\phi \in L^{\infty}(\mathsf{G}_{lc})$ is called
\emph{positive definite} if and only if for all $u \in
(\mathsf{G}^0_{lc})$,
\begin{gather*}
\iint \phi(y^{-1}x)f(y)\overline{f(x)} \, d \lambda^u(x) d
\lambda ^u(y) \geq 0 .
\end{gather*}
Now, one can def\/ine the notion of a
\emph{Fourier--Stieltjes transform} as follows:

\begin{definition}[The Fourier--Stieltjes transform]\label{definition7.1}
Given a \emph{positive definite, measurable function} $f(x)$ on
the interval $(-\infty ,\infty)$ there exists a monotone
increasing, real-valued bounded function $ \alpha (t)$ such that:
\begin{gather*}
f(x)= \int_{-\infty}^{\infty} e^{itx} d \alpha (t),
\end{gather*} for all $x$ except a small set. When $f(x)$ is def\/ined as above
and if $\alpha (t)$ is nondecreasing and bounded then the
measurable function def\/ined by the above integral is called
\emph{the Fourier--Stieltjes transform of} $\alpha (t)$, and it is
\emph{continuous} in addition to being positive def\/inite in the
sense def\/ined above.
\end{definition}

In \cite{Paterson2k3a} is also def\/ined the \emph{continuous}
Fourier--Stieltjes algebra $\mathbf{B}(\mathsf{G})$ as follows. Let us
consider a \emph{continuous} $\mathsf{G}$-Hilbert bundle $\H_\Re$, and
the Banach space $\Delta_b$ of continuous, bounded sections of
$\H_\Re$. For $\xi, \eta \in \Delta_b$,  the coef\/f\/icient $(\xi,
\eta) \in C(\mathsf{G})$ is def\/ined by:
\begin{gather*}
(\xi, \eta)(u) = (L_x \xi(s(x)),\eta(r(x))),
\end{gather*}
where $x \lra L_x$ is the $\mathsf{G}$-action on $\H_\Re$. Then, the
\emph{continuous Fourier--Stieltjes algebra} $\mathbf{B}(\mathsf{G})$ is
def\/ined to be the set of all such coef\/f\/icients, coming from all
possible continuous $\mathsf{G}$-Hilbert bundles. Thus,
$\mathbf{B}(\mathsf{G})$ is an algebra over $\mathbb C$ and the norm of
$\phi \in \mathbf{B}(\mathsf{G})$ is def\/ined to be $\inf \left\| \xi
\right\|  \left\| \eta\right\| $, with the \emph{infimum} $\inf$
being taken over all $\mathsf{G}$ representations $\phi = (\xi, \eta)$.  Then
$\mathbf{B}(\mathsf{G})\subset  C(\mathsf{G})$, and $\left\|\cdot
\right\|_\infty = \left\|\cdot  \right\|$.

\looseness=-1
Paterson in \cite{Paterson2k3a} showed that $\mathbf{B}(\mathsf{G})$ thus def\/ined~-- just
as in the group case~-- is a \emph{commutative Banach algebra}. He
also def\/ined for a general group $G$ the \emph{left regular
representation} $\pi_2$ of $G$ on $L^2(G)$ by: $\pi_2(x)f(t) =
f(x^{-1}t)$. One also has the \emph{universal representation}
$\pi_{2,{\rm univ}}$ of G which is def\/ined on a Hilbert space
$\H_{\rm univ}$. Moreover, every \textit{unitary} representation of G
determines by integration a \textit{non-degenerate}
$\pi_2$-representation of $C_c(G)$. The norm closure of
$\pi_2(C_c(G))$ then def\/ines the \textit{reduced} $C^*$-algebra
$C^*_{\rm red}(G)$ of $G$, whereas the norm closure of
$\pi_{2,{\rm univ}}(C_c(G))$ was def\/ined as the \textit{universal
$C^*$-algebra} of $G$ (\textit{loc.\ cit.}). The algebra  $C^*_
{\rm red}(G)\subset B(L^2(G))$ generates a von Neumann algebra denoted
by $V_N(G)$. Thus, $C^*_ {\rm red}(G)$ representations generate
$V_N(G)$ representations that have a much simpler classif\/ication
through their $V_N$ factors than the representations of general
$C^*$-algebras; consequently, the classif\/ication of $C^*_ {\rm red}(G)$
representations is closer linked to that of $V_N$ factors than in
the general case of $C^*$-algebras. One would expect that a similar
simplif\/ication may not be available when group G symmetries (and,
respectively, their associated $C^*_ {\rm red}(G)$ representations)
are extended to the more general groupoid symmetries (and their
associated groupoid $C^*$-convolution algebra representations relative to
Hilbert {bundles}).

Recently, however, Bos in~\cite{Bos2k6,Bos2k8} reported that
one can extend~-- with appropriate modif\/i\-ca\-tions
and conditions added~-- the Schur's lemma and Peter--Weyl theorems
from group representations to corresponding theorems for
(continuous) internally irreducible representations of
\textit{continuous} groupoids in the case of Schur's lemma, and
restriction maps in the case of two Peter--Weyl theorems, (one of
the latter theorems being applicable only to \textit{compact},
proper groupoids and their isomorphism classes of irreducible
unitary (or internally irreducible) representations $({\rm IrRep}(\mathsf{G})$
and ${\rm IrRep}^i(\mathsf{G})$, respectively)). It is well established that
using Schur's lemma for groups one can prove that if a matrix
commutes with every element in an \textit{irreducible}
representation of a group that matrix must be a multiple of the
\textit{identity}. A continuous groupoid representation $(\pi, \H,
\Delta)$ of a continuous groupoid $\mathsf{G}^{\rightarrow
}_{\rightarrow} M$ was called \textit{internally irreducible} by
Ros if the restriction of $\pi$ to each of the isotropy groups is
an irreducible representation. Thus, in the case of
\textit{continuous} groupoids $\mathsf{G}^{\rightarrow }_{\rightarrow}
M$ (endowed with a Haar system), \textit{irreducible
representations are also internally irreducible} but the converse
does not hold in general (see also the preprints of
R.D.~Bos~\cite{Bos2k8}). Bos also introduced in
\cite{Bos2k6,Bos2k8} the \textit{universal enveloping $C^*$-category} of a
Banach $*$-category, and then used this to def\/ine the
$C^*$-category, $\mathbf{C}^*(\mathsf{G},\mathsf{G})$, of a
groupoid. Then, he found that \textit{there exists a bijection
between the continuous representations of}
$\mathbf{C}^*(\mathsf{G},\mathsf{G})$ \textit{and the continuous
representations of} $\mathsf{G}^{\rightarrow }_{\rightarrow} M$.

\subsection[Categorical representations of categorical groups in monoidal bicategories]{Categorical representations of categorical groups\\ in monoidal bicategories}\label{section7.6}

 Barrett pointed out in \cite{Barrett2k3} that monoidal
categories play an important role in the construction of
\textit{invariants} of three-manifolds (as well as knots, links
and graphs in three-manifolds). The approach is based on
constructions inspired by strict categorical groups which lead to
monoidal categories \cite{Barrett-Mackaay2k6}. A categorical group was thus considered
in this recent report as a group-object in the category of groupoids, and it can also
be shown that categorical groups are equivalent to crossed modules
of groups. (A \textit{crossed module} is a homomorphism of groups
$ \partial : E \rightarrow G$, together with an action
$\triangleright$ of G on E by automorphisms, such that
 $\partial(X \triangleright e) = X(\partial)X^{-1}$, and
$\partial(e)\triangleright e' = ee'e^{-1}$, where $E$ denotes the
\textit{principal} group and $G$ is the \textit{base} group.)

 Specif\/ically, a \textit{categorical group} was def\/ined in~\cite{Barrett-Mackaay2k6} as a groupoid
$\grp$, with a set of objects $G^0 \subset \grp$, together with
functors which implement the group product, $\circ: \grp \times \grp
\lra \grp$, and the inverse $\iota^{-1} : \grp \rightarrow \grp$,
together with an identity object $\textbf{1} \in \grp^0$;  these satisfy the
usual group laws:
\begin{gather*}
a \circ (b\circ c) = (a\circ b) \circ c,\quad a \circ \textbf{1} = \textbf{1} \circ a
= a, \quad  a \circ a^{-1} = a^{-1} \circ a = \textbf{1},
\end{gather*}
for all $a, b, c \in \grp$. Furthermore, a \textit{functorial homomorphism} between two such categorical groups was def\/ined as a strict monoidal functor.

 In particular, $\grp$ is a strict monoidal category (or \textit{tensor category}, that is, a category $\mathcal{C}$ equipped with a bifunctor $\otimes : \mathcal{C} \times \mathcal{C} \longrightarrow \mathcal{C}$ which is associative and an object which is both left and right identity for the bifunctor $\otimes$, up to a natural isomorphism; see also Section~\ref{section8} for further details on tensor products and tensor categories).

  One of many physical representations of such monoidal categories is the \textit{topological order} in condensed matter theory, quantum f\/ield theory~\cite{Witten89} and string models \cite{Gu-Wen2k6,Levin-Wen2k3,Levin-Wen2k6,Wen2k3,Wen2k4}. One of the f\/irst theoretical reports of topological order in metallic glasses with ferromagnetic properties was made by P.W.~Anderson in 1977~\cite{Anderson77}.
In the recent quantum theory of condensed matter, \textit{topological order} is  a pattern of \textit{long-range} entanglement of quantum states (such as the ``string-net condensed'' states \cite{Levin-Wen2k5}) def\/ined by a new set of quantum numbers such as quasi-particle fractional statistics, edge states, ground state degeneracy, topological entropy \cite{Levin-Wen2k6}, and so on; therefore, topological order also introduces new extended quantum symmetries \cite{Ran-Wen2k6} that are beyond the Landau symmetry-breaking model. Such string-net condensed models can have excitations that behave like gluons, quarks and the other particles already present in the standard model
(SUSY).

  On the other hand, \emph{braided} monoidal categories are being applied to both quantum f\/ield theory and string models. A \textit{braided monoidal category} is def\/ined as a monoidal category equipped with a braiding, that is, a natural isomorphism $\gamma_{A,B}: A \otimes B \longrightarrow B \otimes A$ such that the braiding morphisms $\gamma: A\otimes (B \otimes C) \to B \otimes (C \otimes A)$ commute with the associativity isomorphism in two hexagonal diagrams containing all associative permutations of objects~$A$,~$B $ and~$C$. An alternative def\/inition of a braided monoidal category is as a tricategory, or 3-category, with a~single $one$-cell and also a~single $2$-cell \cite{Andre-Street93,Baez99}; thus, it may be thought of as a~`three-dimensional' categorical structure.

As an example of bicategory associated with categorical group representations, consider the closure of $\grp$, denoted by $\overline{\grp}$, to be the 2-category with one object denoted by
$\bullet$, such that $\overline{\grp}(\bullet,\bullet) = \grp$. The horizontal composition is then def\/ined by the monoidal structure in $\grp$.

In the current literature, the notion of categorical group is used in the sense of a
strict monoidal category in which multiplication by an object on either side
is an equivalence of categories, and the def\/inition of a categorical group was seemingly f\/irst published by Brown and Spencer in 1976 \cite{Brown-Spencer76b}).


From a quantum physics perspective, the monoidal categories
determined by quantum groups seen as Hopf algebras, generalize
the notion that the representations of a group form a monoidal
category. One particular example of a monoidal category was reported
to provide a state-sum model for quantum gravity in
three-dimensional spacetime \cite{Turaev-Viro92,Barrett2k3}.
The motivation for such applications was the search for more
realistic categorical models of four-dimensional, relativistic
spacetimes. Thus, it was proposed to construct the monoidal
2-category of representations for the case of the categorical Lie
group determined by the group of Lorentz transformations and its
action on the translation group of Minkowski space.

In higher dimensions than three, the complexity of such algebraic
representations increases dramatically. In the case of four-manifolds
there are several examples of applications of
catego\-ri\-cal algebra \cite{MacLane65} to
four-dimensional topology; these include:
Hopf categories \cite{Crane-Frenkel94}, categorical
groups \cite{Yetter93} or monoidal 2-categories
\cite{Carter-Saito94, Baez-Langford2k3}, and the representation theory
of quasi-triangular Hopf algebras. Invariants of four-manifolds
were derived by Crane and Kauf\/fman, as well as Roberts \cite{Roberts95,Roberts97},
who give information on the homotopy type of
the four-manifold~\cite{Crane-Kauffman-Yetter97, Roberts97, Regge61,Mackaay99}.
Recently, Martin and Porter~\cite{Martin-Porter2k7}
presented results concerning the Yetter invariants~\cite{Yetter93},
and an extension of the Dijkgraaf--Witten invariant to categorical groups.
Other types of catego\-ri\-cal invariants and extended symmetries are also expected to emerge
in higher dimensions as illustrated here in Section~\ref{section9}.
 Barrett \cite{Barrett2k3} also introduced a def\/inition of
categorical representations and the functors between them \cite{Barrett-Mackaay2k6}.  Such def\/initions are analogues of Neuchl's  def\/initions for \textit{Hopf categories} \cite{Neuchl97}. Consider f\/irst the specif\/ic example of  a categorical group $\grp$ and its closure $\overline {\grp}$  as def\/ined above. One can check that a representation of $\grp$ is precisely a functor $\R : \overline {\grp}\rightarrow \textbf{Vect}$,
and that an \textit{intertwiner} is precisely a natural
transformation between two such functors (or representations).
This motivates the categorical representation of categorical
groups in the monoidal bicategory \textbf{2-Vect} of
2-vector spaces. If $\grp$ is an arbitrary categorical group and $\overline {\grp}$
its closure, the \textit{categorical representation} of $\grp$
is a strictly unitary homomorphism
$(R, \tilde{R}): {\overline {\grp}} \rightarrow \textbf{2-Vect}$.

The non-negative integer $R(\bullet) \in \textbf{2-Vect}_0$
was called the \textit{dimension} of the categorical representation.
The categorical group representation can be equivalently described
as a homomorphism between the corresponding crossed modules.
The possibility of generalizing such categorical representations
to monoidal categories other than $\textbf{2-Vect}$ was
also considered.

 A theorem proven by Verdier states that the category of
categorical groups and functorial homomorphisms $\mathbf{C}_ {\grp}$
and the category \textbf{CM} of crossed modules of groups and
homomorphisms between them are equivalent. Based on this theorem,
Barrett in~\cite{Barrett2k3} showed that each categorical group
determines a monoidal bicategory of representations. Typically, such
bicategories were shown to contain representations that are
indecomposable but \textit{not irreducible}.

In the following sections we shall consider even wider classes of
representations for groupoids, arbitrary
categories and functors.

\subsection[General def\/inition of extended symmetries as representations]{General def\/inition of extended symmetries as representations}\label{section7.7}

We aim here to def\/ine extended quantum symmetries as general
representations of mathematical structures that have as many as
possible physical realizations, i.e.\ \textit{via} unif\/ied quantum
theories. In order to be able to extend this approach to very large
ensembles of composite or complex quantum systems one requires
general  procedures for quantum `coupling' of component quantum
systems; we propose to approach this important `composition', or
scale up/assembly problem in a formal manner as described in the
next section.

Because a group \textbf{$G$} can be viewed as a category with a single object, whose morphisms are just the elements of~\textbf{$G$}, a \textit{general representation} of~\textbf{$G$}
in an arbitrary category $\mathbf{C}$ is a functor~$R_G$ from
\textbf{$G$}  to $\mathbf{C}$ that selects an object $X$ in
$\mathbf{C}$ and a group homomorphism from $\gamma$ to $\Aut(X)$,
the automorphism group of $X$. Let us also def\/ine an \emph{adjoint
representation} by the functor $R^*_{\mathbf{C}} : \mathbf{C} \lra
\textbf{G}$. If $\mathbf{C}$ is chosen as the category
$\mathbf{Top}$ of topological spaces and homeomorphisms then
representations of \textbf{$G$} in $\mathbf{Top}$ are homomorphisms
from \textbf{$G$} to the homeomorphism group of a topological space
$X$. Similarly, a \textit{general representation of a groupoid}
$\grp$ (considered as a category of invertible morphisms) in an
arbitrary category $\mathbf{C}$ is a functor $\textbf{R}_\grp$  from
$\grp$ to $\mathbf{C}$, def\/ined as above simply by substituting
$\grp$ for \textbf{$G$}. In the special case of a Hilbert space,
this categorical def\/inition is consistent with the representation of
the groupoid on a bundle of Hilbert spaces.

\begin{remark}\label{remark7.3}
Unless one is operating in super-categories, such as 2-categories
and higher dimensional categories, one needs to distinguish
between the \textit{representations of an (algebraic) object}~-- as
def\/ined above~-- and the \textit{representation of a functor} $\mathcal{S}$
(from $\mathbf{C}$ to the category of sets, $\mathbf{Set}$) by an
object in an arbitrary category $\mathbf{C}$ as def\/ined next.
Thus, in the latter case, a \textit{functor representation} will
be def\/ined by a certain \textit{natural equivalence between
functors}. Furthermore, one needs consider also the following
sequence of functors:
\begin{gather*}
\R_{\textbf{G}}: \ \textbf{G} \lra \mathbf{C}, \qquad  \R^*_{\mathbf{C}}: \
\mathbf{C} \lra \textbf{G} ,
\qquad
\mathcal{S} : \ \textbf{G} \lra \mathbf{Set},
\end{gather*}
where $\R_{\textbf{G}}$ and $\R^*_{\mathbf{C}}$ are adjoint
representations as def\/ined above, and $\mathcal{S}$ is the forgetful
functor which forgets the group structure; the latter also has a
right adjoint $\mathcal{S}^*$. With these notations one obtains the
following commutative diagram of adjoint representations and
adjoint functors that can be expanded to a square diagram to
include either $\mathbf{Top}$~-- the category of topological spaces
and homeomorphisms, or $\mathbf{TGrpd}$, and/or $\mathbf{C}_{\grp}
= \mathbf{CM}$ (respectively, the category of topological
groupoids, and/or the category of categorical groups and
homomorphisms) in a~manner analogous to diagrams~\eqref{(9.29)} 
that will be discussed in Section~\ref{section9} (with the additional, unique
adjunction situations to be added in accordingly)
\begin{gather*}
\xymatrix{ \mathbf{Set} \ar@<-0.5ex> [r]_-{{S}^*}  &\ar
@<-0.5ex>[l]_-{S}
\textbf{G}\ar[d]^{\R^*_C}  \\
&\mathbf{C} \ar [ul]^{F,~F^*} \ar [u]^{\R_G}}
\end{gather*}
\end{remark}

\subsection{Representable functors and their representations}\label{section7.8}

 The key notion of \textit{representable functor} was f\/irst
reported by Grothendieck (also with Dieudonn\'e) during 1960--1962 \cite{Grothendieck-Dieudone60, Grothendieck61, Grothendieck62}
(see also the earlier publication by Grothendieck \cite{Alex57}).
This is a functor $\mathcal{S}: \mathbf{C} \lra  \mathbf{Set}$,
from an arbitrary category $\mathbf{C}$ to the category of sets, $\mathbf{Set}$, if it admits
a (functor) \textit{representation} def\/ined as follows. A~\textit{functor representation} of $\mathcal{S}$ is a pair, $({R}, \phi)$,
which consists of an object ${R}$ of $\mathbf{C}$ and a family
$\phi$ of equivalences $\phi (C): \Hom_{C}(R,C) \cong {\mathcal{S}}(C)$,
which is natural in C. When the functor $\mathcal{S}$ has such a~representation, it is also said to be \textit{represented by the
object $R$} of $\mathbf{C}$. For each object $R$ of $\mathbf{C}$
one writes $h_{R}: \mathbf{C}\lra \mathbf{Set}$ for the covariant
$\Hom$-functor $h_{R}(C)\cong \Hom_{\mathbf{C}}(R, C)$. A
\textit{representation} $(\R, \phi)$ of $\mathcal{S}$ is therefore
\textit{a natural equivalence of functors}
\begin{gather*}
\phi: \  h_{R} \cong {\mathcal{S}}.
\end{gather*}
The equivalence classes of such functor representations (def\/ined
as natural equivalences) obviously determine an \textit{algebraic
groupoid} structure. As a simple example of an \textit{algebraic}
functor representation, let us also consider (cf.~\cite{MacLane65})
the functor ${N}: \mathbf{Gr} \lra \mathbf{Set}$ which assigns to
each group $G$ its underlying set and to each group homomorphism
$f$ the same morphism but regarded just as a function on the
underlying sets; such a functor ${N}$ is called a
\textit{forgetful} functor because it ``forgets'' the group
structure. ${N}$ is a representable functor as it is represented
by the additive group $\mathbb{Z}$ of integers and one has the
well-known bijection  $\Hom_{Gx} (Z,G)\cong  \mathcal{S}(G)$ which assigns
to each homomorphism ${f}: \mathbb{Z}\lra G$ the image ${f}({1})$
of the generator ${1}$ of $\mathbb{Z}$. In the case of groupoids there is also a natural forgetful functor
$F: \mathbf{Grpd} \lra   \mathbf{Directed Graphs}$ whose left adjoint is the
free groupoid on a directed graph, i.e. the groupoid of all paths in the graph.

\textit{Is $F$ representable, and if so, what is the object that represents $F$?}

One can also describe (viz.~\cite{MacLane65}) representable functors
in terms of certain universal elements called \textit{universal
points}. Thus, consider $\mathcal{S}: \mathbf{C} \rightarrow \textbf{Set
}$ and let $\mathbf{C}_{s*}$ be the category whose objects are
those pairs $(A,x)$ for which $x \in \mathcal{S}(A)$ and with morphisms
$f:(A, x) \lra (B, y)$ specif\/ied as those morphisms $f: A
\rightarrow B$ of $\mathbf{C}$ such that $\mathcal{S}(f)x = y$; this
category $\mathbf{C}_{s*}$ will be called the \textit{category of
$\mathcal{S}$-pointed objects} of $\mathbf{C}$. Then one def\/ines a
\textit{universal point} for a functor $\mathcal{S}: \mathbf{C} \lra
\mathbf{Set}$ to be an initial object $(R,u)$ in the category
$\mathbf{C}_{s*}$. At this point, a general connection between
representable functors/functor representations and
universal properties is established by the following,
fundamental functor representation theorem~\cite{MacLane65}.

\begin{theorem}[Theorem 7.1 of MacLane \cite{MacLane65}]\label{theorem7.2} For each functor $\mathcal{S}: {\mathbf{C}}\lra {\mathbf{Set}}$, the formulas $u =(\phi R)1_R$, and $(\phi c)h = (\mathcal{S}h)u$, (with the
latter holding for any morphism $h: R \lra C$), establish a one-to-one correspondence between the functor representations $(R, \phi)$ of $\mathcal{S}$ and the universal points $(R, u)$ for $\mathcal{S}$.
\end{theorem}

\section[Algebraic categories and their representations
in the category of Hilbert spaces.\\ Generalization of tensor products]{Algebraic categories and their representations
in the category\\ of Hilbert spaces. Generalization of tensor products}\label{section8}

Quantum theories of quasi-particle, or multi-particle, systems are
well known to require not just products of Hilbert spaces but
instead their tensor products. On the other hand, symmetries are
usually built through representations of products of groups such
as $\U(1) \times \SU(2) \times \SU(3)$ in the current `standard
model'; the corresponding Lie algebras are of course
$\mathfrak{u}(1)$, $\mathfrak{su}(2)$ and $\mathfrak{su}(3)$. To
represent the more complex symmetries involving quantum groups
that have underlying Hopf algebras, or in general Grassman--Hopf
algebras, associated with many-particle or quasi-particle systems,
one is therefore in need of considering new notions of generalized
tensor products. We have discussed in Sections~\ref{section6} and~\ref{section7} alternative approaches
to extended quantum symmetries involving graded Lie agebroids (in quantum gravity), quantum algebroids, convolution products and quantum algebroid representations.  The latter approaches can be naturally combined with `tensor products of quantum algebroids' if a suitable canonical extension of the tensor product notion is selected from several possible alternatives that will be discussed next.

\subsection{Introducing tensor products of algebroids and categories}\label{section8.1}

Firstly, we note that tensor products of cubical $\omega$-groupoids
 have been constructed by Brown and Higgins \cite{Brown-Higgins87},
thus  giving rise to a tensor product of crossed complexes, which has
been used by Baues and Conduch\'e to def\/ine the \textit{`tensor algebra'}
of a non-Abelian group \cite{HJB-DC91}. Subsequently,  Day and Street in~\cite{Day-Street97}
have also considered \textit{Hopf algebras with many objects} in tensor categories~\cite{Deligne2k2}.
Further work is however needed to explore possible links of
these ideas with the functional analysis and operator algebras
considered earlier. Thus, in attempting to generalize the notion
of Hopf algebra to the many object case, one also needs to consider
what could be the notion of tensor product of two $R$--algebroids
$C$ and $D$.  If this can be properly def\/ined one can then expect
to see the\textit{ composition} in $C$ as some partial functor
$m:C \otimes C \lra C$ and a~\textit{diagonal} as some partial functor
$\Delta: C \lra C \otimes C$. The def\/inition of $C \otimes D$
is readily obtained for categories~$C$,~$D$ by modifying slightly
the def\/inition of the tensor product of groupoids, regarded
as crossed complexes in Brown and Higgins~\cite{Brown-Higgins87}. So we def\/ine $C \otimes D$ as the pushout of categories
\begin{gather*}
\xymatrix{C_0 \times D_0 \ar [d] \ar[r] & C_1 \times D_0 \ar[d] \\
C_0 \times D_1 \ar [r] & C\otimes D}
\end{gather*}

This category may be seen also as generated by the symbols
\begin{gather*}
\{c \otimes y \mid c \in C_1 \} \cup \{x \otimes d \mid d \in D_1 \},
\end{gather*}
for all $x \in C_0$ and $y \in D_0$ subject to the relations given
by the compositions in $C_1$ and on $D_1$.

The category $G \, \# \, H$ is generated by all elements
$(1_x,h)$, $(g,1_y)$ where $g \in G, h \in H$, $x \in G_0$, $y \in H_0$.
We will sometimes write $g$ for $(g,1_y)$ and $h$ for $(1_x,h)$.
This may seem to be willful ambiguity, but when composites are
specif\/ied in $G \, \# \, H$, the ambiguity is resolved; for
example, if $gh$ is def\/ined in $G \, \# \, H$, then $g$ must refer
to $(g,1_y)$, where $y=s h$, and $h$ must refer to $(1_x,h)$,
where $x=t g$. This convention simplif\/ies the notation and there
is an easily stated solution to the word problem for $G \, \# \,
H$. Every element of $G \, \# \, H$ is uniquely expressible in one
of the following forms:
\begin{enumerate}[$i)$]\itemsep=0pt
\item an identity element $(1_x,1_y)$;
\item a generating element
$(g,1_y)$ or $(1_x,h)$, where $x \in G_0$, $y \in H_0$, $g \in G$, $h
\in H$ and $g$, $h$ are not identities;
\item a composite $k_1 k_2
\cdots k_n$ $(n \geq 2)$ of non-identity elements of $G$ or $H$ in
which the $k_i$ lie alternately in $G$ and $H$, and the odd and
even products $k_1 k_3 k_5 \cdots$ and $k_2 k_4 k_6 \cdots$ are
def\/ined in $G$ or $H$.
\end{enumerate}

For example, if $g_1:x \lra y$, $g_2:y \lra z,$ in $G$, $g_2$ is
invertible, and $h_1:u \lra v$, $h_2:v \lra w$ in~$H$, then the word
$g_1 h_1 g_2 h_2 g_2^{-1}$ represents an element of $G \, \# \, H$
from $(x,u)$ to $(y,w)$. Note that the two occurrences of $g_2$
refer to dif\/ferent elements of $G \, \# \, H$, namely $(g_2,1_v)$
and $(g_2,1_w)$. This can be represented as a path in a
$2$-dimensional grid as follows
\begin{gather*}
\xymatrix{(x,u)  \ar[d]^{g_1} & (x,v)  & (x,w)  \\
 (y,u) \ar[r]^{h_1} & (y,v) \ar[d]^{g_2} & (y,w) \\
 (z,u) & (z,v) \ar[r]^{h_2}& (z,w) \ar[u]_{g_2^{-1}} }
\end{gather*}

 The similarity with the free product of monoids is obvious and the
normal form can be verif\/ied in the same way; for example, one can
use `van der Waerden's trick'. In the case when~$C$ and~$D$
are $R$-algebroids one may consider the pushout in the category
of $R$-algebroids.

 Now if $\mathbf{C}$ is a category, we can consider the possibility
of a diagonal morphism
\begin{gather*} \Delta: \ \mathbf{C} \lra \mathbf{C}  \# \mathbf{C}  .
\end{gather*}
We may also include the possibility of a morphism
\begin{gather*} \mu :  \ \mathbf{C}  \# \textbf{}C \lra \mathbf{C}  .
\end{gather*}
This seems possible in the algebroid case, namely the sum of the
odd and even products. Or at least, $\mu$ could be def\/ined on
$\textbf{C }  \# \mathbf{C}((x,x),(y,y))$~

It can be argued that a most signif\/icant ef\/fect of the use of
categories as algebraic structures is to allow for algebraic
structures with operations that are partially def\/ined. These were
early considered by Higgins in `Algebras with a scheme of
operators' \cite{Higgins64,Higgins2k5}. In general, \textit{`higher dimensional
algebra'} (HDA) may be def\/ined as the study of algebraic
structures with operations whose domains of def\/initions are
def\/ined by geometric considerations. This allows for a splendid
interplay of algebra and geometry, which early appeared in
category theory with the use of complex commutative diagrams
(see, e.g., \cite{Grothendieck71, Mitchell65, Popescu68, MacLane-Moerdijk92}). What
is needed next is a corresponding interplay with analysis and
functional analysis (see, e.g., \cite{Patera}) that would extend also to quantum operator
algebras, their representations and symmetries.

\subsection[Construction of weak Hopf algebras  via tensor category classif\/ication\\
(according to Ostrik~\cite{Ostrik2k8})]{Construction of weak Hopf algebras  via tensor category classif\/ication\\
(according to Ostrik~\cite{Ostrik2k8})}\label{section8.2}

 If $\mathbf{k}$ denotes an algebraically closed f\/ield, let
$\mathbf{C}$ be a tensor category over $\mathbf{k}$. The
classif\/ication of all semisimple module categories over
$\mathbf{C}$ would then allow in principle the construction of all
weak Hopf algebras $H$ so that the category of comodules over $H$
is tensor equivalent to $\mathbf{C}$, that is, as realizations of
$\mathbf{C}$. There are at least three published cases where such
a classif\/ication is possible:
\begin{itemize}\itemsep=0pt
\item[(1)] when $\mathbf{C}$ is a group theoretical fusion category (as an
example when $\mathbf{C}_{\gamma}$ is the category of
representations of a f\/inite group $\gamma$, or a Drinfel'd quantum
double of a f\/inite group) (see~\cite{Ostrik2k8});

\item[(2)] when $\textbf{k}$ is a fusion category attached to quantum $\SL(2)$
(see \cite{Ostrik2k8,Etingof-Ostrik2k3, BEK2k, Kirillov-Ostrik2k1,Ocneanu88,Ocneanu2k1,Hazewinkel2k6});

\item[(3)] when $k = \mathbf{C}_q$ is the category of representations
of quantum $\SL_q(2)$ Hopf algebras and ${q}$ is not a root of
unity (see~\cite{Etingof-Ostrik2k3}).
\end{itemize}

This approach was further developed recently for module categories
over quantum $\SL(2)$ representations in the non-simple case (see
also Example~\ref{hopfalgebra} regarding the quantum $\SL_q(2)$
Hopf algebras for further details), thus establishing a link
between weak Hopf algebras and module categories over quantum
$\SL(2)$ representations (viz.~\cite{Ostrik2k8}).

\begin{remark}\label{remark8.1}
One notes the condition imposed here of an \textit{algebraically
closed} f\/ield which is essential for remaining within the bounds
of algebraic structures, as f\/ields~-- in general~-- are not
algebraic.
\end{remark}

\subsection[Construction of weak Hopf algebras from a matched pair $(V, H)$ of  f\/inite groupoids \\ (following Aguiar and Andruskiewitsch in \cite{Aguiar-Andrusk2k4})]{Construction of weak Hopf algebras from a matched pair $\boldsymbol{(\mathbf{V},\mathbf{H})}$\\ of  f\/inite groupoids  (following Aguiar and Andruskiewitsch in \cite{Aguiar-Andrusk2k4})}\label{section8.3}

 As shown in \cite{Aguiar-Andrusk2k4}, the matched pair of groupoids $(\mathbf{V},\mathbf{H})$ can be
represented by the factorisation of its elements in the following diagram:
\begin{gather*}
\begin{CD}
\textit{P }@> x  >> \textit{Q}
\\ @V x \triangleright g VV   @VV g V
 \\ \textit{R }@ > x \triangleleft g >> \textit{S}
\end{CD}
\end{gather*}
where ${g}$ is a morphism or arrow of the vertical groupoid
$\mathbf{V}$ and $x$ is a morphism of the matched horizontal
groupoid $\mathbf{H}$, can be employed to construct a weak Hopf
algebra, or quantum groupoid $\mathbf{k}({\mathbf{V}},
{\mathbf{H}})$. In the above diagram, ${P }$ is the common base
set of objects for both ${\mathbf{V}}$ and $\mathbf{H}$, and
$\triangleright$ and  $\triangleleft$ are respectively the mutual
\textit{actions} of $\mathbf{V}$ and $\mathbf{H}$ on each other
satisfying certain simple axioms (equation~(1.1) on page~3 of~\cite{Aguiar-Andrusk2k4} in \textit{loc.\
cit}). Furthermore, a matched pair of rotations $(\eta,\xi)$ for
$(\mathbf{V}, \mathbf{H})$ gives rise to a
\textit{quasi-triangular weak Hopf} structure ${Q}$ for
$\mathbf{k}(\mathbf{V}, \mathbf{H})$. One can also write
explicitly this structure as a tensor product:
\begin{gather*}
{Q} = \sum  \xi\big({f}^{-1} \triangleleft {g}^{-1}, g\big) \otimes
(\eta({g}),{f}).
\end{gather*}

\begin{remark}\label{remark8.2}
The representation of matched pairs of groupoids introduced in
\textit{loc.\ cit.} is specialized to yield a monoidal structure
on the category $\mathbf{Rep}(\mathbf{V}, \mathbf{H})$ and a
monoidal functor between such restrictive `monoidal'
representations, and is thus not consistent with the genera\-li\-zed
notions of groupoid and functor representations considered, respectively, in
Sections~\ref{section7.3}, \ref{section7.4}, and~\ref{section7.6}--\ref{section7.8}. Nevertheless, the constructions of both weak
Hopf algebras and quasi-triangular Hopf by means of matched pairs
of groupoids is important for extended symmetry considerations as
it suggests the possibility of double groupoid construction of
weak Hopf algebroids (and also bialgebroids and double algebroids).
\end{remark}

The representations of such higher dimensional algebraic
structures will be further discussed in the following Section~\ref{section9}.

\section{Double algebroids and double groupoids}\label{section9}

There is a body of recent non-Abelian algebraic topology results
giving a form of ``higher dimensional group (HDG) theory'' which is
based on intuitive ideas of composing squares or $n$-cubes rather
than just paths as in the case of groups~\cite{Brown-etal2k9}.

Such an HDG theory yielded important results in homotopy theory and the
homology of discrete groups, and seems also to be connected to a
generalized categorical Galois theory introduced by Janelidze
and Brown  \cite{Brown-Janelidze2k4}. The HDG approach has
also suggested other new constructions in group theory,
for example a non-Abelian tensor product of groups.
One of the aims of our paper is to proceed towards a corresponding
theory for associative algebras and algebroids rather than groups.
Then, one also f\/inds that there are many more results and methods in HDG
theories that are analogous to those in the lower dimensional
group theory, but with a corresponding increase in technical
sophistication for the former. Such complications occur mainly at
the step of increasing dimension from one to dimension two; thus,
we shall deal in this section only with the latter case. The
general, $n$-dimensional case of such results presents signif\/icant technical
dif\/f\/iculties but is of great potential, and will be considered in subsequent
publications.

Thus, in developing a corresponding theory for algebras we expect
that in order to obtain a~non-trivial theory we shall have to
replace, for example, $R$-algebras by $R$-algebroids, by which is
meant just an $R$-category for a commutative ring $R$; in the
case when $R$ is the ring of integers, an $R$-algebroid is just a
\textit{`ring with many objects'} in the sense of Mitchell \cite{Mitchell72,Mitchell85}.
(for further details see for example Section~\ref{section4} and other references cited therein). The necessary algebroid concepts were already presented in Section~\ref{section4}. In the following subsections we shall brief\/ly introduce other key concepts needed for such HGD developments. Thus, we begin by considering the simpler structure of double algebras and then proceed to their natural extension to double algebroids.

\subsection{Double algebras}\label{section9.1}

Here we approach convolution and the various Hopf structures that we have already discussed from the point of view of `double structures'. With this purpose in mind, let $A$ be taken to denote one of the following
structures: a Hopf, a weak Hopf algebra or a Hopf algebroid (whose
base rings need not be commutative). Starting with a Frobenius
homomorphism $i : A \lra A^*$, we consider as in \cite{Szlachanyi2k4}
the horizontal~(H) and vertical~(V) components of the algebra along
with a convolution product ($*$). Specif\/ically, we
take unital algebra structures $V = \langle A, \circ, e \rangle$
and $H = \langle A, *, i \rangle$ as leading to a double algebra
structure with axioms as given in \cite{Szlachanyi2k4}. Thus the
basic framework starts with a quadruple $(V,H, *, i)$. With
respect to $k$--linear maps $\vp: A \lra A$, we consider
sublagebras $L, R \subset V$ and $B,T \subset H$ in accordance
with the Frobenius homorphisms (for $a \in A$):
\begin{gather*}
\vp_L(a)  := a * e,\qquad  \vp_R(a) := e * a, \\ \vp_B(a)  := a \circ
i,\qquad  \vp_T(a) := i \circ a.
\end{gather*}
Comultiplication of the `quantum groupoid' arises from the dual
bases of $\vp_B$ and $\vp_T$ with a~$D_4$-symmetry:
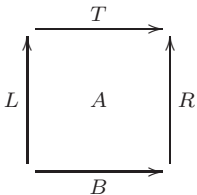
\begin{gather*}
\xymatrix@C=4pc@R=4pc{\ar @{} [dr] |{A}  \ar[r]^T &  \\  \ar[u]^L
\ar[r]_B & \ar[u]_R }
\end{gather*}

\subsection{Double algebroids and crossed modules}\label{section9.2}

In~\cite{Brown-Mosa86} Brown and Mosa introduced the notion of
{\em double algebroid}, and its relationship to crossed modules of
algebroids was investigated. Here we summarize the main results reported so far, but without providing the proofs that can be found in \cite{Mosa86} and \cite{Brown-Mosa86}.

\subsubsection{Crossed modules}\label{section9.2.1}

Let $A$ be an $R$-algebroid over $A_0$ and let $M$ be a
pre-algebroid over $A_0$.  \textit{Actions} of $A$ on $M$ are
def\/ined as follows:

\begin{definition} \label{definition9.1} A {\it left action} of $A$ on $M$ assigns to each
$m \in  M(x,y)$ and $a \in A(w,x)$ an element $^am \in M(w,y)$,
satisfying the axioms:
\begin{enumerate}[$i)$]\itemsep=0pt
\item    $^c(^am) = {}^{(ca)} m$, ${}^1m = m, $
\item $^a(mn)= {}^am n$,
\item $^a(m + m_1) = {}^am + {}^am_1,$
\item $^{ a+b}(m) = {}^am +{} ^bm,$
\item $^a(rm) = r({}^am) = {}^{ra} (m)$,
\end{enumerate}
for all $m,m_1\in M(x,y)$, $n \in M(y,z)$, $a,b \in  A(w,x)$, $c \in
A(u,w)$  and $r \in  R$.
\end{definition}

\begin{definition}\label{definition9.2} A {\it right action} of $A$ on $M$ assigns to each
$m \in M(x,y)$, $a \in  A(y,z)$ an element  $m^a \in  M(x,z)$ satisfying
the axioms:
\begin{enumerate}[$i)$]\itemsep=0pt
\item $(m^a)^c = m^{(ac)}$, $m^1= m$,
\item $ (mn)^a = m n^a$,
\item $(m +m_1)^ a = m^a + m_1^a $,
\item $ m^{(a+b)} = m^a + m^b$,
\item $ (rm)^a = r m^a  = m ^{ra}$
\end{enumerate}
for all $m,m_1 \in M(x,y)$, $n \in  M(y,z)$, $a,b \in  A(y,u)$,  $c
\in A(u,v)$ and $r \in  R$.
\end{definition}
Left and right actions of $A$ on $M$ \emph{commute} if  $(^am)^b =
{}^a(m^b)$, for all $m \in M(x,y)$, $a \in A(w,x)$, $b\in A(y,u)$.

A \emph{crossed module of algebroids} consists of an
$R$-algebroid $A$, a pre-algebroid $M$, both over the same set of
objects, and commuting left and right actions of $A$ on $M$,
together with a pre-algebroid morphism $\mu : M \to A$  over the
identity on $A_0$. These must also satisfy the following axioms:
\begin{itemize}\itemsep=0pt
\item[$i)$] $\mu (^am) = a (\mu m)$, $\mu (m^b) = (\mu m)b $;

\item[$ii)$] $mn = m ^{(\mu n)} = {} ^{(\mu m)} n $,
\end{itemize}
and for all $m \in  M(x,y)$, $n \in  M(y,z)$, $a \in  A(w,x)$, $b
\in A(y,u)$.

A \emph{morphism} $(\alpha,\beta) : (A,M,\mu)\lra (A',M', \mu')$
\emph{of crossed modules} all over the same set of objects is an
algebroid morphism $\alpha  : A \lra A'$ and a pre-algebroid
morphism $\beta : M \lra M'$  such that $\alpha \mu  = \mu' \beta$
and $\beta(^am) = {} ^{\alpha a} (\beta m)$, $\beta (m^b) = (\beta
m)^{\alpha b}$ for all $a,b \in  A$, $m \in M$. Thus one constructs
a \emph{category $\mathbf{CM}$ of crossed modules of algebroids}.

Two basic examples of crossed modules are as follows.
\begin{itemize}\itemsep=0pt
\item[(1)] Let $A$ be an $R$-algebroid over $A_0$ and suppose $I$ is a
two-sided ideal in $A$. Let $i : I\lra A$ be the inclusion
morphism and let $A$ operate on $I$ by $a^c = ac$, ${}^ba = ba$
for all $a \in I$ and $b,c \in  A$ such that these products $ac$,
$ba$ are def\/ined. Then $i : I \to A$ is a crossed module.

\item[(2)] A two-sided module over the algebroid $A$ is def\/ined to be
a crossed module $\mu  : M \to A$ in which  $\mu m = 0_{xy}$ for
all $m \in  M(x,y)$, $x,y \in  A_0$.
\end{itemize}

Similar to the case of categorical groups discussed above, a key
feature of double groupoids is their relation to crossed modules
``of groupoids'' \cite{Brown-Spencer76a}. One can thus establish relations
between double algebroids with thin structure and crossed modules
``of algebroids'' analogous to those already found for double
groupoids, and also for categorical groups. Thus, it was recently
reported  that the \textit{category of double algebroids with
connections} is equivalent to the \textit{category of crossed
modules over algebroids} \cite{Brown-Mosa86}.

\subsubsection{Double algebroids}\label{section9.2.2}

n this subsection we recall the def\/inition of a double algebroid introduced
by Brown and Mosa in \cite{Brown-Mosa86}.
Two functors are then constructed, one from the category
of double algebroids to the category of crossed modules
of algebroids, whereas the other is its unique adjoint functor.

A \textit{double $R$-algebroid} consists of a double category $D$
such that each category structure has the additional structure of
an $R$-algebroid.  More precisely, a double $R$-algebroid $\D$
involves four related $R$-algebroids:
\begin{gather*} \big(D,D_1,\del^0_1 ,\del^1_1 , \vep_1 , +_1 , \circ _1 , ._1\big)
,\qquad   \big(D,D_2,\del^0_2 , \del ^1_2 , \vep_2 , +_2 , \circ _2 ,
._2 \big), \\
 \big(D_1,D_0, \delta^0_1 ,\delta^1_1 , \vep , + , \circ  , .\big) ,\qquad
 \big(D_2 , D_0 , \delta^0_2 , \delta^1_2 , \vep , + , \circ  , .\big)
\end{gather*}
that satisfy the following rules:
\begin{itemize}\itemsep=0pt
\item[$i)$] $\delta^i_2 \del^j_2 = \delta ^j_1 \del ^i_1$  for $i,j \in
\{0,1\}$;

\item[$ii)$]
\begin{gather*} \del^i_2 ( \a +_1 \be) = \del^i_2\a
+\del^i_2\be , \qquad  \del^i_1 ( \a +_2 \be) = \del^i_1 \a
+\del^i_1\be,  \\ \del^i_2 ( \a \circ _1 \be) = \del^i_2 \a \circ
\del^i_2 \be ,\qquad     \del^i_1 (\a \circ _2 \be ) = \del^i_1\a
\circ \del^i_1\be
\end{gather*}
for $i = 0,1$, $\a,\be  \in D$ and both sides are def\/ined;

\item[$iii)$]
\begin{gather*}
r ._1 (\a+_2 \be) = (r ._1 \a) +_2 (r ._1\be ) ,\qquad   r ._2
(\a  +_1 \be  ) = (r ._2 \a ) +_1 (r ._2\be  ), \\ r ._1 (\a  \circ
_2 \be ) = (r ._1\a ) \circ _2  (r ._1 \be )  , \qquad  r ._2 (\a
\circ _1 \be  ) = (r ._2 \a ) \circ _1  (r ._2\be ), \\
 r ._1 ( s ._2 \be  ) = s ._2 ( r._1 \be  )
\end{gather*}
for all $\a ,\be  \in  D$, $r,s \in R$ and both sides are def\/ined;

\item[$iv)$]
\begin{gather*}
(\a  +_1 \be  ) +_2  (\gamma  +_1 \lambda )  = (\a  +_2 \gamma )
+_1 (\be   +_2  \lambda ) ,\\
 (\a  \circ _1 \be ) \circ _2  (\gamma  \circ _1  \lambda )  =
(\a  \circ _2  \gamma ) \circ _1  (\be   \circ _2  \lambda
 ), \\ (\a  +_i \be  ) \circ _j (\gamma  +_i \lambda )  = (\a  \circ _j
\gamma ) +_i (\be   \circ _j \lambda )
\end{gather*}
for  $i \neq j$, whenever both sides are def\/ined.
\end{itemize}

A \textit{morphism $f : \D \to \E$ of double algebroids}
is then def\/ined as a morphism of truncated cubical sets
which commutes with all the algebroid structures. Thus,
one can construct a category $\mathbf{DA}$ of double
 algebroids and their morphisms.
The main construction in this subsection is that
of two functors $\eta$, $\eta'$ from this category
$\mathbf{DA}$ to the category $\mathbf{CM}$ of crossed modules of algebroids.

Let ${D}$ be a double algebroid. One can associate to ${D}$ a
crossed module $\mu : M  \lra {D}_1$. Here $M(x,y)$ will consist
of elements $m$ of ${D}$ with boundary of the form:
\begin{gather*}
\del m = \quadr{a}{1_y}{1_x}{ 0_{xy}},
\end{gather*}
that is
$M(x,y) = \{ m \in D : \del^1_1 m = 0_{xy} , \del^0_2 m =
1_x,\del^1_2  m = 1_y \}$.

The pre-algebroid structure on $M$ is then induced by the second
algebroid structure on ${D}$. We abbreviate $\circ _2$  on $M $
and $\circ _1$  on ${D}_1$ to juxtaposition. The morphism $\mu$ is
def\/ined as the restriction of $\del^0_1 $.

Actions of ${D}_1$ on $M$ are def\/ined by
\begin{gather*}
^am = ( \vep_1 a ) m ,\qquad m^b = m ( \vep_1 b).
\end{gather*}

The only non trivial verif\/ication of the axioms is that $mn = m^{\mu
n} = {}^{\mu m} n $. For this, let~$m$,~$n$ have boundaries
$\quadr{a}{1}{1}{0}$, $\quadr{b}{1}{1}{0}$. Reading in two ways the
following diagram (in which unmarked edges are~1's)  yields $mn =
{}^an$:
\begin{gather*}
\objectwidth{0in} \objectmargin{0in}
\def\labelstyle{\textstyle }
\xymatrix @R=3pc @C=3pc { \ar@{-} [r]^a \ar @{-}[dd] \midsqn{\vep _1 a} &
\ar @{-}[r]^b \ar@{-} [dd] \midsqn{n} & \ar@{-} [dd] \\
\ar@{-} [r]|a \midsqn{m} & \ar@{-} [r] |0 & \\
\ar @{-}[r] _0 & \ar@{-} [r] _0 & }
\end{gather*}
Similarly one obtains that $mn = m^b $. This shows that $\mu : M
\lra {D}_1$ is indeed a crossed module. The construction also
def\/ines $\eta$ which readily extends to a functor from the
\textit{category of double algebroids} $\mathbf{DA}$ to
$\mathbf{CM}$. The second crossed module $\nu  : N \lra D_2$ has
$D_2$ as above, but $N$ consists of elements with boundary of the
form $\quadr{1}{a}{0}{1}$. The actions are def\/ined in a similar
manner to that above for $M$ and one constructs a crossed module in
the manner suggested above. Therefore, an object in the category of
algebroids yields also an \textit{associated} \textit{crossed
module}. In general, the two crossed modules constructed above are
not isomorphic. However, if the double algebroid has a
\textit{connection pair} $(\Gamma,\Gamma')$, then its two associated
crossed modules are isomorphic (cf.~\cite{Brown-Mosa86}).
Furthermore, there is also an associated thin structure $\theta :
\D_1\to \D$ which is a morphism of double categories because $D_1$
also has an associated double algebroid structure derived from that
of~$D_1$.

Next one can construct a functor $\zeta$  from $\mathbf{CM}$ to
double algebroids. Thus, let $\mu : M \lra A$ be a~crossed module.
The double algebroid ${D} = \zeta (\mu  : M \lra A)$ will coincide
with $A$ in dimensions 0 and 1. The set $D$ consists of pairs
$(m,a)$ such that $m \in M$,  $a =\quadr{a_3}{a_1}{a_4}{a_2}$ and
$a_3 a_4 - a_1 a_2 = \mu m$. One def\/ines two algebroid structures on
$\D$ which will turn ${D}$ into a double algebroid. Its additions
and scalar multiplications are def\/ined by:
\begin{gather*}
(m,a)+_i (n,b) = (m + n, a +_i  b ) , \qquad r ._i\, (m,a) = (m,r ._i\,
a).
\end{gather*}

The two compositions are def\/ined by:
\begin{gather*}
(m,a) \circ_i  (n,b) =
 \begin{cases} ( m^{b_4}     +    {}^{a_2}n,  a \circ_i b)&  \text{if } i  = 1, \\
 (m ^{b_2}  + {} ^{a_1}  n, a \circ_i  b)&  \text{if } i  = 2.
 \end{cases}
\end{gather*}

Furthermore, there exists a connection pair
for the underlying double category of $D$  given
by $\Gamma  a =(0,A)$, $\Gamma ' a =(0,A')$, where
\begin{gather*}
A = \quadr{a}{a}{1}{1} , \qquad A'  = \quadr{1}{1}{a}{a}.
\end{gather*}

\begin{proposition}\label{proposition9.1}
The above construction defines a functor  $\zeta $
from the category $\mathbf{CM}$ of crossed modules
to the category of double algebroids with connection
pair $\Gamma  a =(0,A)$, $\Gamma ' a =(0,A')$ on the underlying double category,
where the connection pair satisfies the following
additional properties.

Suppose $u,v,w \in D$  have boundaries \[
 \quadr{c}{a}{d}{b},\qquad
\quadr{c}{e}{f}{b},\qquad \quadr{g}{a}{d}{h}
\] respectively , and $r \in
R$. Then, the following equations must hold:
\begin{itemize}\itemsep=0pt
\item [$i)$]
$\Gamma (a + e)  \circ_2 (u +_1 v )\circ_2 \Gamma'(d+f) =
(\Gamma'a \circ_2 u \circ_2 \Gamma d) +_2 (\Gamma ' e \circ_2 v
\circ_2 \Gamma f)$;

\item[$ii)$]
$\Gamma ' (c + g) \circ_1 (u +_2 w) \circ_1 \Gamma(d + f) =
(\Gamma '\circ_2 u \circ_2 \Gamma b) +_1 (\Gamma'g \circ_2 w
\circ_2  \Gamma h)$;

\item[$iii)$]
$ \Gamma ' (ra) \circ_2  (r ._1 u) \circ_2 \Gamma(rd) = r ._2
(\Gamma ' a \circ_2  u \circ_2  \Gamma d)$;

\item[$iv)$] $\Gamma ' (rc) \circ_1 (r ._2 u) \circ_1 \Gamma(rd) = r ._2
(\Gamma ' c  \circ_1 u \circ_1\Gamma b)$.
\end{itemize}
\end{proposition}

\begin{remark}\label{remark9.1}
Edge symmetric double algebroids were also def\/ined
in~\cite{Brown-Mosa86}, and it was shown that there exist
two functors which respectively associate to a double algebroid its
corresponding horizontal and vertical crossed modules.

 Then, one obtains the result that the categories of crossed
modules ``of algebroids'' and of edge symmetric double algebroids with
connection are \textit{equivalent}. As a corollary, one can
also show that the two algebroids structures in dimension $2$
of a special type of double algebroids are isomorphic, as
are their associated crossed modules \cite{Brown-Mosa86}.
This result will be precisely shown in the next subsection.
\end{remark}

\subsection[The equivalence between the category of crossed modules of algebroids and
the category\\ of algebroids with a connection pair. Quantum algebroid symmetries]{The equivalence between the category of crossed modules\\ of algebroids and
the category of algebroids with a connection pair.\\ Quantum algebroid symmetries}\label{section9.3}

 In order to obtain an equivalence of categories one needs to add
in the extra structure of a~connection pair to a double algebroid.

Let ${D}$ be a double algebroid. A \emph{connection pair} for ${D}$
is a pair of functions $({\GG}, {\GG}') : {D}_1 \to {D}$ which is a
connection pair for the underlying double category of $\mathbf{D}$
as in \cite{Brown-Mosa86,Mosa86}. Thus one has the following
properties:
\begin{itemize}\itemsep=0pt
\item[$i)$]
If $a : x \lra y$  in $D_1$, then $(\GG a, \GG' a)$ have boundaries
given respectively by
\begin{gather*}
\quadr{a}{a}{1_y}{1_y}    , \qquad \quadr{1_x}{1_x}{a}{a}
\end{gather*}
and
\begin{gather*}
\Gamma' a  \circ _2  \Gamma a = \vep_1a  , \qquad \Gamma' a \circ _1 \Gamma
a = \vep _2 a.
\end{gather*}

\item[$ii)$]
If $x \in D_0$, then $\Gamma 1_x = \Gamma'1_x = \vep_1^2 x $.

\item[$iii)$]
The transport laws: if $a\circ b$ is def\/ined in $D_1$ then
\begin{gather*}
\Gamma (a \circ b) = \begin{bmatrix} \Gamma & \vep_1 b \\ \vep_2b &
\Gamma b \end{bmatrix},\qquad
  \Gamma '(a \circ b) = \begin{bmatrix} \Gamma ' a & \vep _2 a \\  \vep_1a & \Gamma' b
  \end{bmatrix}.
\end{gather*}
\end{itemize}

One then def\/ines `folding' operations on elements of $D$ for a
double algebroid with connection. Let $u \in D$ have boundary
$\quadr{c}{a}{d}{b}$. One f\/irst sets
\begin{gather*}
\psi  u  =  { \GG '}a \circ _2  u \circ _2  {\GG} d
\end{gather*}
as in \cite{Brown-Mosa86,Mosa86}.

\begin{proposition}\label{proposition9.2} \qquad
\begin{enumerate}\itemsep=0pt

\item[$i)$]  Let $u$, $v$, $w$ be such that $u\circ _1v$, $u\circ _2 w$ are defined.
Then
\begin{gather*}
\psi ( u \circ _1  v ) = (\psi u \circ _2  \vep_1 \del^1_2
v) \circ _1 (\vep_1 \del ^0_2 u \circ _2  \psi v) , \\
 \psi  (u
\circ _2 w)  = (\vep_1\del^0_1 u \circ _2  \psi w) \circ _1  (\psi
u \circ _2 \vep_1 \del^0_1 w) .
\end{gather*}

\item[$ii)$] If $a \in D_1$ then $ \psi  a  =  \psi  'a = \vep_1
a $.

\item[$iii)$] $ \psi u  = u $  if and only if $\del ^0_ 2 u$, $\del
^1_ 2 u$ are identities.

\item[$iv)$] Let $u$, $v$, $w$ be such that $u+_1v$, $u+_2 w$ are defined. Then
\[
 \psi ( u +_1  v )= \psi u +_2 \psi v, \qquad \psi (u+_2 w) =
\psi u +_2 \psi w .
\]

\item[$v)$] If $r \in R $ and $ u \in D_2$ then $\psi (r._i\, u)
 = r._i\, \psi u$ for $ i=1,2$.
 \end{enumerate}
\end{proposition}

One def\/ines an operation $\phi  : D \to D$  by
\[
 \phi u = \psi u
-_2  \vep_1 \del^1_1 \psi  u .
\] Let $M$ be the set of elements $u
\in D$ such that the boundary $\del u$ is of the form
$\quadr{a}{1}{1}{0}$. We write $0$ for an element of $M$ of the
form $\vep_1 0_{xy}$  where $0_{xy}$ is  the algebroid zero of
$D_1(x,y)$.

\begin{proposition}\label{proposition9.3}
The operation $\phi$ has the following properties:
\begin{itemize}\itemsep=0pt
\item[$i)$] if $\del u = \quadr{c}{a}{d}{b}$ then $\del \phi u = \quadr{cd -
ab}{1}{1}{0}$;

\item[$ii)$] $\phi u = u$ if and only if $u \in M$; in particular $\phi \phi =
\phi$;

\item[$iii)$] If $a \in D_1$ then $ \phi \gamma a = \phi \gamma ' a = \phi
\vep_i a =0$;

\item[$iv)$] $\phi(u+_i v) = \phi u +_2 \phi v$;

\item[$v)$] $ \phi (r._i\, u)= r._2(\phi u)$;

\item[$vi)$] $\phi (u \circ _1 v) = (\phi u)^{\del^1_1 v} +_2 {}^{\del^0_1
u}(\phi v)$;

\item[$vii)$] $ \phi (u \circ _2 w) = (\phi u)^{\del^1_1 w} +_2 {}^{\del^0_1
u}(\phi w)$.
\end{itemize}
\end{proposition}

With the above constructions one can then prove the following, major result of \cite{Brown-Mosa86}.

\begin{theorem}\label{theorem9.1}
The categories of crossed modules of algebroids and of double
algebroids with a~connection pair are equivalent. Let $\mathbf{DA}$
denote the category of double algebroids with connection. One has
the functors defined above:
\begin{gather*}
\eta : \ \mathbf{DA} \lra \mathbf{CM}, \qquad \text{and} \qquad \xi : \
\mathbf{CM} \lra \mathbf{DA}.
\end{gather*}
The functors $\eta \xi$ and $\xi \eta$ are then each naturally
equivalent to the corresponding identity functor for categories
$\mathbf{DA}$ and $\mathbf{CM}$, respectively.
\end{theorem}

\begin{remark}\label{remark9.2}
A word of caution is here in order about the equivalence of
categories in general: the equivalence relation may have more than
one meaning, that is however always a \textit{global} pro\-per\-ty;
thus, categories that are equivalent may still exhibit
substantially dif\/ferent and signif\/icant \textit{local} properties,
as for example in the case of equivalent categories of
semantically distinct, $n$-valued logic algebras \cite{Georgescu2k6}.
\end{remark}

\begin{remark}\label{remark9.3}
The above theorem also has a signif\/icant impact on physical
applications of double algebroid representations with a connection
pair to extend quantum symmetries in the presence of intense
gravitational f\/ields because it allows one to work out such
\textit{higher dimensional representations} in terms of those of
crossed modules of (lower-dimension) algebroids, as for example in
the cases of: Lie, weak Hopf, Grassman--Hopf algebroids or `Lie'
superalgebroids relevant to Quantum Gravity symmetries of intense
gravitational f\/ields and `singularities' that were introduced and
discussed in previous sections.
\end{remark}

The main result of Brown and Mosa \cite{Brown-Mosa86} is now stated as follows.

\begin{theorem}[Brown and Mosa \cite{Brown-Mosa86}]\label{theorem9.2}
The category of crossed modules of $R$-algebroids is equivalent
to the category of double $R$-algebroids with thin structure.
\end{theorem}

 Let  $\mu  : M \lra A$ be a crossed module. Applying $\eta \xi$ to
this yields a crossed module $\nu : N \to B$ say. Then $B = A$ and
N consists of pairs $(m,a)$ where $a=\quadr{\mu m}{1}{1}{0}$ for
all $m \in M$. Clearly these two crossed modules are naturally
isomorphic.

One can now use the category equivalences in the two theorems above
to also prove that \textit{in a~double algebroid with connection}
\textit{the two algebroid structures in dimension two are
isomorphic}. This is accomplished by def\/ining a \textit{reflection}
which is analogous to the rotation employed in~\cite{Brown-Spencer76a} for double groupoids and to the ref\/lection
utilized for those double categories that arise from 2-categories
with invertible 2-cells.

Let $D$ be a double algebroid with a connection pair. One def\/ines
$\rho  : D \lra D$  by the formula in which  assuming $\del  u =
\quadr{c}{a}{d}{b}$
\begin{gather*}
\rho  u = \theta \quadr{a}{1}{b}{ab} \circ _1 (\vep_1 (cd) -_2
\phi  u) \circ _1 \theta \quadr{cd}{c}{1}{d}  .
\end{gather*}

Thus, one has that: $\del \rho  u =\quadr{a}{c}{d}{b}$, and also
the following result of Brown and Mosa~\cite{Brown-Mosa86}:

\begin{theorem} \label{theorem9.3}
The reflection $\rho$ satisfies:
\begin{enumerate}\itemsep=0pt
\item[$i)$] $\rho (a) =  a$, $\rho ('a) = 'a$,  $\rho (\vep_1
a) = \vep_1 a $, $\rho (\vep_2 a) = \vep_2 a$ for  all $a \in
D_1$;

\item[$ii)$] $\rho (u +_1 v) = \rho u +_2 \rho v$, $\rho (w +_2  x ) = \rho w +_1 \rho x$
 whenever $u +_1 v , w +_2 x$ are defined;

\item[$iii)$] $\rho (u \circ _1  v) = \rho u \circ _2 \rho v$, $\rho (w
\circ _2 x) = \rho w \circ _1  \rho x$ $u \circ _1  v$, $w \circ _2
x$ are defined;

\item[$iv)$] $\rho (r ._1 u) = r ._2 \rho u$, $\rho (r ._2 u) = r ._1
\rho u$  where $r \in R$.
\end{enumerate}
\end{theorem}

\begin{remark}\label{remark9.4}
The ref\/lection concept presented above represents the key
\textit{internal symmetries} of double algebroids with connection
pair, and there are also similar concepts for other higher
dimensional structures such as double groupoids, double categories,
and so on.  Therefore, one can reasonably expect that such
ref\/lection notions may also be applicable to all `quantum doub\-les',
including quantum double groupoids and higher dimensional quantum
symmetries that are expected, or predicted, to occur in quantum
chromodynamics, and \textit{via }`Lie' superalgebroids, also in
quantum gravity based on lc-GR theories as proposed in subsequent
sections.
\end{remark}

\subsection{Double groupoids}\label{section9.4}

We can take further advantage of the above procedures by reconsidering the earlier,
double groupoid case \cite{Brown-Spencer76a} in relationship to a $C^*$-convolution
algebroid that links both `horizontal' and `vertical' structures in
an internally consistent manner. The geometry of squares and their
compositions leads to a common representation of a \emph{double
groupoid} in the following form:
\begin{gather*}
 \D= \vcenter{\xymatrix @=3pc {S \ar @<1ex> [r] ^{s^1} \ar @<-1ex> [r]
_{t^1} \ar @<1ex> [d]^{\, t_2}  \ar @<-1ex> [d]_{s_2} & H   \ar[l]
\ar @<1ex> [d]^{\,t}
 \ar @<-1ex> [d]_s \\
V \ar [u]  \ar @<1ex> [r] ^s \ar @<-1ex> [r] _t & M \ar [l] \ar[u]}}
\end{gather*}
where $M$ is a set of `points', $H$, $V$ are `horizontal' and
`vertical' groupoids, and $S$ is a set of `squares' with two
compositions. The laws for a  double groupoid make it also
describable as a groupoid internal to the category of groupoids.
Furthermore, because in a groupoid, any composition of commutative
squares is also commutative, several groupoid square diagrams of the
type shown above can be composed to yield larger square diagrams
that are naturally commutative.

Given two groupoids $H$, $V$  over a set $M$, there is a double
groupoid $\Box(H,V)$ with $H$, $V$ as horizontal and vertical edge
groupoids, and squares given by
 quadruples
$\begin{pmatrix} & h& \\[-0.9ex] v & & v'\\[-0.9ex]& h'&
\end{pmatrix}$
for which we assume always that $h,h' \in H$, $v,v' \in V$ and
that the initial and f\/inal points of these edges match in $M$ as
suggested by the notation, that is for example $sh=sv$, $th=sv'$,
$\ldots$, etc. The compositions are to be inherited from those of
$H$, $V$,
 that is:
\begin{gather*}
\quadr{h}{v}{v'}{h'} \circ_1\quadr{h'}{w}{w'}{h''}
=\quadr{h}{vw}{v'w'}{h''}, \\
\quadr{h}{v}{v'}{h'}
\circ_2\quadr{k}{v'}{v''}{k'}=\quadr{hk}{v}{v''}{h'k'}  .
\end{gather*}

This construction is def\/ined by the right adjoint~$R$ to the forgetful functor~$L$ which takes the
double groupoid as above, to the pair of groupoids $(H,V)$ over~$M$.
Furthermore, this right adjoint functor can be utilized to relate
\textit{double groupoid representations} to the corresponding pairs
of groupoid representations induced by~$L$. Thus, one can
obtain a functorial construction of certain double groupoid
representations from those of the groupoid pairs $(H,V)$ over~$M$.
Further uses of adjointness to classifying groupoid representations
related to extended quantum symmetries can also be made through the
generalized Galois theory presented in the next subsection;
therefore, Galois groupoids constructed with a pair of adjoint
functors and their representations may play a central role in such
future developments of the mathematical theory of groupoid
representations and their applications in quantum physics.

Given a general double groupoid as above, one can def\/ine
$S\quadr{h}{v}{v'}{h'}$ to be the set of squares with these as
horizontal and vertical edges
\begin{gather*}
\label{Rsqu} A\D= \vcenter{\xymatrix @=3pc {AS \ar @<1ex> [r] ^{s^1} \ar @<-1ex> [r]
_{t^1} \ar @<1ex> [d]^{\, t_2}  \ar @<-1ex> [d]_{s_2} & AH \ar[l]
\ar @<1ex> [d]^{\,t}
 \ar @<-1ex> [d]_s \\
AV \ar [u]  \ar @<1ex> [r] ^s \ar @<-1ex> [r] _t & M \ar [l]
\ar[u] }}
\end{gather*}
for which:
\begin{gather*}
AS\quadr{h}{v}{v'}{h'}
\end{gather*}
is the free $A$-module on the set of squares with the given
boundary. The two compositions are then bilinear in the obvious
sense.

Alternatively, one can use the convolution construction $\bar{A}\D$
induced by the convolution $C^*$-algebra over $H$ and $V$.  This
allows us to construct for at least a commutative $C^*$-algebra~$A$
a~double algebroid (i.e., a set with two algebroid structures), as
discussed in the previous  subsection. These novel ideas need
further development in the light of the algebra of crossed modules
of algebroids, developed in~\cite{Mosa86} and~\cite{Brown-Mosa86},
crossed cubes of $C^*$-algebras following~\cite{Ellis88}, as well
as crossed complexes of groupoids~\cite{Brown2k8}.

 The next, natural extension of this \emph{quantum algebroid}
approach to QFT generalized symmetries can now be formulated in
terms of \emph{graded Lie algebroids}, or supersymmetry algebroids,
for the supersymmetry-based theories of \emph{quantum gravity/supergravity} that were discussed in Section~\ref{section6}.

 We shall discuss f\/irst in the next subsection an interesting
categorical construction of a~homotopy double groupoid.

\subsection[The generalized Galois theory construction of a homotopy double groupoid \cite{Brown-Janelidze2k4}\\ and Galois groupoid representations]{The generalized Galois theory construction of a homotopy\\ double groupoid \cite{Brown-Janelidze2k4} and Galois groupoid representations}\label{section9.5}

 In two related papers Janelidze \cite{Janelidze90,Janelidze91} outlined a categorical approach to the Galois theory. In a more recent paper in~2004, Brown and Janelidze~\cite{Brown-Janelidze2k4} reported a \textit{homotopy double groupoid} construction of a surjective f\/ibration of Kan simplicial sets based on a~generalized, \textit{categorical Galois (GCG) theory} which under certain, well-def\/ined conditions gives a \textit{Galois groupoid} from a pair of adjoint functors. As an example, the standard fundamental group arises in GCG from an adjoint pair between topological spaces and sets. Such a homotopy double groupoid (HDG, explicitly given in diagram~1 of \cite{Brown-Janelidze2k4}) was also shown to contain the 2-groupoid associated to a~map def\/ined by Kamps and Porter~\cite{Kamps-Porter99}; this HDG includes therefore the 2-groupoid of a pair def\/ined by Moerdijk and Svenson \cite{Moerdijk-Svensson93}, the  ${\rm cat}^1$-group of a f\/ibration def\/ined by Loday~\cite{Loday82}, and also the classical fundamental crossed module of a pair of pointed spaces introduced by J.H.C.~Whitehead. Related aspects concerning homotopical excision, Hurewicz theorems for $n$-cubes of spaces and van Kampen theorems~\cite{van Kampen33} for diagrams of spaces were subsequently developed in \cite{Brown-Loday87a,Brown-Loday87b}.

 Two major advantages of this generalized Galois theory construction of HDG that were already pointed out are:
\begin{itemize}\itemsep=0pt
\item[(i)] the construction includes information on the map $q: M \rightarrow B$ of topological spaces, and
\item[(ii)] one obtains dif\/ferent results if the topology of M is varied to a f\/iner topology.
\end{itemize}

 Another advantage of such a categorical construction is the
possibility of investigating the global relationships among the
category of simplicial sets, $\mathbf{C}_{{S}} = \mathbf{Set}^{\Delta^{\rm op}}$,
the category of topological spaces, \textbf{Top}, and the category of groupoids,
\textbf{Grpd}. Let $I$ be the fundamental groupoid functor $ I  = \pi_1
: \mathbf{C}_{S} \rightarrow \textbf{X}$  from the category
$\mathbf{C}_{S}$ to the category $\textbf{X} = \textbf{Grpd}$ of
(small) groupoids. Consider next diagram~11 on page~67 of Brown and
Janelidze \cite{Brown-Janelidze2k4}:
\begin{gather}\label{(9.24)}
\xymatrix{\textbf{Top} \ar@<-0.5ex> [r]_-S &\ar @<-0.5ex>[l]_-R
\mathbf{Set}^{\Delta^{\rm op}} \ar @<0.5ex>[r] ^-I & \textbf{Grpd} \ar
@<0.5ex>[l]^-H\\ & \Delta  \ar [ul]^r \ar [u] _y \ar [ur]_i},
 \end{gather}
where:
 \begin{itemize}\itemsep=0pt
\item $\textbf{Top}$ is the category of topological spaces, $S$ is the singular complex functor and $R$ is its left-adjoint, called the geometric realisation functor;
\item $I\vdash  H$ is the adjoint pair introduced in Borceux and Janelidze \cite{Borceux-Janelidze2k1}, with $I$ being the fundamental groupoid functor, and $H$ being its unique right-adjoint \textit{nerve functor};
\item $y$ is the Yoneda embedding, with $ r$ and $i$ being, respectively, the restrictions of $R$ and $I$ respectively along~$y$; thus, $r$ is the singular simplex functor and $i$ carries f\/inite ordinals to codiscrete groupoids on the same sets of objects.
\end{itemize}

 The adjoint functors in the top row of the above diagram are uniquely determined by $r$ and~$i$~-- up to isomorphisms~-- as a result of the universal property of~$y$, the Yoneda embedding construction. Furthermore, one notes that there is a natural completion to a square, commutative diagram of the double triangle diagram~\eqref{(9.24)} reproduced above by three adjoint functors of the corresponding forgetful functors related to the Yoneda embedding. This natural diagram completion, that may appear trivial at f\/irst, leads however to the following Lemma and related Propositions.

\begin{lemma}\label{lemma9.1}
The following diagram \eqref{(9.25)} is commutative and there exist
canonical natural equivalences between the compositions of the
adjoint functor pairs and their corresponding identity functors of
the four categories presented in diagram \eqref{(9.25)}:
\begin{gather}\label{(9.25)}\xymatrix @=2pc{\mathbf{Top}
 \ar @<1.5ex> [r]^{R} \ar @<-1.5ex> [d]_f & \ar @<1.5ex> [d]^I
  \;\; \mathbf{Set}^{\Delta^{\rm op}}  \ar[l]^S  \\
\mathbf{Set} \ar [u]_g  \ar @<1.5ex> [r]^G    & \mathbf{Grpd }\ar
[l]^F \ar[u]^H } \quad \xymatrix@=1.5pc{& \\& \;\\\;&}
\end{gather}

The forgetful functors $f: \mathbf{Top}\lra
\mathbf{Set}$,  $F: \mathbf{Grpd} \lra \mathbf{Set}$ and $\Phi:
\mathbf{Set}^{\Delta^{\rm op}} \lra \mathbf{Set}$ complete this
commutative diagram of adjoint functor pairs. The right adjoint of
$\Phi$ is denoted by  $\Phi*$, and the adjunction pair $[\Phi,
\Phi*]$ has a mirror-like pair of adjoint functors between
$\mathbf{Top}$ and $\mathbf{Grpd}$ when the latter is restricted
to its subcategory $\mathbf{TGrpd}$ of topological groupoids, and
also when $\phi: \mathbf{TGrpd} \lra \mathbf{Top}$ is a functor
that forgets the algebraic structure~-- but not the underlying
topological structure of topological groupoids, which is fully and
faithfully carried over to $\mathbf{Top}$ by~$\phi$.
\end{lemma}

\begin{remark}\label{remark9.5}
Diagram~\eqref{(9.25)} of adjoint functor pairs can be further expanded
by adding to it the category of groups, \textbf{Gr}, and by def\/ining
a `forgetful' functor $\psi:\mathbf{Grpd} \lra \mathbf{Gr}$ that
assigns to each groupoid the product of its `component' groups, thus
ignoring the connecting, internal groupoid morphisms. The
categorical generalization of the Galois theory for groups can be
then related to the adjoint functor $\psi*$ and its pair. As a
simple example of the groupoid forgetful functor consider the
mapping of an extended symmetry groupoid, $\grp_S$, onto the group
product $\U(1)\times \SU(2)\times \SU(3)$ that `forgets' the global
symmetry of~$\grp_S$ and retains only the $\U(1)$, $\SU(2)$ and
$\SU(3)$ symmetries of the `standard model' in physics; here both
the groups and $\grp_S$ are considered, respectively as special,
small categories with one or many objects, and isomorphisms.
\end{remark}


\begin{proposition}\label{proposition9.4}
If  $T: \mathbf{C} \lra \mathbf{Grpd}$ is any groupoid valued
functor then $T$ is naturally equivalent to a functor $\Theta :
\mathbf{C}\lra \mathbf{Grpd}$ which is univalent with respect to
objects.
\end{proposition}

The proof is immediate by taking f\/irst into account the Lemma~\ref{lemma9.1} and diagram~\eqref{(9.25)}, and then by following
the logical proof sequence for the corresponding group category Proposition~10.4 of Mit\-chell~\cite{Mitchell65}. Note that `univalent' is also here employed in the sense of Mit\-chell~\cite{Mitchell65}.

This new proposition for groupoid valued functors can be thus considered as a natural extension of the corresponding theorem for group valued functors.

\begin{remark}\label{remark9.6}
The class of natural equivalences of the type $T
\rightarrow \Theta$ satisfying the conditions in Proposition~\ref{proposition9.4}
is itself a (large) 2-groupoid whose objects are groupoid valued functors. In particular, when
$\mathbf{C} = \mathbf{Top}$ and ${T}$ is the composite functor
${I}\circ {S} : \mathbf{Top}\lra \mathbf{Set}^{\Delta^{\rm op}}\lra
\mathbf{Grpd}$, then one obtains the interesting result (a) that
the class of functors naturally equivalent with ${T }$ becomes the
(large) $2$-groupoid of classical (geometric) \textit{fundamental
groupoid functors} $\pi_1$~\cite{Brown-Janelidze2k4}. Moreover,
according also to Proposition~2.1 of \cite{Brown-Janelidze2k4},
one has that:
\begin{itemize}\itemsep=0pt
\item [(a)] \textit{for every topological space ${X}$, $S({X})$ is a Kan complex, and}
\item [(b)] \textit{the $S$-image of a morphism $p$ in $\mathbf{Top}$ is a Kan
fibration if and only if $p$ itself is a Serre fibration}.
\end{itemize}

(For further details the reader is referred to \cite{Brown-Janelidze2k4}.)
\end{remark}

\subsection{Functor representations of topological groupoids}\label{section9.6}

A representable functor ${S}: \mathbf{C} \lra \mathbf{Set}$ as
def\/ined in Section~\ref{section7.8} is also determined by the equivalent
condition that there exists an object $X$ in $\mathbf{C}$ so that
${S}$ is isomorphic to the $\Hom$-functor~${h}^X$. In the dual,
categorical representation, the $\Hom$-functor $h^X$ is simply
replaced by~$h_X$. As an immediate consequence of the
Yoneda--Grothendieck lemma the set of natural equivalences between~$S$ and $h^X$ (or alternatively ${h}_X$)~-- which has in fact a
groupoid structure~-- is isomorphic with the object ${S}(X)$. Thus,
one may say that if ${S}$ is a representable functor then ${S}(X)$
is its (isomorphic) \textit{representation object}, which is also
unique up to an isomorphism~\cite[p.~99]{Mitchell65}. As an
especially relevant example we consider here the
\textit{topological groupoid representation} as a functor $\gamma
: \mathbf{TGrpd} \lra \mathbf{Set}$, and related to it, the more
restrictive def\/inition of $\gamma: \mathbf{TGrpd} \lra
\mathbf{BHilb}$, where $\mathbf{BHilb}$ can be selected either as
the category of Hilbert bundles or as the category of rigged
Hilbert spaces generated through the GNS construction as specif\/ied
in Def\/inition~\ref{definition5.1} and related equations~\eqref{(5.1)} and~\eqref{(5.2)}.
\begin{gather}\label{(9.26)}
\xymatrix{\textbf{Top} \ar@<-0.5ex> [r]_-M &\ar @<-0.5ex>[l]_-L
\mathbf{BHilb}\ar @<0.5ex>[r] ^-J & \mathbf{TGrpd} \ar
@<0.5ex>[l]^-K\\ &\mathbf{Set}\ar [ul]^g \ar [u] _m \ar [ur]_n}
 \end{gather}

Considering the forgetful functors $f$ and $F$ as def\/ined above, one
has their respective adjoint functors def\/ined by $\textit{g}$ and
$\textit{n}$ in diagram~\eqref{(9.26)}; this construction also leads to a
diagram of adjoint functor pairs similar to the ones shown in
diagram \eqref{(9.25)}. The functor and natural equivalence properties
stated in Lemma~\ref{lemma9.1} also apply to diagram~\eqref{(9.26)} with the
exception of those related to the adjoint pair $[\Phi, \Phi*]$
that are replaced by an adjoint pair $[\Psi,\Psi*]$, with $\Psi :
\mathbf{BHilb} \lra \mathbf{Set}$ being the forgetful functor and
$\Psi*$ its left adjoint functor. With this construction one
obtains the following proposition as a specif\/ic realization of
Proposition~\ref{proposition9.4} adapted to topological groupoids and
rigged Hilbert spaces:

\begin{proposition}\label{proposition9.5}
If $ \R_o : \mathbf{BHilb} \lra  \mathbf{TGrpd}$ is any
topological groupoid valued functor then $\R_o $ is naturally
equivalent to a functor $\rho : \mathbf{BHilb} \lra  \mathbf{TGrpd}$
which is univalent with respect to objects.
\end{proposition}

\begin{remark}\label{remark9.7}
$ \R_o$ and $ \rho$ can be considered, respectively, as adjoint Hilbert-functor representations to groupoid, and respectively, topological groupoid functor representations $\R_o^*$ and $\rho^*$ in the category $\mathbf{BHilb}$ of rigged Hilbert spaces.
\end{remark}

\begin{remark}\label{remark9.8}
The connections of the latter result for groupoid representations
on rigged Hilbert spaces to the weak $C^*$-Hopf symmetry
associated with quantum groupoids and to the generali\-zed
categorical Galois theory warrant further investigation in
relation to quantum systems with extended symmetry. Thus, the
following corollary and the previous Proposition~\ref{proposition9.4} suggest
se\-ve\-ral possible applications of GCG theory to extended quantum
symmetries \textit{via} Galois groupoid representations in the
category of rigged Hilbert families of quantum spaces that involve
interesting adjoint situations and also natural equivalences
between such \textit{functor} representations. Then, considering
the def\/inition of quantum groupoids as \textit{locally compact}
(topological) groupoids with certain extended (quantum)
symmetries, their functor representations also have the unique
properties specif\/ied in Proposition~\ref{proposition9.4} and Corollary~\ref{corollary9.1}, as well
as the unique adjointness and natural properties illustrated in
diagram~\eqref{(9.26)}.
\end{remark}

\begin{corollary}\label{corollary9.1}
The composite functor $\Psi \circ \R_o :\mathbf{TGrpd}\lra
\mathbf{BHilb}\lra \mathbf{Set}$, has the left adjoint ${n}$ which
completes naturally diagram \eqref{(9.26)}, with both $\Psi :
\mathbf{BHilb}\lra \mathbf{Set}$ and $\Psi \circ \R_o$ being
forgetful functors. $\Psi$ also has a left adjoint $\Psi*$, and
$\R_o$ has a defined inverse, or duality functor $\Im$ which
assigns in an univalent manner a topological groupoid to a family
of rigged Hilbert spaces in $\mathbf{BHilb}$ that are specified
\textit{via} the GNS construction.
\end{corollary}

\begin{remark}\label{remark9.9}
The adjoint of the duality functor~-- which assigns in an univalent
manner a family of rigged Hilbert spaces in the category
\textbf{BHilb} (that are specif\/ied \textit{via }the GNS
construction) to a~topological groupoid~-- def\/ines a
\textit{Hilbert-functor adjoint representation} of topological
groupoids; the latter generalizes to dimension~\textbf{2} the
`standard' notion of (object) groupoid representations.  A similar
generalization to higher dimensions is also possible for algebroid
representations, for example by considering functor
representations from the category of double algebroids~\textbf{DA}
(or equivalently from~\textbf{CM}) to the category \textbf{BHilb}
of `rigged' Hilbert spaces.
\end{remark}

\begin{remark}\label{remark9.10}
For quantum state spaces and quantum operators the duality functor
$J= \Im : \mathbf{BHilb} \lra  \mathbf{TGrpd}$ is the
\textit{quantum fundamental groupoid (QFG) functor}, that plays a
similar role for a quantum state space bundle in the category
$\mathbf{BHilb}$ to that of the fundamental groupoid functor $I=
\pi_1 : \mathbf{C}_s \lra \mathbf{X}$ in diagram~\eqref{(9.24)} of the
generalized categorical Galois theory of~\cite{Brown-Janelidze2k4} (in the original, this is diagram (11) on page~67).
 The right adjoint $\R_o$ of the QFG functor $\Im $ thus provides a functor (or `categorical')
representation of \textit{topological} (in fact, locally compact)
quantum groupoids by rigged Hilbert spaces (or quantum Hilbert
space bundles) in a natural manner. Such rigged quantum Hilbert
spaces are the ones actually realized in quantum systems with
extended quantum symmetries described by quantum topological
groupoids and also represented by the QFG functor in the manner
prescribed by the functor $\pi_1$ (or~$I$) in the
generalized (categorical) Galois theory of Brown and Janelidze as shown in diagram~\eqref{(9.24)}.
\end{remark}

Let us consider next two diagrams that include, respectively, two adjoint situation quintets:
\begin{gather}
(\eta; J \circ M, L \circ K; \mathbf{Top, TGrpd}) , \qquad \text{and} \qquad
(\mu ; J \circ M, L \circ K; \mathbf{Top, TGrpd}),\nonumber
\\
 \xymatrix{\mathbf{Top} \ar@<-0.5ex> [r]_-{J \circ
M} &\ar @<-0.5ex>[l]_-{L \circ K} \mathbf{TGrpd} \\ & \mathbf{Set}\ar
[ul]^g \ar [u] _n }
\qquad
\label{(9.29)}
\xymatrix{\mathbf{Top} \ar@<-0.5ex> [r]_-{J \circ M} &\ar
@<-0.5ex>[l]_-{L \circ K} \mathbf{TGrpd} \\ & \mathbf{BHilb}\ar [ul]^L
\ar [u] _J }
\end{gather}
as well as their complete diagram of adjoint pairs:
\begin{gather}\label{(9.30)}
\xymatrix @=2pc{\textbf{Top} \ar @<1.5ex> [r]^{M~~}
\ar @<-1.5ex> [d]_f & \ar @<1.5ex> [d]^J  \;\; \textbf{BHilb } \ar[l]^L  \\
\textbf{Set} \ar [u]_g  \ar @<1.5ex> [r]^n    & \textbf{TGrpd }\ar
[l]^t \ar [u]^K } \quad \xymatrix@=1.5pc{& \\& \;\\\;&}
\end{gather}
where the two natural transformations (in fact, not necessarily
unique natural equivalences) involved in the adjoint situations
are def\/ined between the set-valued bifunctors \textit{via} the
families of mappings:
\[
\eta_{B,A}: [({L \circ K})(B), A] \lra [B,
({J \circ M})(A) ]
\]
  and
  \[
  \mu_{D,C}: [({L \circ K})(D), C]  \lra
[D, ({J \circ M})(C) ]
\]  with $A$, $C$ in ${\rm Obj}(\mathbf{Top})$ and
$B$, $D$ in ${\rm Obj}(\mathbf{TGrpd})$; further details and the notation
employed here are consistent with Chapter~V of~\cite{Mitchell65}.
Then, one obtains the following proposition as a direct consequence
of the above constructions and Proposition~4.1 on page~126 of \cite{Mitchell65}:

\begin{proposition} \label{proposition9.6} 
In the adjoint situations $(\eta; L \circ K, J \circ M ;
\mathbf{Top}, \mathbf{TGrpd})$ and  $(\mu ; L \circ K, J \circ M;$
$\mathbf{Top}, \mathbf{TGrpd})$  of categories and covariant functors
defined in diagrams 
 \eqref{(9.29)} and \eqref{(9.30)} there are
respectively one-to-one correspondences $\eta* : [L \circ K \circ
{n}, {g}] \lra [J, J \circ M \circ {g}]$ $($which is natural in both~$g$ and~$n)$, and $\mu* : [L \circ K \circ J, L ] \lra [ J, J
\circ M \circ  L]$ $($which is natural in both~$J$ and~$L)$.
\end{proposition}

\begin{remark}\label{remark9.11}
One readily notes that a similar adjointness result holds for
$\mathbf{TGrpd}$ and $\mathbf{BHilb}$ that involves naturality in
$\Im = J$, $L$ and $K$, $t$, respectively. Moreover, the functor representations in diagram \eqref{(9.30)} have adjoint functors that in the case of quantum systems link extended quantum symmetries to quantum operator algebra on rigged Hilbert spaces and the locally compact topology of quantum groupoids, assumed here to be endowed with suitable Haar measure systems. In the case of quantum double groupoids a suitable, but rather elaborate, def\/inition of a \textit{double system of Haar measures} can also be introduced (private communication to the f\/irst author from Professor M.R.~Buneci).
\end{remark}

\section{Conclusions and discussion}\label{section10}

 Extended quantum symmetries, recent quantum operator algebra (QOA) developments and also non-Abelian algebraic topology (NAAT) \cite{Brown-etal2k9} results were here discussed with a view to phy\-si\-cal applications in quantum f\/ield theories, general molecular and nuclear scattering theories, symmetry breaking, as well as supergravity/supersymmetry based on a \textit{locally covariant} approach to general relativity theories in quantum gravity.   Fundamental concepts of QOA and quantum algebraic topology (QAT), such as $C^*$-algebras, quantum groups, von Neumann/Hopf algebras, quantum supergroups, quantum groupoids, quantum groupoid/algebroid representations and so on, were here considered primarily with a view to their possible extensions and future applications in quantum f\/ield theories and beyond.

 Recently published mathematical generalizations that represent extended quantum symmetries range from quantum group algebras and quantum superalgebras/quantum supergroups to quantum groupoids, and then further, to quantum topological/Lie groupoids/Lie algebroids \cite{Blaom2k7,Bos2k7a,Bos2k7b} in dipole-dipole coupled quasi-particles/bosons in condensed matter (such as: paracrystals/noncrystalline materials/glasses/topologically ordered systems) and nuclear physics, as well as Hamiltonian algebroids and double algebroid/double groupoid/categorical representations in $W_N$-gravity and more general supergravity theories. We note that supersymmetry was also discussed previously within a dif\/ferent mathematical framework
\cite{Deligne-Morgan99}. Several, algebraically simpler, representations of quantum spacetime than QAT have thus been proposed in terms of causal sets, quantized causal sets, and quantum toposes \cite{Nishimura96,Raptis-Zapatrin2k,Butterfield-Isham2k,Butterfield-Isham2k4,HLS2k8}. However, the consistency of such `quantum toposes' with the real quantum logic is yet to be validated; the `quantum toposes' that have been proposed so far are all clearly inconsistent with the Birkhof\/f--von Neumann quantum logic (see for example, \cite{HLS2k8}). An alternative, generalized \L ukasiewicz topos (GLT) that may allow us avoid such major logical inconsistences with quantum logics has also been developed \cite{Baianu2k4a,Baianu2k4b,Baianu-etal2k6,Baianu-etal2k9a,Georgescu-Vraciu70,Georgescu2k6}.
 We have suggested here several new applications of Grassmann--Hopf algebras/algebroids, graded `Lie' algebroids, weak Hopf $C^*$-algebroids, quantum locally compact groupoids to interacting quasi-particle and many-particle quantum systems. These concepts lead to higher dimensional symmetries represented by double groupoids, as well as other higher dimensional algebraic topology structures \cite{Brown-Mosa86,Mosa86}; they also have potential applications to spacetime structure determination using higher dimensional algebra (HDA) tools and its powerful results to uncover universal, topological invariants of `hidden' quantum symmetries. New, non-Abelian results may thus be obtained through \textit{higher homotopy, generalized van Kampen theorems} \cite{Brown-etal2k2,Brown-Janelidze87}, Lie groupoids/algebroids and groupoid atlases, possibly with novel applications to quantum dynamics and local-to-global problems, as well as quantum logic algebras (QLA). Novel mathematical representations in the form of higher homotopy quantum f\/ield (HHQFT) and quantum non-Abelian algebraic topology (QNAT) theories have the potential to develop a self-consistent quantum-general relativity theory (QGRT) in the context of supersymmetry algebroids/supersymmetry/supergravity and metric superf\/ields in the Planck limit of spacetime \cite{Baianu-etal2k9a,Brown-etal2k7}. Especially interesting in QGRT are global representations of f\/luctuating spacetime structures in the presence of intense, f\/luctuating quantum gravitational f\/ields. The development of such mathematical representations of extended quantum symmetries and supersymmetry appears as a logical requirement for the unif\/ication of quantum f\/ield (and especially AQFT) with general relativity theories in QGRT \textit{via} quantum supergravity and NAAT approaches to determining supersymmetry invariants of quantum spacetime geometry.

  QNAT is also being applied to develop studies of \textit{non-Abelian quantum Hall} liquids and other many-body quantum systems with topological order \cite{Wen91,Blok-Wen92,Wen99,Wen2k4}.

 In a subsequent report \cite{Baianu-etal2k9b}, we shall further consider the development of physical applications of NAAT \cite{Brown-etal2k9} towards a quantum non-Abelian algebraic topology (QNAT) from the standpoints of the theory of categories-functors-natural equivalences, higher dimensional algebra, as well as quantum logics. This approach can also be further extended and applied to both quantum statistical mechanics and complex systems that exhibit broken symmetry and/or various degrees of topological order in both lower and higher dimensions.

\appendix

\section{Appendix}\label{appendixA}
\subsection{Von Neumann algebras}\label{appendixA1}

 Let $\H$ denote a complex (separable) Hilbert space. A \emph{von
Neumann algebra} $\A$ acting on $\H$ is a~subset of the algebra of
all bounded operators $\cL(\H)$ such that:
\begin{itemize}\itemsep=0pt
\item[(1)] $\A$ is closed under the adjoint operation (with the
adjoint of an element $T$ denoted by $T^*$).

\item[(2)]
$\A$ equals its bicommutant, namely:
\begin{gather*}
\A= \{A \in \cL(\H) : \forall \, B \in \cL(\H), \forall \, C\in \A, \
(BC=CB)\Rightarrow (AB=BA)\}.
\end{gather*}
\end{itemize}

If one calls a \emph{commutant} of a set $\A$ the special set of
bounded operators on $\cL(\H)$ which commute with all elements in
$\A$, then this second condition implies that the commutant of the
commutant of $\A$ is again the set~$\A$.

On the other hand, a von Neumann algebra $\A$ inherits a
\emph{unital} subalgebra from $\cL(\H)$, and according to the
f\/irst condition in its def\/inition $\A$ does indeed inherit a
\emph{$*$-subalgebra} structure, as further explained in the next
section on $C^*$-algebras. Furthermore, one has a notable
\emph{Bicommutant Theorem} which states that $\A$ \emph{is a von
Neumann algebra if and only if $\A$ is a $*$-subalgebra of
$\cL(\H)$, closed for the smallest topology defined by continuous
maps $(\xi,\eta)\longmapsto \left\langle A\xi,\eta\right\rangle$ for all~$\left\langle A\xi,\eta \right\rangle$, where $\left \langle\cdot,\cdot\right\rangle$
denotes the inner product defined on~$\H$}. For a well-presented
treatment of the geometry of the state spaces of quantum operator algebras, see e.g.~\cite{Alfsen-Schultz2k3}; the ring structure of operators in Hilbert spaces was considered in an early,
classic paper by Gel'fand and Naimark~\cite{Gelfand-Naimark43}.

\subsection{Groupoids}\label{appendixA2}

Recall that a \textit{groupoid} $\grp$ is a small category in which
all morphisms are invertible, and that has a set of objects $X = \ob(\grp)$.
Thus, a groupoid is a generalisation of a group, in the sense that it is
a \textit{generalized `group with many identities'}, this being possible because
its morphism composition~-- unlike that of a group~-- is, in general, only \textit{partially defined}
(as it is too in the case of abstract categories). One often writes
$\grp^y_x$ for the set of morphisms in $\grp$
from $x$ to $y$.

\subsubsection[Topological groupoid: def\/inition]{Topological groupoid: def\/inition}\label{appendixA2.1}

As is well kown, a \textit{topological groupoid} is just a groupoid
\textit{internal} to the category of topological spaces and continuous
maps.
Thus, a \emph{topological groupoid} consists of a space $\mathsf{G}$, a distinguished subspace $\mathsf{G}^{(0)} =
\obg \subset \mathsf{G}$, called {\it the space of objects} of
$\mathsf{G}$, together with maps
\begin{gather*}
r,s : \  \xymatrix{\mathsf{G} \ar@<1ex>[r]^r \ar[r]_s & \mathsf{G}^{(0)} }
\end{gather*}
called the {\it range} and {\it source maps} respectively,
together with a law of composition
\begin{gather*}
\circ : \ \mathsf{G}^{(2)}: = \mathsf{G} \times_{\mathsf{G}^{(0)}} \mathsf{G} = \{
 (\gamma_1, \gamma_2) \in \mathsf{G} \times \mathsf{G}  : \  s(\gamma_1) =
r(\gamma_2)  \} \  \lra  \ \mathsf{G} ,
\end{gather*}
such that the following hold:
\begin{enumerate}\itemsep=0pt
\item[(1)]
$s(\gamma_1 \circ \gamma_2) = r(\gamma_2)$, $r(\gamma_1 \circ
\gamma_2) = r(\gamma_1)$, for all $(\gamma_1, \gamma_2) \in
\mathsf{G}^{(2)}$.

\item[(2)]
$s(x) = r(x) = x$, for all $x \in \mathsf{G}^{(0)}$.

\item[(3)]
$\gamma \circ s(\gamma) = \gamma$, $r(\gamma) \circ \gamma =
\gamma$, for all $\gamma \in \mathsf{G}$.

\item[(4)]
$(\gamma_1 \circ \gamma_2) \circ \gamma_3 = \gamma_1 \circ
(\gamma_2 \circ \gamma_3)$.

\item[(5)]
Each $\gamma$ has a two-sided inverse $\gamma^{-1}$ with $\gamma
\gamma^{-1} = r(\gamma)$, $\gamma^{-1} \gamma = s (\gamma)$.
Furthermore, only for topological groupoids the inverse map needs be continuous.

It is usual to call $\mathsf{G}^{(0)} = \ob(\mathsf{G})$ {\it the
set of objects} of $\mathsf{G}$. For $u \in \ob(\mathsf{G})$, the
set of arrows $u \lra u$ forms a group $\mathsf{G_u}$, called the
\emph{isotropy group of $\mathsf{G}$ at $u$}.
\end{enumerate}

The notion of \textit{internal groupoid} has proved signif\/icant in a number of f\/ields, since groupoids generalise bundles of groups, group actions, and equivalence relations. For a further, detailed study of
groupoids and topology we refer the reader to the recent textbook by Brown \cite{Brown2k6}.

 Examples of groupoids are often encountered; the following are just a few specialized groupoid structures:
\begin{itemize}\itemsep=0pt
\item[(a)] locally compact groups,  transformation groups , and any group in general,
\item[(b)] equivalence relations,
\item[(c)] tangent bundles,
\item[(d)] the tangent groupoid,
\item[(e)] holonomy groupoids for foliations,
\item[(f)] Poisson groupoids, and
\item[(g)] graph groupoids.
\end{itemize}

 As a simple, helpful example of a groupoid, consider the case (b) above of a groupoid whose morphisms are def\/ined by the equivalence relation in an equivalence class or set. Thus, let $R$ be an \textit{equivalence relation} on a set~$X$. Then $R$ is a groupoid under the following operations:
$(x, y)(y, z) = (x, z)$, $(x, y)^{-1} = (y, x)$. Here, $\grp^0 = X
$, (the diagonal of $X \times X$) and $r((x, y)) = x$,  $s((x, y))
= y$.

 Thus, $ R^2$ = $\left\{((x, y), (y, z)) : (x, y), (y, z) \in R
\right\} $. When $R = X \times X $,  \textit{R} is called a
\textit{trivial} groupoid. A special case of a trivial groupoid is
$R = R_n = \left\{ 1, 2, \dots , n \right\} \times  \left\{
1, 2, \dots , n \right\} $. (So every $i$ is equivalent to
every $j$.) Identify $(i,j) \in R_n$ with the matrix unit
$e_{ij}$. Then the groupoid $R_n$ is just matrix multiplication
except that we only multiply $e_{ij}$, $e_{kl}$ when $k = j$, and
$(e_{ij} )^{-1} = e_{ji}$. We do not really lose anything by
restricting the multiplication, since the pairs $e_{ij}$, $e_{kl}$
excluded from groupoid multiplication just give the 0 product in
normal algebra anyway. For a groupoid~$\mathsf{G}_{lc}$ to be a
\textit{locally compact groupoid }means that $\mathsf{G}_{lc}$ is
required to be a (second countable) \textit{locally compact
Hausdorff} space, and the product and also inversion maps are
required to be \textit{continuous}. Each $\mathsf{G}_{lc}^u$ as
well as the unit space $\mathsf{G}_{lc}^0$ is closed in
$\mathsf{G}_{lc}$. What replaces the left Haar measure on~$\mathsf{G}_{lc}$ is a system of measures $\lambda^u$ ($u \in
\mathsf{G}_{lc}^0$), where $\lambda^u$ is a positive regular Borel
measure on~$\mathsf{G}_{lc}^u$ with dense support. In addition,
the $\lambda^u$ s are required to vary continuously (when
integrated against~$f \in C_c(\mathsf{G}_{lc})$ and to form an
invariant family in the sense that for each~$x$, the map $y \mapsto
xy$ is a measure preserving homeomorphism from
$\mathsf{G}_{lc}^s(x)$ onto~$\mathsf{G}_{lc}^r(x)$. Such a system
$\left\{ \lambda^u \right\}$ is called a \textit{left Haar system}
for the locally compact groupoid~$\mathsf{G}_{lc}$.

This is def\/ined more precisely next.

\subsection{Haar systems for locally compact topological groupoids}\label{appendixA3}

Let
\begin{gather*}
\xymatrix{ \mathsf{G}_{lc} \ar@<1ex>[r]^r \ar[r]_s & \mathsf{G}_{lc}^{(0)}}=X
\end{gather*}
be a locally compact, locally trivial topological groupoid with
its transposition into transitive (connected) components. Recall
that for $x \in X$, the \emph{costar of $x$} denoted
$\rm{CO}^*(x)$ is def\/ined as the closed set $\bigcup\{\mathsf{G}_{lc}(y,x) :
y \in \mathsf{G}_{lc} \}$, whereby
\begin{gather*}
\mathsf{G}_{lc}(x_0, y_0) \hookrightarrow \rm{CO}^*(x) \lra X,
\end{gather*}
is a principal $\mathsf{G}_{lc}(x_0, y_0)$-bundle relative to
f\/ixed base points $(x_0, y_0)$. Assuming all relevant sets are
locally compact, then following~\cite{Seda76}, a (\emph{left}) \emph{Haar
system on $\mathsf{G}_{lc}$} denoted $(\mathsf{G}_{lc}, \tau)$ (for later purposes), is def\/ined to comprise of i) a measure $\kappa$ on $\mathsf{G}_{lc}$, ii)~a~measure $\mu$ on $X$ and iii)~a~measure $\mu_x$ on ${\rm CO}^*(x)$
such that for every Baire set $E$ of $\mathsf{G}_{lc}$, the following hold on
setting $E_x = E \cap {\rm CO}^*(x)$:
\begin{itemize}\itemsep=0pt
\item[(1)] $x \mapsto \mu_x(E_x)$ is measurable;
\item[(2)] $\kappa(E) = \int_x \mu_x(E_x)\, d\mu_x$;
\item[(3)] $\mu_z(t E_x) = \mu_x(E_x)$, for all $t \in \mathsf{G}_{lc}(x,z)$ and $x, z
\in \mathsf{G}_{lc}$.
\end{itemize}

The presence of a left Haar system on $\mathsf{G}_{lc}$ has important
topological implications: it requires that the range map $r :
\mathsf{G}_{lc} \rightarrow \mathsf{G}_{lc}^0$ is open. For such a $\mathsf{G}_{lc}$ with a left Haar system, the vector space $C_c(\mathsf{G}_{lc})$ is a \textit{convolution} \textit{$*$-algebra}, where for $f, g \in
C_c(\mathsf{G}_{lc})$:
\[
f * g(x) = \int f(t)g(t^{-1} x) d \lambda^{r(x)} (t),  \qquad \mbox{with} \quad
f*(x)  = \overline{f(x^{-1})}.
\]

 One has $C^*(\mathsf{G}_{lc})$
to be the \textit{enveloping $C^*$-algebra} of $C_c(\mathsf{G}_{lc})$
(and also representations are required to be continuous in the
inductive limit topology). Equivalently, it is the completion of
$\pi_{\rm univ}(C_c(\mathsf{G}_{lc}))$ where $\pi_{\rm univ}$ is the
\textit{universal representation} of $\mathsf{G}_{lc}$. For example,
if $ \mathsf{G}_{lc} = R_n$, then $C^*(\mathsf{G}_{lc})$ is just
the f\/inite dimensional algebra $C_c(\mathsf{G}_{lc}) = M_n$, the
span of the $e_{ij}$'s.

There exists (viz.~\cite[pp.~91--92]{Paterson99}) a \textit{measurable Hilbert bundle}
$(\mathsf{G}_{lc}^0, \H, \mu)$ with $\H  = \big\{ \H^u_{u \in
\mathsf{G}_{lc}^0} \big\}$ and a G-representation L on $\H$ (see also \cite{Paterson2k3a,Paterson2k3b}).  Then, for every pair $\xi$, $\eta$ of square integrable sections of $\H$,
it is required that the function $x \mapsto (L(x)\xi (s(x)), \eta
(r(x)))$ be $\nu$-measurable. The representation $\Phi$ of
$C_c(\mathsf{G}_{lc})$ is then given by:
\[
\left\langle \Phi(f) \xi
\vert,\eta \right\rangle = \int f(x)(L(x) \xi (s(x)), \eta (r(x)))
d \nu_0(x).
\]

The triple $(\mu, \H, L)$ is called a \textit{measurable
$\mathsf{G}_{lc}$-Hilbert bundle}.

\subsection*{Acknowledgements}

Thanks are due to Dr.~M.A.~Buneci for several helpful comments and suggestions in the f\/inal preparation stages of the manuscript.

\addcontentsline{toc}{section}{References}
\LastPageEnding

\end{document}